\documentclass[12pt]{iopart}

\usepackage{iopams}
\usepackage{float}
\usepackage{amssymb}
\usepackage{graphicx}
\usepackage{bm}

\newcommand{\boldphi}{\mbox{\boldmath$\phi$}}
\newcommand{\bA}{\mbox{\boldmath$A$}}

\renewcommand{\Re}{{\rm Re}}
\renewcommand{\Im}{{\rm Im}}

\begin{document}

\title{Spectral Theory of  Sparse Non-Hermitian Random Matrices}

\author{Fernando Lucas Metz$^{1,2}$, Izaak Neri$^{3,4}$, Tim Rogers$^5$}
\address{$^1$Instituto de F\'{\i}sica, Universidade Federal do Rio Grande do Sul, Caixa Postal 15051, 91501-970 Porto Alegre, Brazil \\
$^2$ London Mathematical Laboratory, 8 Margravine Gardens, London, UK, W68RH \\
$^3$ Department of Mathematics, King's College London, Strand, London, UK, WC2R 2LS\\
$^4$ Centre for Networks and Collective Behaviour, Department of Mathematical Sciences, University of Bath, Bath, UK, BA27AY}
\eads{$^1$\mailto{fmetzfmetz@gmail.com}, $^2$\mailto{izaak.neri@kcl.ac.uk}, $^3$\mailto{t.c.rogers@bath.ac.uk}}

\begin{abstract}
Sparse non-Hermitian random matrices arise in the study of disordered physical systems with asymmetric local interactions, and have applications ranging from neural networks to ecosystem dynamics. The spectral characteristics of these matrices provide crucial information on system stability and susceptibility, however, their study is greatly complicated by the twin challenges of a lack of symmetry and a sparse interaction structure. In this review we provide a concise and systematic introduction to the main tools and results in this field. We show how the spectra of sparse non-Hermitian matrices can be computed via an analogy with infinite dimensional operators obeying certain recursion relations. With reference to three illustrative examples --- adjacency matrices of regular oriented graphs, adjacency matrices of  oriented Erd\H{o}s-R\'{e}nyi graphs, and  adjacency matrices of  weighted oriented Erd\H{o}s-R\'{e}nyi graphs --- we demonstrate the use of these methods to obtain both analytic and numerical results for the spectrum, the spectral distribution,  the location of outlier eigenvalues, and the statistical properties of eigenvectors.   
\end{abstract}

\section{Introduction}
The story of random matrix theory is the story of the search for certainty and predictability in large, complex interacting systems. As any theoretical physicist or applied mathematician knows, when more variables are incorporated into a model, its dimension increases and the analysis of its behaviour can fast become intractable. This problem bit hard in the mid twentieth century as physicists sought to understand the properties of quantum many-body systems such as heavy nuclei. Several major innovations occurred in response, giving rise to influential new fields of study including random matrix theory \cite{Wigner55}.    
The key premise of random matrix theory is that the large-scale behaviour of a complex system should be governed by its symmetries and the statistical properties of its (many) parameters, and should not be sensitive to the precise detail of each interacting element. Freeman Dyson elegantly captured the spirit of the field when he described it as ``a new kind of statistical mechanics, in which we renounce  exact knowledge not of the state of a system but of the nature of the system itself'' \cite{Dyson1962}.

Spectral properties of  classical and  quantum many-body systems reveal important features of complex systems, such as     phases of  matter~\cite{nandkishore2015many} or the stability of  ecosystems~\cite{may1972will, Allesina2015,grilli2016,gibbs2018}.   It remains till today a challenge to determine spectral properties of nonintegrable many-body systems.   Most approaches are either limited to  small system sizes or to approximations that reduce the problem to a single-body one.       Random matrix theory provides an alternative avenue of  approach by    exploiting the randomness and complexity of these systems.  In this paper we present an overview of the research on sparse non-Hermitian random matrices, which are random matrix ensembles that describe the  spectra of non-equilibrium systems that interact through a complex network structure. Our exposition includes detailed discussion of the technical background of the topic, as well as the theoretical challenges posed and links to application areas. A reference table of core topics and key equations is inculded in Table \ref{toc}.

\begin{table}
\caption{\label{toc} Quick reference table for core topics and key equations. }
\begin{indented}
\item[]\begin{tabular}{ll}
\br
Topic& Reference \\
\mr
Basic definitions & Sections 1.1 \& 2.1\\
Resolvent formalism & Section 2.2; Equations (16), (21) \& (26)\\
Hermitisation method & Section 2.4; Equations (33), (36) \& (40)\\
Recursion equations for the spectum & Section 3.2; Equations (56) - (67) \\
Recursion equations for outliers & Section 3.3; Equations (74) - (77)\\
\br
\end{tabular}
\end{indented}
\end{table}

A random matrix $\mathbf{A}$ of size $n$ is a square array of complex-valued random variables $A_{jk} = \left[\mathbf{A}\right]_{j,k}$, with $j,k=1,\ldots,n$. In applications, the entries of $\mathbf{A}$  represent the strength and sign of pair-wise interactions between the $n$ elements of a large complex system. Depending on the context, various properties of this system are encoded by the eigenvalues  $\lambda_j$ of $\mathbf{A}$  and their corresponding left $\langle u_{j}|$ and right $|v_{j} \rangle$ eigenvectors, defined by the relationships
\begin{equation}
\langle u_{j}|  \mathbf{A} = \lambda_j \langle u_j|\,,\quad \mathbf{A} |v_j\rangle = \lambda_j |v_j \rangle\,,\quad j=1,\ldots,n.
\end{equation}
The eigenvalues can be found as the $n$ roots of the characteristic polynomial $p(z)={\rm det}\left(\mathbf{A} - z \mathbf{1}_n\right)$, where $\mathbf{1}_n$ is the $n\times n$ identity matrix.  
It is a common convention to label the eigenvalues in order of size, so that $|\lambda_1|\geq\ldots\geq |\lambda_n|$. Eigenvectors are defined up to multiplication by scalar constants.  We adopt here the  biorthogonal convention   $\langle u_j|v_k \rangle = \delta_{j,k}$ which specifies half of the scalar constants; the remaining constants will  be specified when required.      Broadly speaking, random matrix theory encompasses the study of the statistical properties of the eigenvalues and  eigenvectors of random matrices $\mathbf{A}$, particularly in the large-size limit $n\to\infty$. General introductions to the field can be found in \cite{guhr1998random, mehta2004random, bai2008methodologies, tao2012topics, livan2018introduction}. 

Early work in random matrix theory concerned matrix ensembles with two special properties: Hermitian symmetry and all-to-all interactions (meaning that all degrees of freedom interact symmetrically with each other, similar to mean-field models in statistical physics).  A fundamental result of this theory is that in the limit of large matrix size various statistical properties of eigenvalues and eigenvectors are \emph{universal} in the sense that they are independent of the details of the matrix elements. A famous example in this context is the Wigner semicircle law \cite{Wigner55} which states that the eigenvalue distribution converges for $n \rightarrow \infty$ to a semicircle provided the moments of the matrix elements are finite. Thus, the details of the random interactions are to a large extent unimportant to the global statistical behaviour of the eigenvalues and eigenvectors in symmetric systems with all-to-all interactions.

In the mid 60's, soon after the introduction of the classical ensembles of Hermitian
random matrices \cite{mehta2004random}, Ginibre put forward the natural generalisation of the theory to non-Hermitian
random matrices simply by dropping any symmetry constraints \cite{Ginibre}. The Ginibre ensemble is composed of $n \times n$ random matrices in
which all elements are independently drawn from a common Gaussian distribution.
This ensemble also models complex systems with all-to-all interactions, but with the noticeable property that
the interaction between any two elements is {\it directed}: the influence of a certain degree of freedom $i$ on $j$ has a different
strength than the influence of $j$ on $i$. Non-Hermitian random matrices are as important as their Hermitian counterparts, and the Ginibre
ensemble of random matrices has been an useful tool to study the stability properties of large ecological
systems \cite{may1972will,Allesina2015,grilli2016,gibbs2018}, the spontaneous breaking of chiral symmetry in
quantum chromodynamics \cite{Stephanov1996,Halasz1997,Akemann2004}, disordered quantum systems
with a direction \cite{Efetov1997,Efetov1997a}, and
synchronisation of coupled oscillators \cite{Timme2002,Timme2004}.

An important technical advantage in the ``standard" theory of random matrices, which concerns
systems with all-to-all interactions and Gaussian statistics, lies in the analytical knowledge of the joint probability density of eigenvalues which can be
interpreted as the Boltzmann-Gibbs probability density of charges in a Coulomb gas \cite{Ginibre}. This property allows one to derive a wealth of quantitative information about the eigenvalue statistics, including the  spectral distribution and correlation functions between eigenvalues \cite{Forrester2007,Akemann2007, zabrodin2006large}. Building upon this fact, Ginibre has shown that, for $n \rightarrow \infty$, the eigenvalue distribution
$\rho(z)$ ($z \in \mathbb{C}$) of the Ginibre ensemble with complex Gaussian entries is uniform over the unit disc \cite{Ginibre}
\begin{equation}
  \rho(z) = \frac{1}{\pi} I_{[0,1]}(|z|) d z,
  \label{circ}
\end{equation}
provided the eigenvalues are rescaled as $\lambda_{j} \rightarrow \lambda_{j}/\sqrt{n}$. The indicator function $I_{\Omega}(x)$ in this expression
equals one if $x \in \Omega$ and zero otherwise. Equation (\ref{circ}) is the limiting behaviour of the empirical
spectral distribution for $n \rightarrow \infty$, and it is referred to as the {\it circular law}. Based on the analogy with the Wigner semicircle
law for Hermitian matrices, it was conjectured that Eq. (\ref{circ}) should be an universal result for
non-Hermitian random matrices. There have been several partial results towards validating this conjecture, see \cite{bai2008circular, Tao2008,Tao2009} for a brief survey of these. The most general form, proven only recently \cite{tao2010random}, states  that, if $A_{jk}$ are
independent complex random variables
with zero mean and unit variance, then the empirical spectral distribution
of the rescaled eigenvalues $\lambda_j/\sqrt{n}$  converges for $n \rightarrow \infty$,  both in probability and in the almost sure sense, to Eq. (\ref{circ}). This rigorous result attests the universality of the
circular law: the details of the distribution characterising the individual matrix elements
$A_{ij}$ are immaterial for the asymptotic form of the eigenvalue distribution of very
large random matrices. 

Relaxing the all-to-all assumption introduces {\it topological} disorder and leads to ensembles of {\it sparse} random matrices, whose
main defining property is the existence of an enormous number of zero matrix entries. More precisely, an $n \times n$ sparse
random matrix has an average total number of nonzero elements proportional to $n$.   In contrast to ``standard" random-matrix
ensembles, sparse matrices model complex systems where a given degree of freedom interacts with a small number of others. 
Such systems can be pictured in terms of random graphs \cite{BarratBook}, where the nodes of the graph represent
the individual entities composing
the system, while the links or edges connecting them denote their interactions. Typical real-world complex systems, such
as neural networks and ecosystems, contain an abundance of asymmetric or directed interactions. This sort of system is modelled using {\it sparse non-Hermitian random matrices}.

In the last twenty years, sparse non-Hermitian random matrices have proven to be as important as the  Ginibre ensemble.   The spectra of sparse random matrices do not  obey the circular law --- the matrix elements of sparse matrices are often correlated random variables and the variance of the matrix elements scales as $1/n$---  which makes them more versatile to model real-world systems.      
They have a central role in the stability and dynamics of large biological systems, such as ecosystems  \cite{Allesina2015} and neural networks  \cite{rajan2006eigenvalue, ahmadian2015properties, aljadeff2016low}. Other applications include
the robustness of synchronisation in coupled dynamical systems \cite{Pecora98}, the localisation of
eigenfunctions in non-Hermitian quantum mechanics \cite{Hatano96}, the performance
of spectral algorithms for community detection in graphs \cite{krzakala2013spectral, bordenave2015non, kawamoto2018algorithmic}, the behaviour
of random walks on complex networks \cite{Noh2004}, and the evaluation of search algorithms such as PageRank \cite{langville2011google, ermann2015google}.

In spite of their importance to different fields, the spectral properties of sparse non-Hermitian random matrices have received significant attention 
only in the last ten years. This situation may be due to the technical difficulties arising from the combined effects of sparseness and asymmetry, which prevents
the immediate application of the standard tools of random matrix theory.   Rigorous results for sparse non-Hermitian matrices are almost non-existent since it is very difficult to prove the convergence of properties of eigenvalues and eigenvectors to a deterministic limit at large matrix sizes \cite{bordenave2012around}.   Interesting questions in this area include how spectral properties depend on the structural details of the corresponding random graph, such as the
distribution of interaction strengths and the distribution of the number of links per node.  In this topical review, we discuss
recent theoretical progress in the study of the spectra
of sparse non-Hermitian random matrices, with a focus on exact approaches which are based on the fruitful analogy between random-matrix calculations and the statistical mechanics of disordered spin systems.  For simple models, these methods give access to analytical results of spectral properties of sparse non-Hermitian random matrices, and for more complicated models  spectral properties can be computed in the limit  $n \rightarrow \infty$ using  simple
numerical algorithms.   

The review is organised as follows. In the next section we discuss general mathematical aspects of non-Hermitian random matrices; define the main mathematical quantities of interest; discuss the main differences with respect to Hermitian random matrix theory; explain the main difficulties due to lack of Hermiticity symmetry, and how to circumvent these problems following the Hermitisation procedure.    
 The third section
presents the general recursion equations that determine spectral properties of sparse non-Hermitian random matrices, such as the spectral distribution,  the    outliers of non-Hermitian 
random matrices and spectral properties of eigenvectors; in this section we also interpret the recursive equations and explain how to solve them. In the
fourth section, we illustrate the general theory at work on three specific examples: adjacency matrices of regular oriented random graphs and Erd\H{o}s-R\'{e}nyi graphs,   and adjacency matrices of weighted random graphs. In both cases we derive analytical
and numerical results from our theory and confirm their exactness by comparisons with
numerical diagonalisation of large random matrices.
The last section provides an outlook on open problems in the field. The paper is complemented by
an appendix, where we carefully explain  how to derive the main recursive equations using
the cavity method of disordered spin systems. 

\subsection{Notation}
We summarise here the notational conventions used throughout the paper.  
We use lowercase letters ($a,b,c$) to denote deterministic variables  and uppercase letters ($A,B,C$) to denote random variables.    We use the letters $j$, $k$ and $\ell$ to  denote discrete indices that are elements of the set $\left\{1,2,\ldots,n\right\}$ with $n$ a natural number, which we often use for the size of a matrix.  The symbol $i$ is the imaginary unit.   Complex variables are denoted by $z$ (or sometimes $\lambda$) and we decompose them as  $z = x + iy$, with $x= \Re(z)$ and $y = \Im(z)$ representing the real and imaginary parts of $z$, respectively.  The complex conjugate is $z^\ast = x - iy$. We denote the Dirac distribution on the complex plane by $\delta(z) = \delta(x)\delta(y)$, and we use the derivatives $\partial_z = (\partial_x-i\partial_y)/2$ and $\partial_{z^\ast} = (\partial_x+i\partial_y)/2$. We also write $d^2z = dxdy$, $dz = dx+idy$, and $dz^\ast= dx-idy$ in integrals.  
We use the big-O-notation $O(n)$ for the class of functions $f$ of $n$ for which there exists a constant $m\geq 0$ such that  $|f(n)|\leq m \,n$ when $n$ is large enough.   
We use  boldface letters  ($\mathbf{A},\mathbf{a}$)  to denote large matrices of size $O(n)$  and   sans serif letters  ($\mathsf{A},\mathsf{a}$) to denote $2 \times 2$ square matrices. Matrix elements are expressed as $A_{kl} = [\mathbf{A}]_{kl}$.    We write the conjugate transpose of matrices as $\mathbf{A}^\dagger$ and $\mathsf{A}^\dagger$.  Probability density functions or probability mass functions of a random variable $A$ are denoted by $p_A(a)$.  Column matrices or vectors in $\mathbb{C}^n$ are  denoted by $|u\rangle$ and row matrices or adjoint vectors are denoted by $\langle u| =|u \rangle^{\dagger}$. The scalar product in $\mathbb{C}$ reads $\langle u|v\rangle =|u \rangle^{\dagger} |v\rangle$.   We use the symbol $\rho$ for distributions.   Expectation values with respect to all random variables are denoted by $\langle \cdot \rangle$, for example $\langle A \rangle = \int da \: p_A(a) a$ for a continuous variable and $\langle A \rangle = \sum_a p_A(a) a$  for a discrete variable.    
\section{Non-Hermitian Random Matrices}

\subsection{Spectrum,  spectral distribution and statistical properties of eigenvectors}
Random matrix theory encompasses the study of the statistical properties of the eigenvalues and  eigenvectors of random matrices $\mathbf{A}$ in the large-size limit $n\to\infty$.   Therefore, our first task is to identify mathematical quantities that remain meaningful in the limit of  large $n\gg 1$. We will see that distributions often capture the mathematical properties of the large $n$ limit.  

The {\it spectrum} of a matrix $\mathbf{A}$ is the set $\sigma\left(\mathbf{A}\right) = \left\{\lambda_1(\mathbf{A}), \lambda_2(\mathbf{A}),\ldots,\lambda_n(\mathbf{A}) \right\}\subset \mathbb{C}$ of its eigenvalues.   The {\it empirical spectral distribution} determines the fraction of eigenvalues of $\mathbf{A}$ contained in subsets of the complex plane; it is defined by \cite{tao2012topics}
\begin{eqnarray}
\rho_{\mathbf{A}}(z)= \frac{1}{n}\sum^n_{j=1}\delta\big(z-\lambda_j(\mathbf{A})\big)\,,
\end{eqnarray} 
with $\delta$ denoting the Dirac delta distribution on the complex plane and $z\in\mathbb{C}$. To interpret this object, notice that for any smooth function $f$ of compact support on $\mathbb{C}$,
\begin{equation}
\int_{\mathbb{C}}d^2z f(z)\rho_{\mathbf{A}}(z)=\frac{1}{n}\sum^n_{j=1}f(\lambda_j(\mathbf{A}))\,, 
\end{equation} 
where the integral is the Lebesgue integral on the complex plane.  
In words, integrating $f$ with respect to the empirical spectral distribution computes the arithmetic average of $f$ applied to the eigenvalues of $\mathbf{A}$. In particular, if $f=I_{\Omega}$ is itself an indicator function, for some subset $\Omega$ of the complex plane, then, as stated above, this procedure will compute the fraction of eigenvalues lying in $\Omega$. 

In the large dimension limit $n\to \infty$, certain properties of the random set $\sigma\left(\mathbf{A}\right)$ and the random distribution $\rho_{\mathbf{A}}(z)$  may become predictable.
Depending on the choice of random matrix ensemble (i.e. on the law of the entries of $\mathbf{A}$), as $n\to\infty$ the set spectrum and the  empirical spectral distribution may converge to a deterministic distribution on the complex plane, 
\begin{eqnarray}
\sigma\left(\mathbf{A}\right)\rightarrow \sigma, 
\quad \rho_{\mathbf{A}}\to\rho \quad .
\end{eqnarray}
 In the case of non-Hermitian random matrices, the most famous example of this phenomenon is the circular law, but other limiting distributions are possible depending on the structure of the matrix ensemble. 

Questions around the mode of convergence of spectral measures  are an important part of the mathematics of random matrices, see \cite{tao2012topics, bordenave2012around} for reviews.   For non-Hermitian matrices it is difficult to prove the convergence of  spectral distributions because it is not guaranteed that the spectral set $\sigma$ of a random matrix is stable under small perturbations in the asymptotic limit of $n\rightarrow \infty$.     It is not part of the scope of this article to deal directly with this issue because few  results about the convergence of spectral properties of sparse non-Hermitian matrices exist, and because the mathematical methods used to prove convergence are  different than the methods used to compute the spectra of random matrices.  Instead, we  we will present methods to calculate  properties of the spectrum $\sigma$,  the spectral distribution $\rho$, as well as other spectral properties of matrices,  under the presumption of convergence.  Such an approach is sufficient when using random matrices to model systems in physical and life sciences.  

\begin{figure}[htb]
\centering
\includegraphics[width=0.90\textwidth]{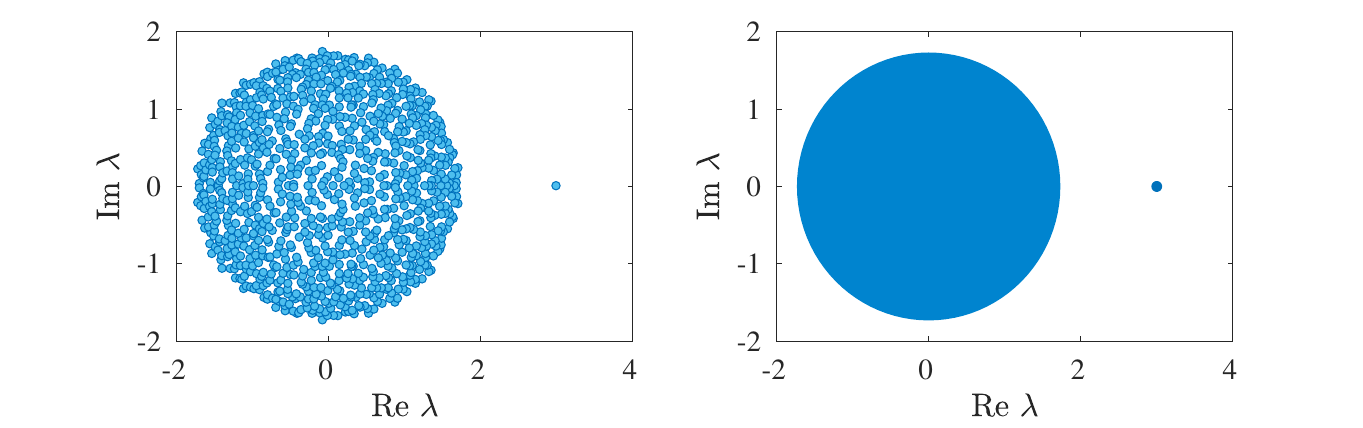}
\caption{Left: eigenvalues of a random matrix of dimension $n=10^3$, with three non-zero entries per row and column, placed uniformly at random. Right: spectrum of the non-Hermitian operator $\mathcal{A}$ acting on the space of sequences indexed by elements of the free group $\mathcal{F}_3$ on generators $\{\alpha,\beta,\gamma\}$, defined by $\mathcal{A}|x\rangle_{w}=|x\rangle_{w\alpha}+|x\rangle_{w\beta}+|x\rangle_{w\gamma}$. The group $\mathcal{F}_3$ consists of all finite length strings of symbols $\{\alpha,\beta,\gamma\}$, or their inverses $\{\alpha^{-1},\beta^{-1},\gamma^{-1}\}$, after cancellation of adjacent reciprocal pairs. To see that $\mathcal{A}$ is non-Hermitian, note that $\mathcal{A}^{\dag}|x\rangle_{w}=|x\rangle_{w\alpha^{-1}}+|x\rangle_{w\beta^{-1}}+|x\rangle_{w\gamma^{-1}}$. The spectrum of $\mathcal{A}$ comprises a circle of radius $\sqrt{3}$ around the origin, and the eigenvalue $\lambda=3$.}  
\label{fig:spectraIll}
\end{figure}

To compute spectral properties of a sparse non-Hermitian random-matrix ensemble for large values of $n$, we will seek to identify  {\it non-Hermitian linear operators}  whose spectral properties are close to those of  large matrices; these operators act on an infinite dimensional vector space.  For example, in Section \ref{kreg} we will compute the limiting spectral distribution of the adjacency matrix of $3$-regular oriented random graphs via an analogy with an operator acting on the free group with three generators; see Fig.~\ref{fig:spectraIll} for an illustration.  The methods in this paper work since sparse random  matrix ensembles are often locally tree-like, and therefore their asymptotic properties are captured by a tree operator in the limit of $n\rightarrow \infty$. As we will show, the spectra of tree operators are mathematically tractable using  recursive approaches.
 
Before discussing how to compute spectral properties of sparse non-Hermitian random matrices, we briefly discuss generic properties of  spectra of  operators, which are considerably different than those of matrices: spectra of matrices are discrete sets  of eigenvalues whereas spectra of operators  can also contain continuous components and eigenvalues with infinite multiplicity.   
The Lebesgue decomposition theorem states that spectra of  operators consist of an absolute continuous component, a singular continuous component and a pure point component \cite{reed1980methods}.    Accordingly, the   
 spectral distribution $\rho$ of an operator  can be decomposed into three mutually singular components, namely
 \begin{eqnarray}
 \rho = \rho_{\rm ac} + \rho_{\rm sc} + \rho_{\rm pp}, 
 \end{eqnarray}  
 where $\rho_{\rm ac}$ is the absolute continuous distribution,  $\rho_{\rm sc}$ is the singular continuous distribution, and $\rho_{\rm pp}$ is the pure point distribution.    
 Analogously, we can decompose the spectrum into three mutually disjoint sets 
 \begin{eqnarray}
 \sigma = \sigma_{\rm ac} \cup \sigma_{\rm sc}\cup \sigma_{\rm pp}.
 \end{eqnarray}   
The absolutely continuous part $\rho_{\rm ac}$ is a function supported on the set $\sigma_{\rm ac}$, which is a subset of non-zero Lebesgue measure in $\mathbb{C}$. The singular continuous part $\rho_{\rm sc}$ is a continuous distribution supported on the set $\sigma_{\rm sc}$, which is a subset of the complex plane with zero Lebesgue measure. Finally, $\rho_{\rm pp}$ is a countable sum of weighted Dirac delta distributions supported on a collection of points $\sigma_{\rm pp}$.  The pure point part of the spectrum can be further decomposed into eigenvalues of finite multiplicity, which form the discrete spectrum, and eigenvalues of infinite multiplicity.      We call  the non-degenerate eigenvalues in the discrete spectrum of an operator {\it outliers}.       Figure~\ref{fig:spectraIll} provides an illustration of the spectrum of a finite-dimensional matrix (left) and of an infinite operator (right); the spectrum of the matrix is discrete, whereas the spectrum of the operator contains a continuous component and an outlier.  

The empirical spectral distribution $\rho(\mathbf{A})$ captures the spectral properties of both matrices and operators and is therefore a very useful quantity to study random matrices.  For matrices  $\mathbf{A}$    of finite size the empirical spectral distribution $\rho_{\mathbf{A}}$ is discrete.  However, in the infinite size limit  the empirical spectral distribution can converge  to  a spectral  distribution $\rho$ that may have continuous components and eigenvalues of infinite multiplicity.  
It is important to note that, depending on the choice of random matrix ensemble, the different contributions to the spectrum may have different scalings in the limit $n\to\infty$. For example, for matrices with identically and independently distributed (IID) positive entries, order $n$ eigenvalues contribute to $\rho_{\rm ac}$, order $\sqrt{n}$ contribute to the singular continuous part, and there is at most a single outlier eigenvalue lying beyond the main part of the spectrum \cite{Akemann2007, Forrester2007, tao2013outliers}.

We are also interested in the statistical properties of eigenvectors of random matrices and infinite operators.   Let $\lambda$ be an eigenvalue of $\mathbf{A}$ with right eigenvector $|v\rangle$ and left eigenvector $\langle u|$.  We define the  {\it distribution  of eigenvector elements} associated with the eigenvalue $\lambda$ of $\mathbf{A}$ by \cite{Neri2016}
\begin{eqnarray}
p_{R,L}(r,l;\mathbf{A},\lambda) = \frac{1}{n}\sum^n_{k=1}\delta(r-\langle k|v\rangle)\delta(l-\langle k|u\rangle) ,\label{eq:eigenvectorDistri}
\end{eqnarray}  
where $\delta$ is again the Dirac distribution on the complex plane, and $\{  | j \rangle \}_{i=1,\dots,n}$ is the standard site basis with $  \langle k | j \rangle = \delta_{kj}$.  If $\lambda$ is an outlier, then the distribution (\ref{eq:eigenvectorDistri}) has often a  meaningful limit $n\rightarrow \infty$.    In particular, in this paper we  study the deterministic limits of the marginal distributions 
\begin{eqnarray}
 \int_{\mathbb{C}} d^2l\: p_{R,L}(r,l;\mathbf{A},\lambda)   \rightarrow p_{R}(r)  ,\quad  \int_{\mathbb{C}} d^2r \: p_{R,L}(r,l;\mathbf{A},\lambda)  \rightarrow p_{L}(l).
\end{eqnarray}

Another interesting quantity for  eigenvector statistics is the correlation between right and left eigenvectors of $\mathbf{A}$ associated to
eigenvalues around a given $\lambda \in \mathbb{C}$.  We characterise eigenvector correlations with the quantity
\begin{eqnarray}
  C_{\mathbf{A}}(\lambda) &=&  \frac{1}{n}\sum^n_{j=1} \langle u_j|u_j\rangle \langle v_j|v_j\rangle\delta(\lambda-\lambda_j).  \label{eq:CDef}
\end{eqnarray}   
In the limit $n \rightarrow \infty$,  $C_{\mathbf{A}}(\lambda)$ characterises the non-normality of matrices, see references \cite{chalker1998eigenvector, mehlig2000statistical, fyodorov2018statistics, gudowska2018synaptic}.  
 Indeed, for normal matrices we have $|v_j\rangle = |u_j\rangle$ such that $ C_{\mathbf{A}}(\lambda)= 1$.    On the other hand, for  non-normal matrices $C_{\mathbf{A}}(\lambda)$ can scale as  $O(n)$.      We  therefore analyse the limit 
\begin{eqnarray}
 \lim_{n \rightarrow \infty}\frac{1}{n}C_{\mathbf{A}}(\lambda) = \mathcal{C} (\lambda).
\end{eqnarray}
If  $\mathcal{C}(\lambda) =0$, the spectrum of $\mathbf{A}$   is stable  around  the eigenvalue $\lambda$ under small perturbations, whereas if $\mathcal{C}(\lambda)  >0$ the spectrum can be unstable under small perturbations.  Indeed, if we add a perturbation $\mathbf{A} + \epsilon \mathbf{d}$ to the matrix $\mathbf{A}$, with  $\epsilon$  a small positive number  and   $\mathbf{d}$   an arbitrary matrix of  norm $\|\mathbf{d}\|=1$, then (see Chapter 2 of \cite{wilkinson1965algebraic}) 
 \begin{eqnarray}
\kappa_j = \lambda_j + \epsilon \langle u_j | \mathbf{d} |v_j \rangle+ O(\epsilon^2), \quad j\in \left\{1,2,\ldots,n\right\} ,\label{eq:perturb}
\end{eqnarray}  
where $\kappa_j$ are the eigenvalues of $\mathbf{A} + \epsilon \|\mathbf{d}\|$.  We have used the matrix norm $\|\cdot\|$  induced by the scalar product, namely,  $\|\mathbf{d}\| = {\rm sup}\left\{\langle \mathbf{d} w  |\mathbf{d} w\rangle  : |w\rangle    \in \mathbb{C}^n, \langle w,w\rangle = 1 \right\} $. 
We apply the Cauchy-Schwartz inequality to (\ref{eq:perturb}) and  find that \begin{eqnarray}
|\langle u_j| \mathbf{d} |v_j \rangle|^2 \leq  \langle u_j|u_j\rangle \langle v_j|v_j\rangle \|\mathbf{d}\| .  \label{eq:overlapPerturb}
\end{eqnarray}
On the right hand side of (\ref{eq:overlapPerturb}) we recognise the correlations between left and right eigenvectors that define $C_{\mathbf{A}}(\lambda)$ in (\ref{eq:CDef}).
Hence, the quantity $C_{\mathbf{A}}(\lambda)$ bounds perturbations of the spectrum around $\lambda$.  If $\mathcal{C}(\lambda) =0$, then we can bound arbitrary perturbations of $\mathbf{A}$ if we choose $\epsilon$  small enough.   However, if $\mathcal{C}(\lambda)  > 0$, then the spectrum in the limit of $n\rightarrow\infty$ can be unstable to arbitrary small perturbations of $\mathbf{A}$.

\subsection{Resolvent of a matrix}
The resolvent of matrix $\mathbf{A}$ is defined by
\begin{eqnarray}
  \mathbf{G}_{\mathbf{A}}(z) = \left( \mathbf{A} - z\mathbf{1}_n  \right)^{-1}, \qquad z\in\mathbb{C} \setminus \sigma(\mathbf{A}),
  \label{kpqa}
\end{eqnarray}
with $\mathbf{1}_n$ denoting the $n \times n$ identity matrix. The resolvent is a useful matrix, since spectral properties of random matrices can be studied from the resolvent  with  basic  mathematical manipulations.  

In the following paragraph, we show that the  spectral distribution $\rho_{\mathbf{A}}$ can be expressed in terms of the trace of the resolvent $\mathbf{G}_{\mathbf{A}}$.
Let $\Omega$ be a compact connected subset of the complex plane that contains $m$ eigenvalues of $\mathbf{A}$, i.e., $|\Omega \cap \sigma| = m$ and  no eigenvalues of $\mathbf{A}$ are located at the boundary $\partial\Omega$ of $\Omega$.  
Integrating the trace of $\mathbf{G}_{\mathbf{A}}(z)$ over the boundary  $\partial \Omega$ we obtain
\begin{eqnarray} 
\fl \frac{1}{n}\int_{\partial \Omega} \frac{dz}{2\pi i}   \sum^n_{k=1} G_{kk}(z)  &=& \frac{1}{n}\sum^{n}_{k=1}\sum^n_{j=1}\langle u_j| k \rangle\langle k |v_j\rangle 
  \int_{\partial \Omega} \frac{dz}{2\pi i} \frac{1 }{\lambda_j(\mathbf{A}) - z}  = -\frac{m}{n},  \label{eq:idx}
\end{eqnarray}
where we have used Cauchy's integral formula,  the notation  $ [ \mathbf{G}_{\mathbf{A}}(z)]_{kk} = G_{kk}(z)$, the resolution of the  identity $\mathbf{1}_n = \sum^n_{k=1}| k \rangle\langle k |$, and the biorthogonality condition $\langle u_j|v_k\rangle = \delta_{j,k}$. We  thus identify 
\begin{eqnarray} 
-\frac{1}{n}\int_{\partial \Omega} \frac{dz}{2\pi i}   \sum^n_{k=1} G_{kk}(z)  =  \int_{\Omega} d^2z \: \rho_{\mathbf{A}}(z).\label{eq:rel1ab}
\end{eqnarray}
We apply Stokes theorem 
\begin{eqnarray}
 \int_{\partial \Omega} \left( f_1dz + f_2 dz^\ast \right) = -2i \int_{\Omega} \left(\partial_z  f_2 - \partial_{z^\ast}f_1\right) d^2z \label{eq:Stokes}
\end{eqnarray} 
to the functions  $f_1 = G_{kk}(z)$ and $f_2=0$ and obtain for each term in the left-hand side of (\ref{eq:rel1ab}) 
\begin{eqnarray}
\int_{\partial \Omega} \frac{dz}{2\pi i}   G_{kk}(z) = \frac{1}{\pi} \int_{\Omega}d^2z\: \partial_{z^\ast} G_{kk}(z) .\label{eq:rel2ab}
\end{eqnarray}
Since relations (\ref{eq:rel1ab}) and (\ref{eq:rel2ab}) hold for arbitrary subsets $\Omega$ of the complex plane, we  find the following expression 
\begin{eqnarray}
   \rho_{\mathbf{A}}(\lambda)=  -\frac{1}{\pi n} \partial_{z^{*}} {\rm Tr}\,\mathbf{G}_{\mathbf{A}}(z)\Bigg|_{z=\lambda} \,,
  \label{ff}
\end{eqnarray}
which should be understood as an identity between two distributions on the space of smooth and compactly supported
complex-valued functions on $\mathbb{C}$. 
Hence, the spectral distribution    $\rho_{\mathbf{A}}(\lambda)$ can be derived from the diagonal elements of the resolvent $\mathbf{G}_{\mathbf{A}}(z)$.   The spectral distribution is equal to zero whenever the  diagonal elements of $\mathbf{G}_{\mathbf{A}}(z)$ are analytic, which is the case for  all  $z\in \Omega\setminus \sigma$.

The resolvent $G_{\mathbf{A}}$ also contains useful information about the eigenvectors of $\mathbf{A}$.  
In this regard, it is insightful to write $\mathbf{G}_{\mathbf{A}}(z)$ in its canonical form 
\begin{eqnarray}
\mathbf{G}_{\mathbf{A}}(z) = \sum^{n}_{j=1} \frac{|v_j\rangle \langle u_j|}{ \lambda_j(\mathbf{A}) - z   } ,
\end{eqnarray}
and its matrix elements are thus 
\begin{eqnarray} 
  G_{kl}(z) =  \langle k|\mathbf{G}_{\mathbf{A}}(z)|\ell \rangle = \sum^{n}_{j=1}
  \frac{\langle k |v_j\rangle \langle u_j| \ell \rangle }{\lambda_j(\mathbf{A}) - z}.
  \label{Gv}
\end{eqnarray}
Evaluating $G_{\mathbf{A}}(z)$ in the neighbourhood $z=\lambda_j - \epsilon$ of a (non-degenerate) eigenvalue $\lambda_j\in \sigma(\mathbf{A})$  we obtain
\begin{equation}
\fl G_{\mathbf{A}}(\lambda_j -\epsilon)  | v_j \rangle = \frac{1}{\epsilon}  | v_j \rangle+ O(|\epsilon|^0), \qquad  \langle u_j |  G_{\mathbf{A}}(\lambda_j-\epsilon) = \frac{1}{\epsilon} \langle u_j | + O(|\epsilon|^0), 
  \label{eigB0x}
\end{equation}   
where $|\epsilon|\ll 1$, that is, $1/\epsilon$ is to leading order an eigenvalue of $G_{\mathbf{A}}(\lambda_j-\epsilon) $. In general, all other eigenvalues of  $G_{\mathbf{A}}(\lambda_j-\epsilon) $ are of order $O(|\epsilon|^0)$, hence for small $|\epsilon|$ we obtain
\begin{equation}
  \epsilon G_{\mathbf{A}}(\lambda_j-\epsilon)  = | v_j \rangle\langle u_j |  + O(|\epsilon|).
  \label{eigBx}
\end{equation}   
The resolvent $G_{\mathbf{A}}$ also contains useful information about  correlations between left and right eigenvectors of $\mathbf{A}$.   Indeed, the  correlator  $\mathcal{C}(\lambda)$  is given by 
\begin{equation}
  \fl \mathcal{C}(\lambda) = \lim_{\eta\rightarrow 0^+}{\rm Tr} \left[ \frac{\eta}{\eta^2 + (  \mathbf{A} -  \lambda\mathbf{1}_n)(  \mathbf{A} -  \lambda \mathbf{1}_n)^{\dagger}} \right]
      {\rm Tr} \left[ \frac{\eta}{\eta^2 +(  \mathbf{A} -  \lambda \mathbf{1}_n)^{\dagger} (  \mathbf{A} -  \lambda\mathbf{1}_n)} \right] . \label{eq:Corr}
\end{equation}
We refer to the references  \cite{chalker1998eigenvector, mehlig2000statistical, PhysRevE.60.2699} for a derivation of (\ref{eq:Corr}).

In this way, the computation of statistical properties of both the eigenvalues and  eigenvectors of a matrix  boils down to  the calculation of the resolvent $G_{\mathbf{A}}$.      Unfortunately, as we discuss in the next subsection, for non-Hermitian ensembles the resolvent does not provide useful information about the continuous part of the spectrum of non-Hermitian matrices in the infinite size limit.

\subsection{Differences between Hermitian and Non-Hermitian Ensembles}\label{sec:grav}
Until now, nothing we have described has been specialised to non-Hermitian matrices. However, the spectral properties of general non-Hermitian matrices are considerably different than those of their Hermitian cousins. All eigenvalues of Hermitian matrices are real, and the corresponding eigenvectors are orthogonal. An important consequence of these facts is that, for Hermitian matrices, evaluating the trace of the resolvent produces a regularised form of the spectral distribution. Specifically, if $\mathbf{H}$ is Hermitian, then for $x\in\mathbb{R}$ and $\eta>0$ one can show that
\begin{equation}
\frac{1}{\pi} \Im \left[ \frac{1}{n}{\rm Tr}\mathbf{G}_{\mathbf{H}}(x+i\eta) \right] = \frac{1}{\pi}\int_{\mathbb{R}} \frac{\eta}{\eta^2+(x-\lambda)^2}  \rho_{\mathbf{H}}(\lambda)d\lambda.
\label{HermReg}
\end{equation}
The right hand side here is simply the convolution of the empirical spectral measure with a Lorentzian of width $\eta$; sending $\eta\to0$, one recovers the spectral measure itself. This identity is a cornerstone of Hermitian random matrix theory, since the limit $\eta\to0$ can be safely exchanged with the limit $n\to\infty$, allowing one to abandon the empirical measure and instead work entirely with the well-behaved resolvent defined on the upper half complex plane.

The intuition behind the above regularisation (\ref{HermReg})  is that, for Hermitian matrices in  the limit $n\rightarrow \infty$, the distribution  $\rho$ of eigenvalues on the real line can be inferred from the the trace of the  resolvent (\ref{kpqa}), which only exists  on the part of the complex plane that is not real.     Indeed,  since for Hermitian matrices all eigenvalues are real, the resolvent is well defined  on  $\mathbb{C}\setminus\mathbb{R}$. In this case the  resolvent contains sufficient information to infer the spectral distribution on the real line.     Unfortunately, this approach does not work  when the spectrum $\sigma$ has a  nonzero Lebesgue measure on the complex plane. This is often the case for non-Hermitian operators or non-Hermitian matrices in the limit of $n\rightarrow \infty$.

An  analogy between non-Hermitian random matrix theory and  Newtonian gravity in two dimensions reveals why it is not possible to infer the  spectral distribution $\rho$ inside  $\sigma_{\rm ac}$ from  the resolvent $\mathbf{G}_{\mathbf{A}}(z)$,   which only exists for $z$ located outside $\sigma_{\rm ac}$.   
 Consider the two-dimensional ``gravitational" field $\vec{g} = (g_x,g_y)$  with components  \cite{PhysRevLett.60.1895}
 \begin{eqnarray}
 g_x  =  \frac{2}{n} \Re\left[{\rm Tr}\mathbf{G}_{\mathbf{A}}(z)\right] \quad {\rm and} \quad g_y = -\frac{2 }{n}\Im\left[{\rm Tr}\mathbf{G}_{\mathbf{A}}(z)\right] \label{eq:traceResolvent}
 \end{eqnarray} 
 and its gravitational potential $\phi$ defined by
   \begin{eqnarray}
g_x = -\frac{\partial \phi}{\partial x} ,\quad  g_y = -\frac{\partial \phi}{\partial y}.
 \end{eqnarray}
The derivative of the trace  resolvent is related to the Laplacian of the gravitational potential,
 \begin{eqnarray}
 \frac{4}{n}\partial_{z^\ast}{\rm Tr}\mathbf{G}_{\mathbf{A}}(z) &=&  -\left(\frac{\partial^2}{\partial x^2}+\frac{\partial^2}{\partial y^2}\right)\phi. \label{eq:1xaq}
 \end{eqnarray} 
According to (\ref{ff}) the derivative of the trace  resolvent is  the spectral distribution.  Hence,    (\ref{ff}) and (\ref{eq:1xaq})  imply the Poisson equation 
  \begin{eqnarray}
   \left(\frac{\partial^2}{\partial x^2}+\frac{\partial^2}{\partial y^2}\right)\phi = 4\pi \rho(z) . \label{eq:poisson}
    \end{eqnarray}
Equation~(\ref{eq:poisson}) is the Poisson equation and the trace  resolvent ${\rm Tr}\mathbf{G}_{\mathbf{A}}(z)$ is mathematically equivalent to a two-dimensional gravitational field $\vec{g} = (g_x, g_y)$ generated by a clump of mass  distributed according to $\rho$  in the region $\sigma$ of the complex plane.   It is a well-known fact,  expressed by Gauss's law of gravity, that different mass distributions $\rho$ can generate the same gravitational field $\vec{g}$ in a region of space that does not contain mass.   Therefore, it  is  not possible to infer $\rho$ from the resolvent $\mathbf{G}_{\mathbf{A}}(z)$.     

The potential $\phi$  also appears in proofs of the convergence of the spectral distribution $\rho_{\mathbf{A}}$ for $n\rightarrow \infty$, see for instance \cite{bordenave2012around}, where $-\phi/2$ is called the logarithmic potential. 

 Since for non-Hermitian random matrices the trace resolvent is not useful to study the continuous component of the spectrum, we present in  the next section  a regularisation procedure that works for non-Hermitian matrices and allows us to compute spectral distributions of sparse non-Hermitian matrices in the limit of $n\rightarrow \infty$.

\subsection{Hermitisation method} \label{HermMet}

The general idea
of the Hermitisation method is to introduce a regularisation to non-Hermitian spectra that can function in place of formula (\ref{HermReg}), which does not hold for non-Hermitian matrices. This can be achieved by enlarging the dimension of the problem, introducing a $2 n \times 2 n$ block matrix including a regularising  parameter $\eta>0$. There are several different choices for implementation of the enlarged matrix in the literature
(see e.g.~\cite{feinberg1997non,Janik99,Tim2009}), all of which are essentially equivalent. The name of the method refers to early implementations in which the constructed $2 n \times 2 n$  matrix is Hermitian, gaining access to standard tools of Hermitian random matrix
theory \cite{feinberg1997non, feinberg1997nonx}. 

Here we introduce the $2 n \times 2 n$ normal block matrix 
\begin{eqnarray}
  \mathbf{B}(z,\eta) = \left(\begin{array}{cc} \eta \mathbf{1}_n &  -i (  \mathbf{A} -  z \mathbf{1}_n)
    \\   -i ( \mathbf{A}^{\dagger} - z^* \mathbf{1}_n )   & \eta \mathbf{1}_n\end{array}\right),
    \label{hjpq}
\end{eqnarray}
for  $z \in \mathbb{C}$ and a regularizer $\eta > 0$. To see that $\mathbf{B}(z,\eta)$ is normal, observe that $i(\mathbf{B}(z,\eta)-\eta\mathbf{1}_{2n})$ is Hermitian, hence $\mathbf{B}(z,\eta)$ is diagonalised by an unitary transformation. The inverse of $\mathbf{B}(z,\eta)$ can be written as 
\begin{equation}
  \mathbf{B}^{-1}(z, \eta) =  \left(\begin{array}{cc}  \eta\mathbf{X} & i \mathbf{X}(\mathbf{A}-z\mathbf{1}_n)
    \\ i \mathbf{Y}(\mathbf{A}-z\mathbf{1}_n)^\dag  &  \eta \mathbf{Y} \end{array} \right), \label{eq:matrixB}
\end{equation}
where $\mathbf{X}=(\eta^2+(\mathbf{A}-z\mathbf{1}_n)(\mathbf{A}-z\mathbf{1}_n)^\dag)^{-1}$
 and $\mathbf{Y}=(\eta^2+(\mathbf{A}-z\mathbf{1}_n)^\dag(\mathbf{A}-z\mathbf{1}_n))^{-1}$ are the Schur complements. Importantly, notice that the lower left block of the inverse  obeys 
\begin{equation}
i\mathbf{Y}(\mathbf{A}-z\mathbf{1}_n)^\dag=i\mathbf{G}_{\mathbf{A}}(z)+O(\eta^2)
\end{equation}
for small $\eta$.
Following Eq.~(\ref{ff}), the empirical spectral distribution is given by
\begin{eqnarray}
  \rho_{\mathbf{A}}(z) =  \lim_{\eta\rightarrow 0^+} \frac{i}{n\pi}\partial_{z^{*}}\sum_{j=1}^n \big[\mathbf{B}(z, \eta)^{-1}\big]_{j+n,j}\,.
  \label{ghaap}
\end{eqnarray} 
For fixed $\eta>0$, $\rho^{(\eta)}_{\mathbf{A}}(z)$ is a continuous  function of $z$, since the matrix  $\mathbf{B}(z, \eta)$ is invertible for all $z\in \mathbb{C}$ and all  $\eta >0$.  In this way the parameter $\eta$ acts to regularise the spectral distribution. 

The matrix function $\mathbf{B}^{-1}(z,\eta)$ also contains useful information about eigenvectors of $\mathbf{A}$.   Evaluating $\mathbf{B}(z,\eta)$ at a (non-degenerate) eigenvalue $z=\lambda_j\in\sigma(\mathbf{A})$, we find 
\begin{equation}
  \mathbf{B}(\lambda_j,\eta)  | b_j^+ \rangle = \eta  | b_j^+ \rangle, \qquad  \mathbf{B}(\lambda_j,\eta)   | b_j^- \rangle = \eta  | b_j^- \rangle,
  \label{eigB0}
\end{equation}  
where
\begin{equation}
| b_j^+ \rangle =
 \left(\begin{array}{cc}  |u_{j} \rangle  \\  |v_{j} \rangle \end{array} \right)\,, \quad 
 | b_j^- \rangle =
 \left(\begin{array}{cc}  -|u_{j} \rangle  \\  |v_{j} \rangle \end{array} \right).
\end{equation}
That is, $\eta$ is an eigenvalue of $\mathbf{B}(\lambda_j,\eta)$ of geometric multiplicity two, with eigenvectors $| b_j^\pm \rangle$. In general all other eigenvalues of $\mathbf{B}(\lambda_j,\eta)$ are of order one relative to $\eta$, hence for small $\eta$ we have
\begin{equation}
  \eta\mathbf{B}^{-1}(\lambda_j,\eta)  =  | b_j^+ \rangle \langle b_j^+ | +  | b_j^- \rangle \langle b_j^- |  + O(\eta).
  \label{eigB}
\end{equation}
The off-diagonal elements of the matrix $\mathbf{B}$ also carry interesting information about the eigenvectors of $\mathbf{A}$.  We recognise in $\mathbf{X}$ the eigenvector correlator $\mathcal{C}$, viz., 
\begin{equation}
\mathcal{C}(\lambda)=  \lim_{\eta\rightarrow 0^+} \frac{1}{\pi n^2}\sum_{j=1}^n \big[\mathbf{B}(\lambda, \eta)^{-1}\big]_{jj}\sum_{j=1}^n \big[\mathbf{B}(\lambda, \eta)^{-1}\big]_{j+nj+n}. \label{ghaapxsd}
\end{equation}  
Hence, the diagonal elements of $\mathbf{B}(z, \eta)^{-1}$ characterise  the  sensitivity of the spectrum to small perturbations.

We have demonstrated that both the computation of the spectral distribution function and the statistical properties of eigenvectors boils down to being able
to invert $\mathbf{B}(z,\eta)$. In the next sections we  develop methods to compute this inverse for sparse matrices directly in the infinite size limit.   To draw firm conclusions about the limiting spectra it is therefore necessary to consider how the limits $n\to\infty$ and $\eta \to 0$ interact.
We discuss equation (\ref{eigB}) first, which technically holds only for $\eta$ much smaller than the typical separation between simple eigenvalues. For large matrices and inside the bulk of the spectrum, the separation between eigenvalues is of order $1/n$, hence one cannot take the $n\to\infty$ limit with fixed $\eta$. For outlier eigenvalues, however, this is not a problem as the separation does not shrink with $n$. As we will see in the later sections, the location of outlier eigenvalues in the limiting pure point spectrum can in fact be inferred from the existence of non-trivial solutions to (\ref{eigB0}), whose defining property is a finite non-zero norm.  

\begin{figure}[htb]
\centering
\includegraphics[width=0.90\textwidth]{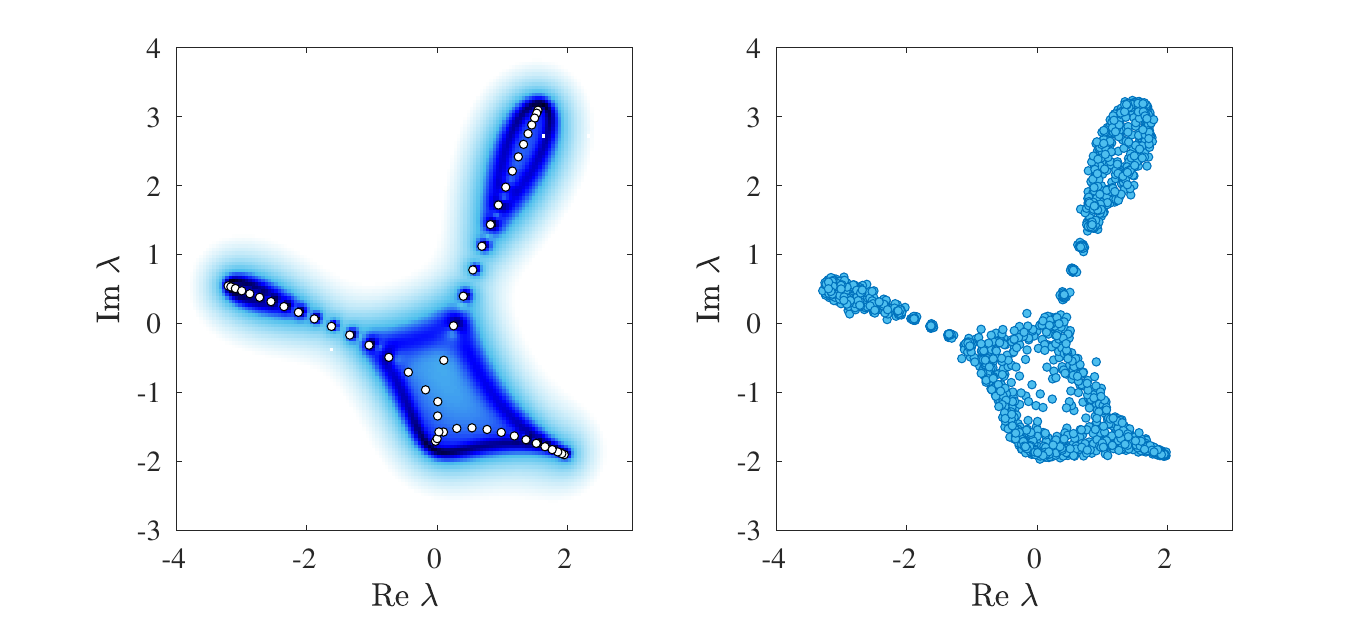}
\caption{Left: white circles show the eigenvalues of the ``bull's head'' matrix $\mathbf{A}$ of dimension $n=50$, colour map shows the regularised spectral distribution of $\mathbf{A}$ with $\eta=0.01$. Right: eigenvalues of 50 realisations of the perturbed bull's head matrix $\mathbf{A}+\eta \mathbf{P}\mathbf{Q}^{-1}$, where $\mathbf{P}$ and $\mathbf{Q}$ have independent complex Gaussian entries. The bull's head matrix is a banded Toeplitz matrix (i.e. constant along its diagonals) with rows $(\cdots 0,\, 2i,\, 0,\, 0,\, 1,\, 7,\, 0 \cdots)$. As illustrated here, its spectrum is highly sensitive to perturbations, even at  large values of $n$.   This implies that  the limits $\eta\to0$ and $n\to\infty$ do not commute for this example.}  
\label{fig:perturbation}
\end{figure}

The computation of the spectral distribution function is also sensitive to the relationship between $\eta$ and $n$. The spectral regularisation induced by $\eta$ can in fact be shown to correspond to calculating the average empirical spectral distribution of the perturbed matrix $\mathbf{A}\mapsto\mathbf{A}+\eta \mathbf{P}\mathbf{Q}^{-1}$, where $\mathbf{P}$ and $\mathbf{Q}$ are $n\times n$ random matrices having independent complex Gaussian entries \cite{rogers2010}. In contrast to normal matrices, it is well-known that the spectra of  non-normal matrices can be highly unstable with respect to small perturbations to the elements, as illustrated in Fig.~\ref{fig:perturbation} (see \cite{trefethen2005spectra} for many more beautiful examples). To rigorously justify an exchange of limits between $\eta$ and $n$ it is therefore necessary to quantify how the stability of the spectrum of the random matrix scales with $n$, and thus to analyse how the two limits $\eta$ and $n$ interact. This has been achieved in only a handful of cases (see \cite{bai2008circular, tao2012topics, bordenave2011} for examples), and it remains an important open problem to develop more generally applicable proof techniques.  

\section{Theory for Sparse Non-Hermitian Ensembles}
In this section we present the theory for the spectral properties of {\it sparse} non-Hermitian random matrices that are locally tree-like. The theory is based  on the derivation of recursion relations for the  elements of the inverse of the matrix $\mathbf{B}(z, \eta)$. These equations are derived with the cavity approach of disordered systems \cite{MezardBook,rogers2008cavity,Tim2009,Metz2010}.     We obtain a simple set of recursion equations that determine the different aspects of the spectrum of the random ensemble in the limit $n\rightarrow \infty$, including the empirical spectral distribution,  the outliers in the spectrum, and the statistics of their corresponding eigenvectors.   The recursion relations exploit the local tree-like structure of sparse non-Hermitian random matrices. By using techniques such as those in \cite{rogers2010a, Metz2011, bolle2013spectra}, we note that it is possible to extend the methods discussed here to more complex structures including plaquette models with many short cycles. However, this topic lies beyond the scope of the present review.

\subsection{Sparse random matrices, random graphs and ensembles of dynamical systems}
\label{SparseDef}
Before presenting the main equations and  applying them to specific models, it is useful to introduce
some definitions and discuss the connection between sparse random matrices,  random graphs, and ensembles of dynamical systems.    

Sparse matrices are those in which almost all entries are zero; they contain a typical number of non-zero entries that is proportional to $n$. This property makes them fundamentally different from ensembles with independent and identical distributed elements.     Therefore, we 
 parametrise sparse matrices as $A_{jk} = C_{jk}J_{jk} + \delta_{jk}D_{j}$, where  $C_{jk}$ are connectivity variables,  $J_{jk}$ are off-diagonal weight variables, and $D_j$ are diagonal weight variables.  The connectivity variables $C_{jk}\in \left\{0,1\right\}$ determine which off-diagonal elements  of $\mathbf{A}$ are non-zero, and the structure of the connectivity variables characterises the sparseness of the matrix.   The weight variables $J_{jk}\in\mathbb{C}$  and $D_{j}\in\mathbb{C}$ specify the values of the non-zero elements.
  The probability densities of $\mathbf{C}$,   $\mathbf{J}$ and $\mathbf{D}$ fully determine the random matrix ensemble.

It is instructive to associate a sparse non-Hermitian random matrix  $\mathbf{A}$ to a random graph \cite{Bollobas}, defined
informally as a collection of nodes or vertices mutually interconnected by edges or links in a random fashion. Random graphs
are extremely useful in the study of complex systems, since they allow us to build models for the causal relationships between a
large number of entities, being them proteins, individuals, or neurons. From this perspective, $\mathbf{C}$
encodes all information regarding the structure of a graph with a total number of $n$ nodes. In fact, $\mathbf{C}$ is the so-called {\it adjacency matrix} of the graph.
If there is a directed link pointing from node $k$ to $j$, the $k$-th degree of freedom causally influences
the $j$-th degree of freedom and we set $C_{kj}=1$; if such link is absent, we set $C_{kj}=0$. The random variables 
\begin{eqnarray}
K^{\rm in}_{j} \equiv \sum^n_{k=1, (k\neq j)}C_{kj} , \quad K^{\rm out}_{j}  \equiv   \sum^n_{k=1, (k\neq j)}C_{jk} , 
\end{eqnarray} 
are the {\it indegree} and the {\it outdegree} of node $j$, respectively.   Note that $K^{\rm in}_{j}$ is the number of non-zero off-diagonal elements in the $j$-th column of $\mathbf{A}$ and $K^{\rm out}_{j}$ is the number of non-zero off-diagonal elements in the $j$-th row of $\mathbf{A}$.    The local connectivity of a node $j$ is determined by the sets
\begin{eqnarray}
\partial^{\rm in}_{j} \equiv \left\{k\neq j: C_{kj} =1   \right\}, \quad \partial^{\rm out}_{j} \equiv \left\{k\neq j: C_{jk} =1   \right\},
\end{eqnarray} 
and 
\begin{eqnarray}
\partial_j \equiv \partial^{\rm in}_{j} \cup \partial^{\rm out}_{j}, 
\end{eqnarray} 
which are, respectively, called the {\it in-neighbourhood}, the {\it out-neighbourhood}  and the {\it neighbourhood} of the $j$-th node.

A sparse matrix or graph is called {\it oriented} if $C_{jk}C_{kj} = 0$  for $k\neq j$, i.e., an oriented random
graph has no bidirected edges.  A matrix or graph  is $c${\it -regular} if $K^{\rm in}_{j} = K^{\rm out}_{j}  = c$ for any node $j$.     
If $D_{j}=0$ ($j=1,\dots,n$) and  $J_{jk} = 1$ ($j \neq k$), then the matrix $\mathbf{A}$ is  the
adjacency matrix of the corresponding graph.   A sequence of graphs with adjacency matrix $\mathbf{C}$ is {\it locally tree-like} if 
for $n\rightarrow \infty$ the finite 
neighbourhood of a randomly chosen vertex is with probability one connected like a tree graph. \cite{bordenave2010}.

It is also instructive to associate an ensemble of   {\it dynamical systems} with a random matrix   $\mathbf{A}$.  Let   $\langle x_t| \in \mathbb{R}^n$ be a   state variable or  let   $\langle x_t| \in \mathbb{C}^n$ be a complex amplitude, where $t\geq 0$ is a continuous time variable.   We can associate a  dynamics to the vector $\langle x_t|$  through a set of linear differential equations
\begin{eqnarray}
 \partial_t \langle x_t| = \langle x_t|  \mathbf{A},  \label{eq:linearEq}
\end{eqnarray}  
 with a given initial state $\langle x_0|$.  
Componentwise, equation (\ref{eq:linearEq}) reads
\begin{eqnarray}
\partial_t \langle x_t|j\rangle = \sum_{k\in \partial^{\rm in}_{j}} \langle x_t|k\rangle J_{kj}   + D_j  \langle x_t|j\rangle , \quad j\in\left\{1,2,\ldots, n\right\}.
\end{eqnarray}

Despite its linearity, equation (\ref{eq:linearEq}) is general enough to describe interesting systems in physical and life sciences.  For example, the Master equation, which describes the evolution of  probability mass functions of  stochastic processes, and the Schr\"{o}dinger equation, which describes the dynamical evolution of the wavefunction amplitudes in quantum systems,  are of the form (\ref{eq:linearEq}).    On the other hand, models of many interesting systems are non-linear, such as the firing rates of  neurons in a neural network or the population dynamics of species in ecosystems.  Nevertheless,  in  the vicinity of a fixed point we can  often approximate the dynamics of a non-linear system   by a linear set of differential equations of the form~(\ref{eq:linearEq}).

The relation (\ref{eq:linearEq}) provides us with a concrete picture of  random matrix ensembles.  The connectivity matrix determines  the network of causal interactions between  degrees of freedom of a dynamical system and  the weight matrix $\mathbf{J}$ determines the strength of these interactions ($|J_{jk}|$ large implies a strong interaction whereas $|J_{jk}|$ small implies a weak interaction), and whether they are excitatory ($J_{kj}>0$) or inhibitory ($J_{kj}<0$).   The diagonal weight matrix $\mathbf{D}$  can be interpreted as a source ($D_{jj}>0$) or a sink term  ($D_{jj}<0$).

The randomness in $\mathbf{A}$  reflects our ignorance about the specific interactions between the degrees of freedom. Instead we may know the statistical properties of these interactions, such as their average strength or the average number of direct interactions between degrees of freedom.   It is here that random matrices become useful, since they  describe the typical behaviour of the dynamics of an ensemble of dynamical systems under certain statistical constraints on the interactions.   This becomes evident when we express the solution of  (\ref{eq:linearEq}) in terms of the spectral properties of $\mathbf{A}$:
\begin{eqnarray} 
\langle x_t| = \sum^n_{j=1}e^{\lambda_{j}t} \langle x_0|v_j\rangle \langle u_j| . \label{eq:dyn}
 \end{eqnarray}  
The long time behaviour  $t\gg 1$ of the state $\langle x_t|$ is thus given by
   \begin{eqnarray} 
\langle x_t| =  \sum_{j\in \sigma_{\rm max}} e^{\lambda_j t} \langle x_0|v_j\rangle \langle u_j|  + O\left(e^{-\zeta t}\right), \label{eq:asympto}
    \end{eqnarray}  
    where $\sigma_{\rm max}$ is the set of eigenvalues $\lambda_j$ whose 
real part is larger or equal than the real part of all the other eigenvalues in the spectrum $\sigma$ of $\mathbf{A}$, i.e., for $j\in \sigma_{\rm max}$ we have that $\Re(\lambda_j) = {\rm max}_{k\in \left\{1,2,\ldots,n\right\}} \Re(\lambda_k)$.    The constant $\zeta$ is the  
{\it spectral gap}, which is $\zeta = \lambda_{\rm max}-\lambda'_{\rm max}$, where $\lambda_{\rm max} ={\rm max}_{k\in \left\{1,2,\ldots,n\right\}} \Re(\lambda_k)$ is the maximum value of the real parts, and where $\lambda'_{\rm max}$ is the second largest real part of the eigenvalues (the maximum value of the real parts in the set $\sigma\setminus \sigma_{\rm max}$).   
Relation (\ref{eq:asympto}) reveals that the study of spectral properties of random  matrices provides a mean to access the stability and instability of dynamical systems in the vicinity of a fixed-point.   When  all eigenvalues have negative real parts ($\lambda_{\rm max}<0$) then the fixed point  is {\it stable}, whereas if there is at least one eigenvalue with  a positive real part then the fixed point is {\it unstable} ($\lambda_{\rm max}>0$).   Thus, in the context of dynamical systems, we are particularly interested in the eigenvalue with the largest real part and the spectral gap of a random matrix. 

The state vector (\ref{eq:asympto}) either converges to zero for large $t$ (stable system) or its norm  $\langle x_t|x_t \rangle$ diverges (unstable system).  
The asymptotics of the dynamics provided by the linear model  (\ref{eq:dyn})  are thus not so interesting,  and the eigenvectors of  $\mathbf{A}$ do not provide information about the asymptotic state of the dynamics.  There is one exceptional case for which linear systems of the form (\ref{eq:dyn})  do provide a nontrivial asymptotic state,  namely  when $\lambda_{\rm max } = 0$.  This situation is realised for  {\it Laplacian} or {\it stochastic} matrices~$\bA$.   Stochastic matrices have the property that all $J_{kj}>0$ and that  
 $D_j = -\sum^n_{k=1(\neq j)} J_{kj}$.   For stochastic matrices the Perron-Frobenius theorem states that $\lambda_{\rm max} = 0$ \cite{horn1990matrix}.   This eigenvalue is unique if the matrix is positive definite or irreducible \cite[Section 8.3 and 8.4]{horn1990matrix}, and is called the Perron-Frobenius eigenvalue.  As a consequence, the asymptotic solution is given by  $\lim_{t\rightarrow \infty}\langle x_t | = \langle u_{\rm max}|$, with  $\langle u_{\rm max}|$ the eigenvector associated with the Perron-Frobenius eigenvalue.     The Perron-Frobenius theorem also states that the 
 elements  $\langle u_{\rm max}|j\rangle$ are positive for all $j\in \left\{1,2,\ldots,n\right\}$.

Another way to define a dynamical system with a  normalisable steady state solution $\lim_{t\rightarrow \infty}\langle x_t| $ is to introduce a  time-dependent   degradation term $\gamma(t)$  that constraints the norm of the state vector to be constant, namely 
\begin{eqnarray}
 \partial_t \langle x_t| = \langle x_t| \left[ \mathbf{A}  - \gamma(t)\mathbf{1}_{n}\right],   \label{eq:linearEqxx}
 \end{eqnarray}  
 with 
\begin{eqnarray}
\gamma(t) =  \frac{\langle x_t|\mathbf{A}|x_t\rangle}{\langle x_t|x_t\rangle },  \label{eq:gamma}
\end{eqnarray} 
where, for simplicity, we consider here $\mathbf{A}$ a real matrix. 
The constraint (\ref{eq:gamma}) implies that  $\partial_t \langle x_t|x_t\rangle = 0$.   Therefore, the vector $|x_t\rangle$  is constrained to the surface of an $n$-dimensional sphere of fixed radius  $\sqrt{\langle x_t|x_t\rangle}$. This model is called the   {\it  spherical model}, and
it plays an important role in statistical physics because it is an  exactly solvable model of a many-body system that exhibits phase transitions and glassy dynamics  \cite{cugliandolo1995full, castellani2005spin}.   The solution of the spherical model can be decomposed into the basis  $\left\{\langle u_j|\right\}_{j=1,\ldots, n}$ of the left eigenvectors of $\mathbf{A}$, 
\begin{eqnarray}
\langle x_t| = \sum^n_{j=1}\langle x_0|v_{j}\rangle  e^{\lambda_j t-\int^t_0 dt' \gamma(t')}  \langle u_j|\label{eq:solxadd}.
\end{eqnarray} 
 Substitution of (\ref{eq:solxadd}) in (\ref{eq:gamma}) and solving towards $\gamma(t)$ gives
  \begin{eqnarray}
  \fl \gamma(t) =  \frac{\sum^n_{j=1}\sum^n_{\ell=1}\langle x_0|v_j \rangle \langle v_\ell|x_0\rangle  \lambda_j    e^{ \lambda_j t + \lambda^\ast_\ell t   } \langle u_j| u_\ell\rangle}{ \langle x_0 |x_0\rangle+  2\sum^n_{j=1}\sum^n_{\ell=1}\langle x_0|v_j \rangle \langle v_\ell|x_0\rangle \frac{ \lambda_j }{ \lambda_j +  \lambda^\ast_\ell}  \left( e^{ \lambda_j t +  \lambda^\ast_\ell t   } -1 \right)\langle u_j| u_\ell\rangle} .  \label{eq:gammaSol}
  \end{eqnarray} 
  Notice that $\gamma$ is a real number which follows from the following two properties: (i) if $\lambda$ is an eigenvalue of $\mathbf{A}$, then also $\lambda^\ast$ is an eigenvalue of $\mathbf{A}$; (ii) if $\langle u|$ is an eigenvector of $\lambda$, then  
   $\langle u|^\ast$ is an eigenvector of $\lambda^\ast$.     If the matrix $\mathbf{A}$ is normal, then 
 $\langle u_\ell|u_k\rangle  = \delta_{\ell,k}$ and  $\langle v_\ell|v_k\rangle  = \delta_{\ell,k}$, such that 
   \begin{eqnarray}
  \fl \gamma(t) =  \frac{\sum^n_{j=1}|\langle x_0|v_j \rangle|^2   \Re[\lambda_j]     e^{ 2  \Re[\lambda_j] t   } }{   \sum^n_{j=1}|\langle x_0|v_j \rangle|^2 e^{ 2  \Re[\lambda_j] t   }  } . 
  \end{eqnarray} 
The steady state solution  $t\gg 1$ is determined by the eigenvalues with the largest real part, which   dominate the sum in (\ref{eq:gammaSol}), and thus 
\begin{eqnarray}
\gamma(t) = \Re\left(\lambda_{\rm max}\right)  + O\left(e^{-\zeta t}\right),
\end{eqnarray}
with $\zeta = \Re(\lambda_{\rm max}) - \Re(\lambda'_{\rm max})$ the spectral gap as defined below equation (\ref{eq:asympto}).   The steady state of the system for large $t$ is 
\begin{eqnarray}
\langle x_t| =\sum_{\ell \in \sigma_{\rm max}}\langle x_0|v_{\ell}\rangle  e^{i\:\Im(\lambda_{\ell})t }   \langle u_\ell| + O\left(e^{-\zeta t}\right) \label{eq:solxaddxx},
\end{eqnarray} 
with $\sigma_{\rm max}$  again the set of eigenvalues whose 
real part is larger or equal than the real part of all the other eigenvalues in the spectrum $\sigma$ of $\mathbf{A}$.   If   the eigenvalue with the largest real part is thus unique, then 
\begin{eqnarray}
\langle x_t| =  \langle x_0|v_{\rm max}\rangle   \langle u_{\rm max}| + O\left(e^{-\zeta t}\right) 
\end{eqnarray} 
with $|v_{\rm max}\rangle$ and   $\langle u_{\rm max}|$ the right and left eigenvectors associated with  $\lambda_{\rm max}$.

We have shown that the dynamics of complex systems can be defined in terms of a sparse non-Hermitian random matrix.  The dynamics of the state  $\langle x_t|$ of the system is determined by  the spectral properties of the random matrix ensemble and    the steady state solution at large times $t\gg 1$  is governed by the eigenvalue with the largest real part and its associated eigenvector; the relaxation time towards the steady state is given by the spectral gap.    We will thus be interested in developing mathematical methods to study these spectral properties for sparse non-Hermitian random matrices, and this is the topic of the next section.  
 
\subsection{Recursion relations  for   $\mathbf{B}^{-1}$ and the spectral distribution of locally tree-like random matrices}   \label{sec32}
In section \ref{HermMet} we have introduced the $2n\times 2n$ matrix $\mathbf{B}$ as a device to regularise the spectral statistics of the non-Hermitian matrix under study and we have seen how the computation of spectral quantities --- such as the spectral distribution $\rho$, the correlator $\mathcal{C}$, and the outlier eigenvectors --- boils  down to inverting $\mathbf{B}$. In what follows, we will show how the graph structure associated to \emph{sparse} non-Hermitian matrices $\bA$ may be exploited to invert $\mathbf{B}$.   We consider sparse random matrices $\bA$ that are locally tree-like, i.e., in the limit of $n\rightarrow \infty$  the finite neighbourhood of a randomly chosen vertex is almost surely tree-like.  

   In order to reveal the  locally tree-like structure of  $\mathbf{B}$ we  permute the rows and columns of $\mathbf{B}$.   There is no harm in permuting the rows and columns of $\mathbf{B}$, since the permutation operation commutes with matrix inversion.  In other words, if $\mathbf{P}$ is the orthogonal matrix that permutes the rows and columns of $\mathbf{B}$, then $\mathbf{B}^{-1} = \mathbf{P}^{-1} [\mathbf{P} \mathbf{B} \mathbf{P}^{-1}]^{-1}\mathbf{P}$.   
   We consider the permutation $(j,j+n)\rightarrow (2j,2j+1)$ (${\rm mod}\,\, 2n$) for all $j\in\left\{1,2,\ldots,n\right\}$ and thus   collect together elements according to the site to which they are associated.    We  obtain a new matrix $\tilde{\mathbf{B}} = \mathbf{P} \mathbf{B} \mathbf{P}^{-1}$ that is a locally tree-like block matrix built out of $2 \times 2$ blocks or submatrices.   The off-diagonal submatrices are 
   \begin{eqnarray}
 \fl  \tilde{\mathsf{B}}_{jk} = \left( \begin{array}{cc} \left[\tilde{\mathbf{B}}\right]_{2j,  2k}& \left[\tilde{\mathbf{B}}\right]_{2j, 2k+1} \\  \left[\tilde{\mathbf{B}}\right]_{2j+1, 2k}  & \left[\tilde{\mathbf{B}}\right]_{2j+1, 2k+1}  \end{array}\right)  = \left( \begin{array}{cc} \left[\mathbf{B}\right]_{j,  k}& \left[\mathbf{B}\right]_{j, k+n} \\  \left[\mathbf{B}\right]_{j+n, k}  & \left[\mathbf{B}\right]_{j+n, k+n}  \end{array}\right) 
     \end{eqnarray} 
   for all  $j,k\in\left\{1,2,\ldots,n\right\}$.   Using the definition (\ref{hjpq}) of $\mathbf{B}$  we obtain  for   $j\neq k$ 
   \begin{eqnarray}
  i \, \tilde{\mathsf{B}}_{jk} =  \mathsf{A}_{jk}\quad {\rm with} \quad  \mathsf{A}_{jk} = \left(\begin{array}{cc} 0 & A_{jk} \\ A^\ast_{kj} & 0  
\end{array}\right),\label{eq:offd}
  \end{eqnarray} 
  and 
  \begin{eqnarray}
    i \,\tilde{\mathsf{B}}_{jj} = \mathsf{A}_{jj} - \mathsf{z} - i\eta \mathbf{1}_2 
        \end{eqnarray} 
with 
\begin{eqnarray}
\mathbf{1}_2 = \left(\begin{array}{cc} 1 & 0 \\
0& 1
\end{array}\right) , \quad\mathsf{z} = \left(\begin{array}{cc} 0 & z \\
  z^\ast & 0  
\end{array}\right). \nonumber  \label{eq:offdx}
\end{eqnarray}

 Since permuting the  rows and columns of a matrix commutes with matrix inversion, the following expression holds 
\begin{eqnarray}
\fl \mathsf{G}_{jk} =   i \left( \begin{array}{cc} \left[\tilde{\mathbf{B}}^{-1} \right]_{2j,  2k}& \left[\tilde{\mathbf{B}}^{-1} \right]_{2j, 2k+1} \\  \left[\tilde{\mathbf{B}}^{-1} \right]_{2j+1, 2k}  & \left[\tilde{\mathbf{B}}^{-1} \right]_{2j+1, 2k+1}  \end{array}\right) =  i   \left(\begin{array}{cc} [\mathbf{B}^{-1}]_{j,k} & [\mathbf{B}^{-1}]_{j,k+n} \\ 
 {}[\mathbf{B}^{-1}]_{j+n,k} & [\mathbf{B}^{-1}]_{j+n,k+n}  
\end{array}\right).   \label{llpxx}
\end{eqnarray}
Note that we have dropped the dependence on $\eta$ and $z$ from the notation and we have set   $ \mathsf{G}_{jj} = \mathsf{G}_{j} $ for
simplicity. We call $\mathsf{G}_{j} $ the diagonal entries of the {\it generalized resolvent}.     
The limiting spectral distribution can be expressed in terms of the off-diagonal elements of $\mathsf{G}_{j} $ as follows  
 \begin{equation}
   \rho(z) =  \lim_{\eta \rightarrow 0^+} \lim_{n \rightarrow \infty}  \frac{1}{n \pi} 
   \sum_{j=1}^n  \partial_{z^{*}} \left[ \mathsf{G}_{j} \right]_{21} .
   \label{llp}
 \end{equation}
  The diagonal elements of $\mathsf{G}_{j}$ are also interesting, since they provide the correlations between left and right eigenvectors:
 \begin{equation}
\mathcal{C}(z) =  -\frac{1}{\pi}\lim_{\eta \rightarrow 0^+} \lim_{n \rightarrow \infty}  \frac{1}{ n^2} 
   \sum_{j=1}^n  \left[ \mathsf{G}_{j} \right]_{11}  \sum_{j=1}^n  \left[ \mathsf{G}_{j} \right]_{22}.
   \label{off}
 \end{equation}  
Thus, the off-diagonal elements of $\mathsf{G}_{j}$ characterise the spectral distribution, whereas the diagonal elements of $\mathsf{G}_{j}$ characterise  the sensitivity of the spectrum to small perturbations.  
 
We derive recursion formulae for $\mathsf{G}_{j}(z)$ at an arbitrary node $j$ using the Schur complement inversion formula  
 \begin{eqnarray}
 \left(\begin{array}{cc}\mathbf{a} & \mathbf{b} \\ \mathbf{c} & \mathbf{d} \end{array}\right)^{-1} =  \left(\begin{array}{cc} \mathbf{s}_\mathbf{d} & -\mathbf{s}_\mathbf{d} \:\mathbf{b}\mathbf{d}^{-1} \\ -\mathbf{d}^{-1}\mathbf{c}\: \mathbf{s}_\mathbf{d}&\mathbf{s}_\mathbf{a}\end{array}\right), \label{SchurInversion}
 \end{eqnarray}
 with $\mathbf{s}_\mathbf{d}  = (\mathbf{a}-\mathbf{b}\mathbf{d}^{-1}\mathbf{c})^{-1}$ the inverse of the Schur complement of $\mathbf{d}$ and $\mathbf{s}_\mathbf{a} = (\mathbf{d}-\mathbf{c}\mathbf{a}^{-1}\mathbf{b})^{-1}$ the inverse of the Schur complement of $\mathbf{a}$.      Applying formula (\ref{SchurInversion}) to the diagonal $2$ by $2$ submatrices of  $-i\tilde{\mathbf{B}}$  we find the recursion formula 
 \begin{eqnarray}
\mathsf{G}_{j}(z) &= \left(
\mathsf{z} -i\eta \mathbf{1}_2 - \mathsf{A}_{jj} - \sum^n_{k=1, (k \neq j)} \sum^n_{\ell=1, (\ell \neq j) }\mathsf{A}_{jk}\mathsf{G}^{(j)}_{k\ell}\mathsf{A}_{\ell j} \right)^{-1},
\label{circ_rec1}
\end{eqnarray}
where $\mathsf{G}_{k \ell}^{(j)}$ denotes the matrix 
\begin{eqnarray}
\mathsf{G}^{(j)}_{k\ell} = i\left(\begin{array}{cc} [(\mathbf{B}^{(j)})^{-1}]_{k,\ell} & [(\mathbf{B}^{(j)})^{-1}]_{k,\ell+n} \\ 
 {}[(\mathbf{B}^{(j)})^{-1}]_{k+n,\ell} & [(\mathbf{B}^{(j)})^{-1}]_{k+n,\ell+n}  
\end{array}\right), \nonumber 
\end{eqnarray}with $\mathbf{B}^{(j)}$  the $2(n-1)\times 2(n-1)$ submatrix of $\mathbf{B}$ obtained by deleting the rows and columns with index $j$ and $j+n$.     Note that  (\ref{circ_rec1}) reads as  $ \mathsf{G}_{j}(z) = \mathbf{s}_{\mathbf{d}} = (\mathbf{a}-\mathbf{b}\mathbf{d}^{-1}\mathbf{c})^{-1}$ with $\mathbf{a} =   -i\,\tilde{\mathsf{B}}_{jj}$, $\mathbf{b} =  -i\,\tilde{\mathsf{B}}_{j\ast}$, $\mathbf{c} = -i\,\tilde{\mathsf{B}}_{\ast j}$ and $\mathbf{d} =-i \tilde{\mathbf{B}}^{(j)}$.

In order to proceed we need to simplify the relation  (\ref{circ_rec1}) using the specifics of the ensemble under study.    We first briefly discuss the behaviour of equation (\ref{circ_rec1}) for  ensembles of {\it dense} matrices, such as matrices for which the elements $A_{jk}$ are independently and identically distributed with zero mean and variance $1/n$. In the limit of large $n$ several simplifications occur.  First,  by the strong law of large numbers, the sum on the right hand side of (\ref{circ_rec1}) approaches $(n-1)(n-2)$ times the mean of its off-diagonal terms ($k\neq\ell$) and $(n-1)$ times the mean of its diagonal terms ($k=\ell$).   Second, the decorrelation between entries implies that only diagonal terms  will contribute to the sum. Third, the order one sum dominates $\mathsf{A}_{jj}$, which is of order $1/\sqrt{n}$. Finally, since the row $j$ is arbitrary and the matrix large, we identify $\mathsf{G}_{jj}=\mathsf{G}^{(j)}_{kk}=\mathsf{g}$, with $\mathsf{g}$ a deterministic $2\times 2$ matrix. Together, these observations reduce equation (\ref{circ_rec1}) to
\begin{eqnarray}
\mathsf{g}(z) &= \left(\begin{array}{cc} g_{11} & g_{12} \\ g_{21} & g_{22}\end{array}\right)=\left(
\mathsf{z} -i\eta \mathbf{1}_2 - \left(\begin{array}{cc} g_{22} & 0 \\ 0 & g_{11}  
\end{array}\right) \right)^{-1}.
\label{circ_rec2}
\end{eqnarray}
Taking $\eta\to0$, two solutions are possible: either $g_{11}=g_{22}=0$ and $g_{12}=g_{21}^*=-z^*$, or  $g_{11}, g_{22}\neq0$ and $g_{12}=g_{21}^*=z$, the latter solution appearing if $|z|<1$. The latter case yields the uniform distribution on the unit circle when inserted into equation (\ref{llp}). This procedure has been detailed in the literature, for example in reference \cite{Tim2009}, where it arises as the high-connectivity limit of sparse non-Hermitian matrices. 

For {\it locally tree-like sparse}  matrices   with connectivity matrix  $\mathbf{C}$, off-diagonal weights $\mathbf{J}$  and  diagonal weights $\mathbf{D}$, equation (\ref{circ_rec1}) simplifies into
 \begin{eqnarray}
   \mathsf{G}_{j} &= \left(
    \mathsf{z} -i\eta \mathbf{1}_2 - \mathsf{D}_{j} - \sum_{k\in\partial_j}\mathsf{J}_{jk}\mathsf{G}^{(j)}_{k}\mathsf{J}_{kj} \right)^{-1}, \label{eq:res1}
 \end{eqnarray}
 with 
 \begin{eqnarray}
 \mathsf{J}_{jk} = \left(\begin{array}{cc} 0 & J_{jk} \\ J^\ast_{kj} & 0  
 \end{array}\right),\quad  \mathsf{D}_{j} = \left(\begin{array}{cc} 0 & D_{j} \\ D^\ast_{j} & 0  
 \end{array}\right). \nonumber 
 \end{eqnarray} 
 Note that to obtain (\ref{eq:res1}) we have set the off-diagonal terms $\mathsf{A}_{jk}\mathsf{G}^{(j)}_{k\ell}\mathsf{A}_{\ell j} = 0$.  This condition is met  when the graph with $\mathbf{C}$ as adjacency matrix  is a tree, because then the matrix $\tilde{\mathbf{B}}^{(j)}$ is a diagonal block matrix with a $|\partial_j|$ number of blocks, one for each of the neighbours of $j$.   Hence, if both $\mathsf{A}_{jk}$ and $\mathsf{A}_{jk}$  are non-zero matrices,  then $\mathsf{G}^{(j)}_{k\ell}$ is a zero matrix.   If the graph with  $\mathbf{C}$ as adjacency matrix  is locally tree-like, then    $\mathsf{A}_{jk}\mathsf{G}^{(j)}_{k\ell}\mathsf{A}_{\ell j} = 0$ holds in the limit of $n\rightarrow \infty$.

Since the sum on the right hand  side of (\ref{eq:res1}) has a number of terms of order $O(n^{0})$, the statistical arguments of the previous paragraph do not hold. It is therefore necessary to develop a theory for the objects $\mathsf{G}^{(j)}_{k}$. 
\begin{figure}[ht]
\begin{center}
\includegraphics[width=360pt]{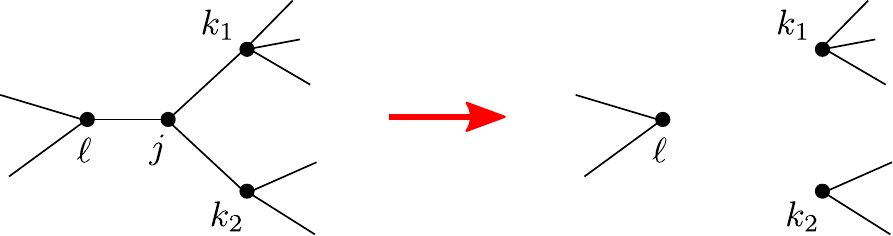}
\caption{Illustration of the local tree-like structure of sparse graphs (adapted from \cite{rogers2008cavity}). In large tree-like networks, removal of a site de-correlates the neighbours, leading to an almost exact statistical recursion relation. Here, nodes $k_1$ and $k_2$ become disconnected from each other after removal of node $j$, moreover, they are also insensitive to the additional removal of node $\ell$.}  \label{fig:cavity}  
\end{center}
\end{figure}  
Performing a second row/column removal, one finds the recursion formula
\begin{eqnarray}
\mathsf{G}^{(\ell)}_{j}  &= \left( \mathsf{z}  -i\eta \mathbf{1}_2   - \mathsf{D}_{j} -
    \sum_{k\in\partial_j\setminus \left\{\ell\right\}}\mathsf{J}_{jk}\mathsf{G}^{(j,\ell)}_{k}\mathsf{J}_{kj} \right)^{-1} ,
\label{Gil}
 \end{eqnarray}
 
where $\mathsf{G}^{(j,\ell)}_{k}$ now refers to the problem with nodes $j$ and $\ell$ both removed. Continuing this process would lead to a hierarchy of equations. For many sparse ensembles it is possible to circumvent this computational difficulty by exploiting the {\it locally tree-like} nature of the connectivity graph for large $n$ (see Figure~\ref{fig:cavity})  and closing the recursion by setting $\mathsf{G}^{(j,\ell)}_{k} \approx \mathsf{G}^{(\ell)}_{k} $ to find \cite{Tim2009}
\begin{eqnarray}
       \mathsf{G}^{(\ell)}_{j}  &= \left( \mathsf{z}  -i\eta \mathbf{1}_2   - \mathsf{D}_{j} -
    \sum_{k\in\partial_j\setminus \left\{\ell\right\}}\mathsf{J}_{jk}\mathsf{G}^{(j)}_{k}\mathsf{J}_{kj} \right)^{-1} \label{eq:res2}.
 \end{eqnarray}

Note that this relation is equivalent to simply setting $\mathsf{G}^{(j,\ell)}_{k}=\mathsf{G}^{(j)}_{k}$ in equation (\ref{Gil}), which would be exact in  a tree, as illustrated in Fig.~\ref{fig:cavity}.   The closure of the recursion formula is also closely related to the cavity method  of statistical physics  \cite{abou1973selfconsistent, mezard1988spin, mezard2001bethe, MezardBook, Tim2008, Tim2009} and Gaussian belief propagation \cite{weiss2000correctness, bickson2008gaussian}.    In the appendix we elaborate on this link and show how the relations (\ref{eq:res1}) and (\ref{eq:res2}) can be derived using statistical physics methods.

The relations (\ref{eq:res1}) and (\ref{eq:res2}) are called the {\it cavity equations}  or {\it generalised resolvent equations} for sparse  non-Hermitian  random matrices that are locally tree-like.  Solving these equations in the limit $\eta\rightarrow 0^+$ provide us an expression for the spectral distribution using (\ref{llp})  and for the correlation between right and left eigenvectors using (\ref{off}).    We now provide some insights into these relations by analysing generic properties of their solutions.  The relations (\ref{eq:res1}) and (\ref{eq:res2})  admit in the limit $\eta\rightarrow 0^+$ two solution branches which we call the {\it trivial} solution and the {\it non-trivial} solution.   In general, the non-trivial solution is associated with the absolute continuous part of the spectrum and the trivial solution is associated with all the rest.   We first discuss the trivial solution and then the non-trivial solution.  

The {\it trivial} solution is given by 
\begin{eqnarray}
    \mathsf{G}_j =  \left(\begin{array}{cc} 0 & - G^\ast_j\\ -G_j& 0  \end{array}\right) + O(\eta), \quad        \mathsf{G}^{(\ell)}_{j} =  \left(\begin{array}{cc} 0 &- (G^{(\ell)}_j)^\ast\ \\-G^{(\ell)}_j & 0 \end{array}\right)+ O(\eta), \nonumber\\ \label{eq:diagSol}
 \end{eqnarray}
where the  variables $G_j$ and  $G^{(\ell)}_{j}$ solve
 \begin{eqnarray}
G_{j}(z) &=& \frac{1}{ D_{j} - z - \sum_{k\in\partial_j}A_{jk}G^{(j)}_{k} A_{kj}}, \label{jpq} \\ 
G^{(\ell)}_{j}(z) &=& \frac{1}{D_{j} - z -\sum_{k\in \partial_j\setminus \{ \ell \} }A_{jk}G^{(j)}_{k}A_{kj}}. \label{jpq1}
 \end{eqnarray}    
 The relations (\ref{jpq}) and (\ref{jpq1}) solve the equations (\ref{eq:res1}) and (\ref{eq:res2})  for $\eta=0$.  We are however interested in the solution of the relations   (\ref{eq:res1}) and (\ref{eq:res2})   for small but nonzero values of $\eta>0$, because of the limit $\eta\rightarrow 0^+$ in  (\ref{llp}) and (\ref{off}).    We therefore must check the  stability of the trivial solution under small perturbations along the diagonal elements of $\mathsf{G}^{(\ell)}_{j}$.  
If the trivial solution   (\ref{eq:diagSol}-\ref{jpq1})  is stable, then it is a solution  the  generalised resolvent  relations    (\ref{eq:res1}) and (\ref{eq:res2}).     
 
We  discuss the meaning of the trivial solution.      Because of the  zero diagonal elements in (\ref{eq:diagSol}),  the correlator $\mathcal{C} = 0$.    Using (\ref{llp}) we find 
 \begin{eqnarray}
 \rho(z) = -\frac{1}{n \pi} 
   \sum_{j=1}^n  \partial_{z^{*}} G_{j}(z) . \label{eq:specTriv}
 \end{eqnarray}
 for the spectral distribution of the trivial solution.
In order to interpret (\ref{eq:specTriv}), we identify the 
 relations (\ref{jpq}-\ref{jpq1}) with the recursion relations for the diagonal elements of the resolvent $\mathbf{G}_{\mathbf{A}}(z)$ of a non-Hermitian matrix for $z\notin \sigma$; this follows from the Schur complement inversion formula (\ref{SchurInversion})  applied to the diagonal elements of $\mathbf{G}_{\mathbf{A}}(z)$, see \cite{bordenave2010} for a derivation for symmetric matrices.      As a consequence,  the trivial solution provides a spectral distribution $\rho$ that is a discrete sum of Dirac delta distributions $\delta(z-\lambda_{\alpha})$ supported on the pure point part of the spectrum $\sigma_{\rm pp}$ and on the singular continuous part of the spectrum $\sigma_{\rm sc}$.    For example, consider the case for  Hermitian matrices, i.e., for $A_{jk}= A^{\ast}_{kj}$.  The  relations (\ref{jpq}-\ref{jpq1}) are then the recursion relations for the diagonal elements of the  resolvent, see for instance   references \cite{abou1973selfconsistent, rogers2008cavity, biroli2010anderson, bordenave2010, aizenman2011extended}.    The spectral distribution  (\ref{eq:specTriv})  is thus equal to zero for $z\in \mathbb{C}\setminus \mathbb{R}$ and for $z\in \mathbb{R}$ we get the spectral distribution of the matrix supported on the real line.  

It is possible that (\ref{eq:diagSol}) solves the resolvent relations on a subset of all the graph nodes  $\left\{1,2,\ldots, n\right\}$.  This is the case when the graph is not strongly connected and instead consists of several connected subgraphs.  The trivial solution may then apply to   some of these subgraphs.    This happens for example for adjacency matrices of  the oriented Erd\H{o}s-R\'{e}nyi ensemble, which we will discuss in this paper.

 For small values of $|z|$ we get another solution branch to the cavity equations  (\ref{eq:res1}) and (\ref{eq:res2}), which we call the  {\it non-trivial} solution.  This solution branch has  non-zero diagonal elements, and it appears when the trivial solution becomes unstable under small perturbation along the diagonal elements of $\mathsf{G}_j$ and  $\mathsf{G}^{(k)}_j$.    
  The non-trivial solution is in general characterised by a correlator $\mathcal{C}$ that is non-zero, and therefore this solution appears in the region of the complex plane where the spectrum may be more sensitive to perturbations in the matrix elements.   In general, the non-trivial solution solves the relations  (\ref{eq:res1}) and (\ref{eq:res2})  in a set $\tilde{\sigma}$ that contains the absolute continuous part of the spectrum $\sigma_{\rm ac}$.      The boundary of  $\tilde{\sigma}$ can be determined through a linear stability analysis of the  generalized resolvent equations  (\ref{eq:res2}) around the trivial solution (\ref{eq:diagSol}).     In most cases, the two sets  $\sigma_{\rm ac}$ and  $\tilde{\sigma}$ coincide, and a linear stability analysis provides us with the boundary of the spectrum $\sigma_{\rm ac}$.  
 
  As an illustration, let us consider the case of   oriented  matrix ensembles, i.e., ensembles for which $A_{jk}A_{kj} = 0$ for all $k\neq j$.  For oriented locally tree-like ensembles the trivial solution to the cavity equations, obtained
  from equations (\ref{jpq}) and (\ref{jpq1}), takes the form
 \begin{eqnarray}
 G_j = G^{(\ell)}_j  =  \frac{1}{D_{j}-z}. \label{eq:trifvialoriented}
 \end{eqnarray}
 As a consequence,  $G_j$ is an analytic function of $z$ for $|z|>{\rm max}_j {D_j}$, and therefore from (\ref{llp})  it follows that $\rho(z)=0$  for $|z| >{\rm max}_j {D_j} $.   The trivial solution holds as long as (\ref{eq:trifvialoriented}) is a stable solution of the cavity equations (\ref{eq:res2}) under small perturbations along the diagonal elements of $\mathsf{G}^{(\ell)}_j$.    A stability analysis of the cavity equations around the trivial solution (\ref{eq:trifvialoriented}) provides us the following two criteria \cite{Neri2016}
 \begin{eqnarray}
   \frac{\sum^n_{\ell=1} \sum_{j\in \partial^{\rm out}_{\ell}}  \sum_{k\in\partial^{\rm out}_{j}} |J_{jk}|^2/|z-D_{j}|^2}{\sum^n_{\ell=1}\sum_{j\in \partial^{\rm out}_{\ell}}  }< 1    ,   \nonumber \\
  \frac{ \sum^n_{\ell=1} \sum_{j\in\partial^{\rm in}_{\ell}}
  \sum_{k\in\partial^{\rm in}_{j}} |J_{kj}|^2/|z-D_{j}|^2}{ \sum^n_{\ell=1} \sum_{j\in\partial^{\rm in}_{\ell}}}< 1.\label{eq:boundary}
 \end{eqnarray}
 The subset $\tilde{\sigma}$ of the complex plane where the non-trivial solution solves the cavity equations (\ref{eq:res1}) and (\ref{eq:res2}) is determined by the values $z\in\mathbb{C}$ for which one of the two conditions (\ref{eq:boundary}) is violated.  
   Interestingly, the relations (\ref{eq:boundary}) provide us in many cases with the boundary of the absolute continuous part of the spectrum $\sigma_{\rm ac}$, see reference  \cite{Neri2016} for a detailed analysis.

   Remarkably, the result  (\ref{eq:trifvialoriented})  implies that for oriented locally tree-like ensembles  the trace of the resolvent $1/n {\rm Tr}[\mathbf{G}_{\mathbf{A}}]$   is  independent of the  off-diagonal elements $\mathbf{J}$, provided $z \notin \sigma_{\rm ac}$. This is a signature of  universality of certain spectral properties of sparse oriented ensembles, strictly valid outside the continuous part of the spectral distribution. Equation (\ref{eq:trifvialoriented}) can be interpreted using the analogy between spectral theory and Newtonian gravity.      For simplicity, we consider the case of $D_j = 0$.   If $D_j=0$, the trace of the resolvent $\frac{1}{n} {\rm Tr}[\mathbf{G}_{\mathbf{A}}]$,  according to equations (\ref{eq:trifvialoriented}) and (\ref{eq:diagSol}), is equal to $-1/z$.  
As discussed in Section~\ref{sec:grav}, the trace of the resolvent  can be interpreted as a gravitational field  $\vec{g} = (g_x, g_y)$ that is generated by a mass distributed on the complex plane according to the spectral distribution $\rho$. The relation between the resolvent and the gravitational field  is provided by equation (\ref{eq:traceResolvent}).   For an oriented locally tree-like ensemble with $D_j = 0$, the spectral distribution $\rho$  is circularly symmetric, i.e., $\rho(e^{i\theta}z) = \rho(z)$, which follows from the fact that ${\rm Tr}[\mathbf{A}^m] = 0$ for all values of $m$.  
According to Gauss's law of gravity, a circularly symmetric mass distribution generates, in the region of space where there is no mass,   a gravitational field as if all mass would be concentrated at the central point.   This explains the universality of the expression (\ref{eq:trifvialoriented}) for the resolvent of oriented locally tree-like random matrix ensembles: the circular symmetry of the spectral distribution is sufficient to specify the form of the trace resolvent.   Moreover, since a gravitational field generated by a circular symmetric distribution of unit mass  is directed radially inwards and obeys an inverse distance law, we have that $\vec{g} = -2\vec{r}/|r|^2$, with $\vec{r} = x\vec{e}_x + y\vec{e}_y$ the radial vector that points outwards.   Hence, we have  derived   the trivial solution  (\ref{eq:trifvialoriented})  using arguments from Newtonian gravitation in two dimensions.   
We can now also understand why  universality in the trace of the resolvent is lost for non-oriented ensembles or ensembles that are not locally tree-like: in these cases the spectral distribution is not circularly symmetric and hence the form of the spectral distribution will  affect the expression of the trace resolvent.    

\subsection{Recursion relations for the eigenvectors associated with outliers}
In this subsection we present recursion relations for the eigenvector elements associated with  outlier eigenvalues of operators that are locally tree-like  \cite{Neri2016, kabashima2010cavity, kabashima2012first}.   Outlier eigenvalues are non-degenerate eigenvalues that form the discrete spectrum of an operator.    

We use the notation $R_j$ for the $j$-th element $\langle j|v \rangle$ of the  right eigenvector associated to an outlier eigenvalue $\lambda_{\rm isol}$ of a very large random matrix; analogously, we use $L_j = \langle j|u \rangle$ for the complex conjugate of the $j$-th element of the  left eigenvector associated with the outlier $\lambda_{\rm isol}$.  In the appendix we derive the following recursion relation for the eigenvector elements of outlier eigenvalues 
\begin{eqnarray}
  R_j = - G_{j}(z)  \sum_{k \in \partial_{j}} A_{jk} R_{k}^{(j)}, \label{jk1} \\
  L_{j} = -  G^{*}_{j} (z) \sum_{k \in \partial_{j}} L_{k}^{(j)}  A^{*}_{kj} ,  \label{jk2}
\end{eqnarray}   
where $z\notin \sigma_{\rm ac}$. The quantities $G_j(z)$ are the  diagonal elements of the resolvent matrix $\mathbf{G}_{\mathbf{A}}$ that solve the recursion relations (\ref{jpq}-\ref{jpq1}).   The random variables $R_{k}^{(j)}$ and $L_{k}^{(j)}$ are the elements of the right and left eigenvectors  associated with the same eigenvalue $\lambda_{\rm isol}$ of the submatrix $\mathbf{A}^{(j)}$, obtained from $\mathbf{A}$ by deleting the $j$-th row and $j$-the column. We thus require that the outlier is stable under small perturbations of the original matrix $\mathbf{A}$.    For sparse matrices with a {\it local tree-like} structure the  elements $R_{k}^{(j)}$ and $L_{k}^{(j)}$ solve the recursion relations
\begin{eqnarray}
  R_j^{(\ell)} = - G_{j}^{(\ell)}(z)  \sum_{k \in \partial_{j} \setminus \{ \ell \} } A_{jk} R_{k}^{(j)},  \label{jk3}  \\
  L_{j}^{(\ell)}  = - \left( G_{j}^{(\ell)}(z) \right)^{*} \sum_{k \in \partial_{j} \setminus \{ \ell \}  }  L_{k}^{(j)}  A^{*}_{kj},  \label{jk4}
\end{eqnarray}    
with $G_{j}^{(\ell)}(z)$ the diagonal elements of the resolvent $\mathbf{G}_{\mathbf{A}^{(j)}}$ that solve the relations~(\ref{jpq1}).

The relations (\ref{jk1}-\ref{jk4}), together with (\ref{jpq}-\ref{jpq1}), form a closed set of self-consistent equations, which we call the {\it cavity equations for outlier eigenvalues} or the {\it recursion relations for eigenvector elements}.  We can solve these cavity equations either on a given large graph or for an ensemble of random graphs. In the latter case, all the above quantities are random variables, and the relations (\ref{jk1}-\ref{jk4}) and (\ref{jpq}-\ref{jpq1}) should be interpreted as recursive distributional equations. The value of an outlier $\lambda_{\rm isol}$ is found when the above equations have a non-trivial solution, i.e., a solution for which the expected values $\langle |R_j|^2\rangle$ and $\langle |L_j|^2\rangle$are positive (non-zero) numbers. We will show in the worked examples that identifying solutions to these equations provides us precisely with the value of the outlier.   Note that the relations (\ref{jk1}-\ref{jk4}) are invariant under the rescaling by arbitrary constants, $R_j \rightarrow R_j\alpha$ and $L_j\rightarrow L_j\beta$, which reflects that if $|v_{\rm isol}\rangle$ and $\langle u_{\rm isol}|$ are left and right eigenvectors of $\mathbf{A}$, then also $|v_{\rm isol}\rangle\rightarrow \alpha |v_{\rm isol}\rangle$ and $\langle u_{\rm isol}|\rightarrow \beta \langle u_{\rm isol}|$  are left and right eigenvectors of $\mathbf{A}$.

We end this subsection by showing that Eqs. (\ref{jk1}-\ref{jpq1}) simplify in the case
of oriented random graphs.    In the previous section, we have shown that the relations (\ref{jpq}-\ref{jpq1})  are solved by
equation (\ref{eq:trifvialoriented}).
The recursion relations (\ref{jk1}-\ref{jk2})  for the eigenvector elements thus become  
\begin{eqnarray}
  R_j = \frac{1}{\left( z - D_{jj} \right)}   \sum_{k \in \partial_{j}^{\rm out}} J_{jk} R_{k}^{(j)}, \label{jk11} \\
  L_{j} = \frac{1}{\left( z - D_{jj} \right)^{*}} \sum_{k \in \partial_{j}^{\rm in}} L_{k}^{(j)}  J^{*}_{kj} ,  \label{jk21}
\end{eqnarray}  
where  $\partial_{j}^{\rm in}$ is the set of neighbouring nodes of $j$ that have a directed
link pointing towards $j$, while $\partial_{j}^{\rm out}$ is the set of neighbours of $j$ that
receive a directed edge from $j$.
In order to close the set of equations, we need expressions for the variables $\{ R_{k}^{(j)} \}_{k=1,\dots,n}$ and $\{ L_{k}^{(j)} \}_{k=1,\dots,n}$, with $j \in  \partial_{k}^{\rm in}$
and  $j \in  \partial_{k}^{\rm out}$, respectively. Comparing Eqs. (\ref{jk3}) and (\ref{jk4}) with
Eqs. (\ref{jk11}) and (\ref{jk21}) yield
\begin{eqnarray}
  R_k^{(j)} = R_k \quad {\rm if} \,\,\,\,  j \in  \partial_{k}^{\rm in} , \label{Rk11}\\
  L_k^{(j)} = L_k \quad {\rm if} \,\,\,\,  j \in  \partial_{k}^{\rm out},\label{Lk11}
\end{eqnarray}  
for $k = 1,\dots,n$.     
The above relations have two important consequences.  First, using (\ref{Rk11}-\ref{Lk11}) we find that equations (\ref{jk11}) and (\ref{jk21}) form a closed set of equations for the eigenvector components $R_j$ and $L_{j}$. Second, since the random variables $\{ R_{k}^{(j)}, L_{k}^{(j)} \}_{k=1,\dots,n}$  are statistical independent from  $\{ R_{k} , L_{k} \}_{k=1,\dots,n}$,  equations (\ref{jk11}) and (\ref{jk21})  are also closed in a distributional sense.  
This property allows to calculate ensemble averaged quantities from Eqs.~(\ref{jk11}) and (\ref{jk21}) in
a simple way, as we will see in the following sections.      
For oriented    
  random matrix  ensembles our formalism becomes thus very powerful as the recursion relations simplify considerably and generic analytical results can be derived. This simplification is reminiscent of the mathematics of  the  dynamics of  Markov process of many-body systems on graphs, where the case of oriented ensembles is also exactly solvable, see e.g.~references \cite{derrida1987exactly, neri2009cavity, aurell2011three}.  

  In the derivation of the recursion relations for the eigenvector elements of outliers we have assumed that the outlier is stable  under small perturbations of the original matrix $\mathbf{A}$.   This assumption can  be verified with the correlator $\mathcal{C}(z)$: if $\mathcal{C}(\lambda_{\rm isol}) = 0$ the outlier is stable, whereas if $\mathcal{C}(\lambda_{\rm isol})\neq 0$ the outlier can be unstable.   As  a consequence,  the outlier is for sure  stable when  the trivial solution solves the cavity equations   (\ref{eq:res1}) and (\ref{eq:res2}) for $z=\lambda_{\rm isol}$.    For oriented matrix ensembles that are locally tree-like, this amounts to verifying whether  the two conditions   (\ref{eq:boundary}) hold for  $z = \lambda_{\rm isol}$.  
  
  To summarise, in this subsection we have presented a generic formalism to compute the location of outliers in the spectrum of large sparse random matrix ensembles and to compute the statistics of the corresponding eigenvector elements.

\subsection{Interpreting and solving the cavity equations}\label{eq:sovling}
What is happening mathematically when we make the tree approximation to derive a closed set of cavity equations (\ref{eq:res2}) or (\ref{jk3})?  
By writing down eqs. (\ref{eq:res2}) and (\ref{jk3}) we have effectively replaced the non-Hermitian matrix of interest $\mathbf{A}$ with an operator $\mathcal{A}$ acting on a vector space $\mathcal{W}$.   In order to see the link between the cavity equations and the operator $\mathcal{A}$, notice   first that the quantities $\mathsf{G}^{(j)}_{k}$ appearing on the right hand side of the expression for $\mathsf{G}^{(\ell)}_{j}$ are exactly those indexed by directed edges $j\to k$ in which it is possible to traverse the underlying graph taking edges $\ell \to j$ and $j\to k$ in order, without backtracking. Repeatedly applying this argument, we see that the key object on which the cavity equations operate is the set  $\mathcal{T}$ of non-backtracking walks $w$ of the graph with adjacency matrix $\mathbf{C}$.  We label  the non-backtracking walks of length $m-1$ in terms of a directed sequence of nodes $w = j_1\rightarrow j_2\rightarrow j_3\ldots \rightarrow j_{m}$.   The sequences $w\in \mathcal{T}$ can be seen as words in a language.   A word $w' = w \rightarrow j_{m+1}$ of length $m$  can be constructed from the word $w$ of length $m-1$ using the following two rules: (i)  two consequent indices are connected by a link on the graph, i.e., either  $C_{j_{m}j_{m-1}} = 1$ or $C_{j_{m-1}j_{m}} = 1$, and (ii) indices cannot backtrack, i.e.,    $j_{m}\neq j_{m-2}$.    We now expand the scope of our language by building a vector space.   To each  word $w$  we associate a vector $|w\rangle$  and we  construct   the vector space  $\mathcal{W}$   spanned by the vectors   $|w\rangle$.  The  operator $\mathcal{A}$ acts on the  vector space $\mathcal{W}$  by the operation
\begin{equation}
\mathcal{A}|w\rangle  = \sum_{k\,:\,w\rightarrow k\in\mathcal{W}} A_{\ell k}|x\rangle_{wk}\,.
\end{equation} 
The spectral properties of  $\mathcal{A}$ are described by the cavity equations.  
If $\mathbf{C}$ is the adjacency matrix of a tree graph, i.e., the graph has no cycles, then the vector space $\mathcal{W}$ is finite dimensional.  In this case the cavity equations compute the spectral properties of a matrix and provide a discrete spectrum.   If on the other hand the graph contains at least one cycle, then the vector space 
$\mathcal{W}$ is infinite dimensional and the cavity equations compute the spectral properties of this operator,  which in general will contain both continuous  components  and discrete components.

\begin{figure}[H]
\begin{center}
\includegraphics[width=350pt]{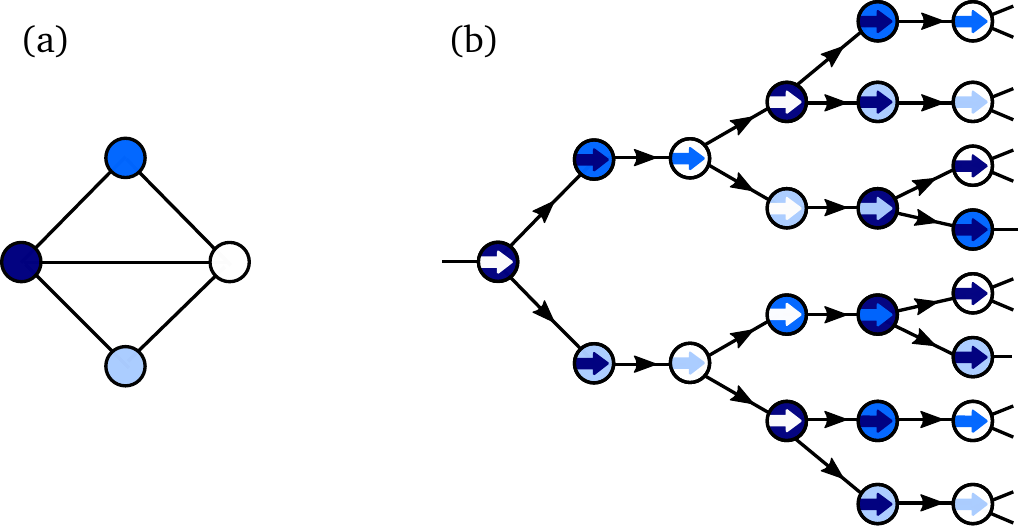}
\caption{{\bf(a)} A simple four-node graph $G$. {\bf(b)} A piece of the infinite directed tree $\mathcal{T}$ of non-backtracking walks on $G$. The five edges of $G$ give rise to ten equivalence classes of nodes in $\mathcal{T}$ indexed by the two different directions each edge may be traversed, illustrated here by the five distinct node designs.}  \label{fig:walks}  
\end{center}
\end{figure}

If there are $cn$ edges in the original graph, then the elements of the set $\mathcal{T}$ of non-backtracking walks  fall into $2cn$ equivalence classes indexed by the two different directions each edge may be traversed. The closed and finite system of $2cn$ cavity equations have an one-to-one correspondence with these equivalence classes. In the case of $k$-regular oriented random graphs we observe a further simplification: there are in fact only two equivalence classes, determined by whether the final step of the walk goes with or against the orientation of the final edge. In this special case, the tree $\mathcal{T}$ is exactly the Caley graph (or Bethe lattice) defined by the connectivity graph of the free group on three elements, see Figure \ref{fig:walks}. 
We remark that solving the cavity equations (\ref{eq:res2}) is equivalent to computing the Brown spectral measure \cite{Brown1983} of the operator $\mathcal{A}$. This topic is explored in the free probability literature, which links random matrices to operators in finite von Neumann algebras, see for example \cite{haagerup2002,sniady2002, belinschi2018eigenvalues}.  

The interest in non-backtracking walks has grown recently amongst the complex networks community, where they have applications including community detection \cite{krzakala2013}, percolation and epidemic dynamics \cite{karrer2014,rogers2015,kuhn2017}. Indeed, the so-called non-backtracking matrix, which encodes the dependencies of the cavity equations, is itself a non-Hermitian matrix whose spectral properties have been studied using the techniques we have discussed in this review \cite{saade2014}.

Having understood what the cavity equations truly represent, we move on now to discuss their practical use. There are essentially four options: exact solution for special cases, numerical iteration for fixed instances, ensemble average approximations, and population dynamics to sample from the $n\to\infty$ limit. We will discuss these solution methods in turn.  We focus here specifically on the generalised resolvent equations (\ref{eq:res2}); the recursion equations  (\ref{jk3}-\ref{jk4}) for eigenvector elements can be treated analogously.

As exemplified above for the case of $k$-regular oriented graphs, the cavity equations may be significantly reduced in number in the event that there are symmetries between the neighbourhoods of nodes in the graph. Occasionally, the reduction is so great that the number of algebraic degrees of freedom is manageable, and an exact solution can be found by hand; we demonstrate this procedure in detail in Section \ref{section4}. Unfortunately, due to the Abel--Ruffini theorem on the non-existence of algebraic solutions to general quintic polynomials, more complicated cases are very difficult to treat in generality. 

It is also possible to reduce the effective number of cavity equations by performing an ensemble average. In certain cases it is possible to perform this average exactly (see Section \ref{oripois}), however, in general it is usually necessary to make an approximation in order to close the averaged system. A simple example is the effective medium approximation, introduced for Hermitian matrices in \cite{dorogovtsev2003}. Replacing all quantities in (\ref{eq:res2}) by their ensemble averages and ignoring their fluctuations, we have 
\begin{eqnarray}
       \langle\mathsf{G}'\rangle  &= \left\langle \Big(\mathsf{z}  -i\eta \mathbf{1}_2   - \langle\mathsf{D}\rangle -
    k'\langle\mathsf{J}\langle\mathsf{G}'\rangle\mathsf{J}^\dag\rangle \Big)^{-1}\right\rangle \label{ema}\,,
 \end{eqnarray}
where $k'$ is the effective branching degree, having distribution $q_k=kp_k/\langle k\rangle$. This equation for $\langle\mathsf{G}'\rangle$ can be solved for a particular choice of disorder. This is a crude approximation, but nonetheless it is often useful for providing an estimate of bulk properties such as the diameter of support of the spectrum, particularly when the mean degree is not small.  

Alternative to attempting an analytical solution, another approach to solve the system of cavity equations is the numerical iteration of these equations on finite, but large instances of random graphs. Considering that to extract the spectral distribution via (\ref{llp}) requires taking the anti-holomorphic derivative, it is convenient to also iterate the equations for the derivatives. Differentiating on both sides of (\ref{eq:res2}) we obtain a second set of recursion relations:
\begin{eqnarray}
       \partial_{z^*}\mathsf{G}^{(\ell)}_{j}  &= -\mathsf{G}^{(\ell)}_{j} \left( \partial_{z^*}\mathsf{z} -    \sum_{k\in\partial_i\setminus \left\{\ell\right\}}\mathsf{J}_{jk}\Big(\partial_{z^*}\mathsf{G}^{(j)}_{k}\Big)\mathsf{J}_{kj} \right) \mathsf{G}^{(\ell)}_{j} \label{eq:dG},
 \end{eqnarray}
where we note that 
\begin{equation}
\partial_{z^*}\mathsf{z} = \left(\begin{array}{cc} 0 & 0 \\
  1 & 0  
\end{array}\right)\,.
\end{equation}
An analogous equation holds for     $\partial_{z^*}\mathsf{G}_{j}$.  
By iterating these equations in tandem with (\ref{eq:res2}), one can effectively compute an approximation to the spectral distribution of the underlying non-Hermitian matrix. The computational cost of this procedure should be compared to that of simply computing the eigenvalues directly, which grows as $n^3$ (although faster methods exist for computing only a subset of the eigenvalues).  

\begin{figure}[H]
\begin{center}
\includegraphics[width=300pt]{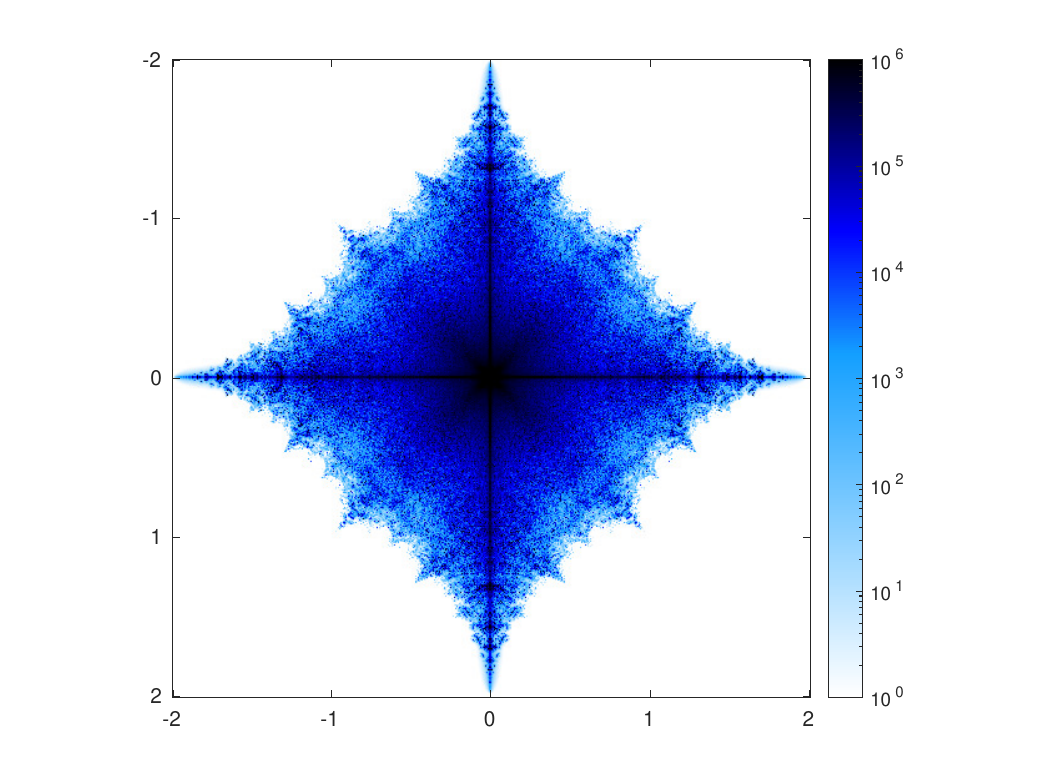}
\caption{Spectral distribution of the one-dimensional lattice with IID random weights $J_{kj},J_{jk}\in\{-1,1\}$, computed using population dynamics simulation of the cavity equations. }  \label{fig:pop}  
\end{center}
\end{figure}

It is also possible to numerically sample directly from the $n\to\infty$ limit of a sparse non-Hermitian ensemble. In this limit the tree approximation becomes exact, and we may view the cavity equations as in fact describing the distributional equations
\begin{eqnarray}
       \mathsf{G}&\stackrel{D}{=} \left( \mathsf{z}  -i\eta \mathbf{1}_2   - \mathsf{D} -
    \sum_{\ell=1}^{K}\mathsf{J}_{i\ell}\mathsf{G}'_{\ell}\mathsf{J}_{j\ell}^\dag \right)^{-1}\,,
    \label{Gpopxx}
 \end{eqnarray} 
 and 
\begin{eqnarray}
       \mathsf{G}' &\stackrel{D}{=} \left( \mathsf{z}  -i\eta \mathbf{1}_2   - \mathsf{D} -
    \sum_{\ell=1}^{K'-1}\mathsf{J}_{i\ell}\mathsf{G}'_{\ell}\mathsf{J}_{j\ell}^\dag \right)^{-1}\,,
    \label{Gpop}
 \end{eqnarray}
where $\stackrel{D}{=}$ denotes equality in distribution.  The random variables $\mathsf{D}$ and $\{\mathsf{J}_{i\ell},\}_{\ell=1}^k$
are drawn from their distributions, while $K'$ is drawn from the branching distribution, i.e., the degree distribution obtained by randomly picking an edge from the graph. Equations (\ref{Gpopxx}) and (\ref{Gpop}) are solved to obtain the distributions of   $\mathsf{G}$ and  $\mathsf{G}'$. 
One can find the distributions of $\mathsf{G}$ and  $\mathsf{G}'$ by  using a population dynamics algorithm that initialises a large sample $\{\mathsf{G}'_{\ell}\}$ and repeatedly apply equation (\ref{Gpop}) to randomly  update the entries of this sample \cite{abou1973selfconsistent, mezard2001bethe}.  For practical applications, this should be performed concurrently with updating the sample of derivatives, according to the distributional recursion equation~\cite{Tim2009}
\begin{eqnarray}
       \partial_{z^*}\mathsf{G}' &\stackrel{D}{=} -\mathsf{G}'\left( \partial_{z^*}\mathsf{z} -    \sum_{\ell=1}^{K'-1}\mathsf{J}_{j\ell}\Big(\partial_{z^*}\mathsf{G}'_{\ell}\Big)\mathsf{J}_{j\ell}^\dag \right) \mathsf{G}' \label{dGpop},
 \end{eqnarray}
 and the corresponding equation for  $\partial_{z^*}\mathsf{G}$.
In Figure~\ref{fig:pop} we show the result of population dynamics sampling for an interesting example in which the underlying graph is the one-dimensional lattice, with no diagonal disorder, but $J_{jk}$ and $J_{kj}$ chosen uniformly independently from $\{-1,1\}$. See \cite{amir2016} for further analysis of this model. 
It is also insightful to write the relations (\ref{Gpop})-(\ref{dGpop}) explicitly in terms of the densities of the random variables.   This is however only possible if we specify a model for $\mathbf{A}$.  We will therefore present such relations in Section~\ref{section4} 

The cavity equations contain information about the continuous, singular, and pure-point components of the spectrum.  However, in the limit $\eta\rightarrow 0^+$, the cavity equations loose  information about the  pure-point  component of the spectrum, since this component  exists on a discrete subset of the complex plane.  Hence, for a general $z\in\mathbb{C}$, only the continuous part of the distribution $\rho$ survives in the limit $\eta\rightarrow 0^+$.  It is therefore necessary to keep a small but finite $\eta$ in order to see the pure-point component of the spectrum.

An important question concerns the existence and uniqueness of solutions to the recursive distributional equat ion  (\ref{Gpop}).   Unfortunately, this question has not been studied in detail for non-Hermitian matrices.    Nevertheless, for Hermitian matrices, Bordenave and Lelarge have shown that the recursive distributional equation (\ref{Gpop})  admits a unique solution \cite{bordenave2010}.   Their proof relies on the Banach-fixed point theorem: they show that the recursion (\ref{Gpop}) is a contraction in a  complete metric space of distributions.  

  In the next section we apply our insights about sparse non-Hermitian random matrices to worked examples.  

\section{Worked Examples}\label{section4}
For the sake of illustration, here we solve the recursive equations presented in the last section for three simple examples of  sparse random matrices: the adjacency matrix of oriented regular random graphs,  the adjacency matrix of oriented Erd\H{o}s-R\'{e}nyi graphs, which we also call Poisson graphs or irregular graphs, and the adjacency matrix of weighted oriented Erd\H{o}s-R\'{e}nyi graphs.      More complicated examples have been treated in the literature, such as the adjacency matrix of  non-oriented regular graphs \cite{Neri2012, Neri2016},  or adjacency matrices of oriented graphs with cycles \cite{rogers2010a, Metz2011, bolle2013spectra}.  Here  we  restrict ourselves to the simplest ensembles in order to clearly demonstrate the methods presented in this paper.  

\subsection{Adjacency Matrices of Oriented Regular Random Graphs}
\label{kreg}
   The elements of adjacency matrices of simple graphs are either $1$ or $0$, corresponding to the presence or absence of an edge between two vertices.  Therefore, in our parametrisation, $D_{j} = 0$ ($j=1,\dots,n$) and $J_{jk} = 1$ for all pairs $(j,k)$.   Moreover, since the graph is oriented it holds that $C_{jk}C_{kj} = 0$  for all pairs $(j,k)$, and the $c$-regularity of the graph implies that the indegree and outdegree of all nodes is equal to $c$. In other words,  $\sum^n_{j=1}C_{jk} =  \sum^n_{j=1}C_{kj} = c$ for all values of $k$ and every node is thus connected to $2c$ edges.  

Oriented regular random graphs with $c>1$ are homogeneous in the sense that each node experiences the same local 
neighbourhood, and differences between nodes are determined by cycles of length $O(\log(n))$.  For $n\rightarrow \infty$, regular graphs with $c>1$ typically have one giant strongly connected component that contains all vertices of the graph.  The case of  $c=1$ is particular, since graphs from this ensemble typically  consist of   one giant ring together with  several disconnected smaller rings.     
 
\begin{figure}[H]
\begin{center}
\includegraphics[scale=0.28]{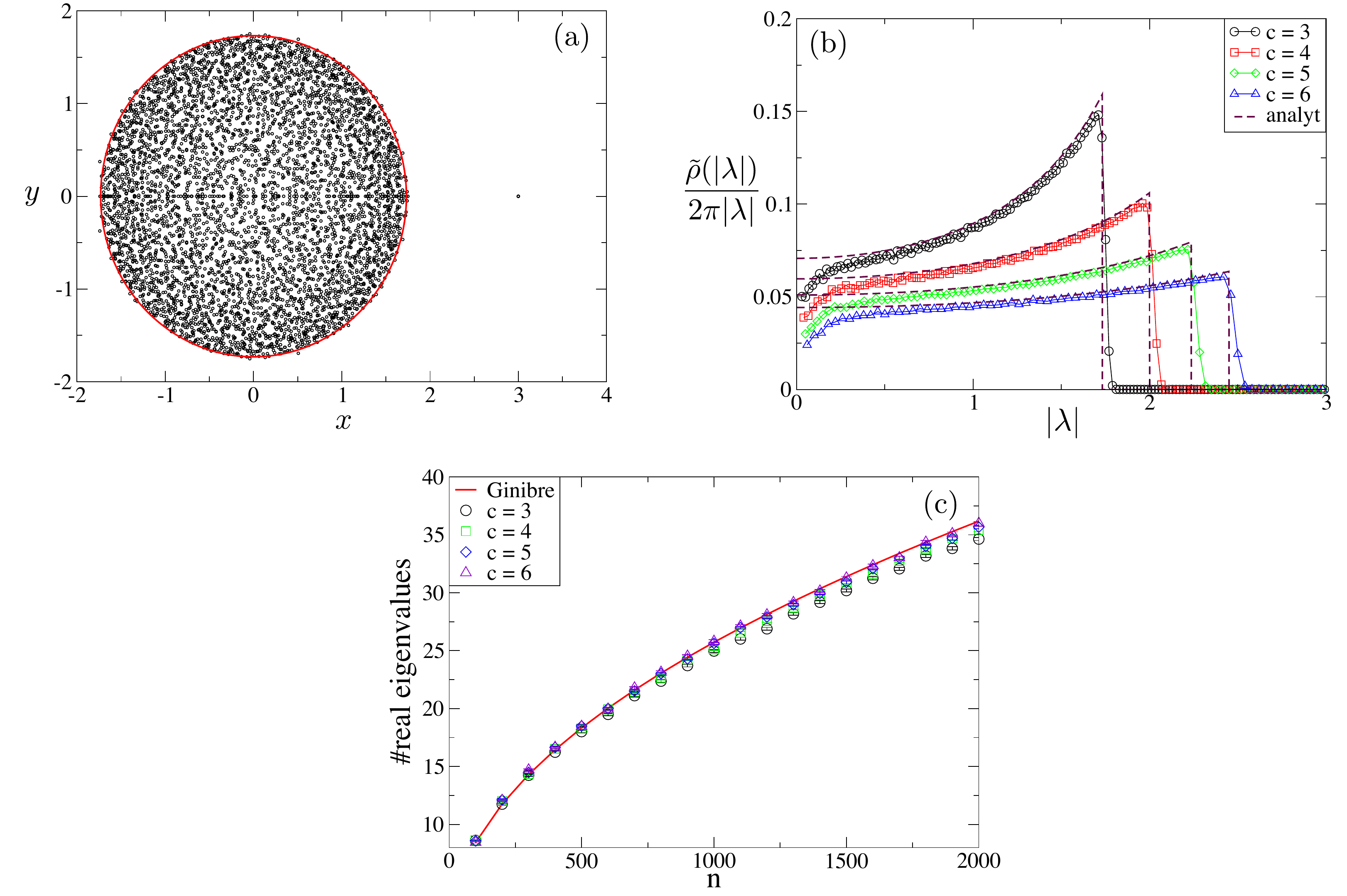}
\caption{(a) Spectrum of the adjacency matrices of oriented  random regular graphs with $c=3$ and $J_{ij}=1$, for any pair of adjacent nodes $i$ and $j$.  We show the eigenvalues of five matrix realisations of size $n=1000$ drawn from the ensemble, together with the theoretical result $|z|=\sqrt{c}$ (red solid line) for the boundary of the support. Note the presence of an outlier at $z = 3$, in agreement with the theoretical results.
 (b)  The spectral distribution as a function of $|\lambda|$ is plotted for different values of $c$.   Because of the circular symmetry of the spectral distribution $\rho(\lambda)$, we have that $\rho(\lambda) = \tilde{\rho}(|\lambda|)/(2\pi |\lambda|)$, where $\tilde{\rho}(|\lambda|)$ is the distribution of the absolute values of the eigenvalues in $\mathbb{R}$ and $\rho(\lambda)$ is the eigenvalue distribution in $\mathbb{C}$.  The numerical data is a histogram generated from $1000$ matrix realisations of size $n=2000$. The brown dashed lines are the theoretical curves given by  Eq.~(\ref{sqp11}). (c) Average number of real eigenvalues in a given matrix  from the ensemble as a function of $n$.   The red line is the theoretical result $\sqrt{2n/\pi}(1-3/(8n))+1/2$ for the Ginibre ensemble \cite{edelman1994many}.     Each data point is an   average over $1000$  realisations.  Matrices are diagonalised using the  subroutine {\it gsl\_eigen\_nonsymmv} of the  GNU Scientific Library (GSL)  (https://www.gnu.org/software/gsl/).    }  \label{fig:Poisson}  
\end{center}
\end{figure}

First insights in the spectral properties of the    ensemble of  adjacency matrices of $c$-regular oriented random graphs can be gained from brute force direct diagonalisation routines.  In Figure~\ref{fig:Poisson}(a) we present the eigenvalues of five matrices which we have drawn uniformly and at random from the regular ensemble with $c=3$ and $n = 1000$.  Each dot in Figure~\ref{fig:Poisson}(a) represents one eigenvalue.  We  notice three components in the spectrum.   First,  we note the presence of the outlier $\lambda_{\rm isol} = 3$, which appears in fact five times, but for each matrix realisation it is exactly equal to 
$3$, and therefore we only see one point in Figure~\ref{fig:Poisson}(a).   The  the uniform vectors $\langle u_{\rm isol}|j\rangle = 1/n$   and $\langle j| v_{\rm isol}\rangle = 1$  are the  left and right eigenvectors  associated with the outlier.       A second  noticeable feature in the spectrum is that all other  eigenvalues randomly occupy the space of a disc  of radius~$\sqrt{3}$, and there is no apparent structure in the locations of the eigenvalues on the disc.   We will show  that for large matrices this part of the spectrum becomes the   absolute continuous spectrum,  and we will derive the associated spectral distribution~$\rho$.   The third component of the spectrum consists of eigenvalues that accumulate on the real line: this is the singular continuous component of the spectrum.   For finite values of $c$ the number of real eigenvalues 
seems to scale as $\sqrt{n}$, as shown in Figure~\ref{fig:Poisson}(c), with a prefactor approaching the analytical result $\sqrt{2n/\pi}(1-3/(8n))+1/2$ for the Ginibre
ensemble when $c \rightarrow \infty$~\cite{edelman1994many}.

We now derive an analytical expression for the spectral distribution $\rho(z)$ of the absolute continuous part of the spectrum
in the limit $n\rightarrow \infty$. 
To this aim we direct our attention to the 
  the relations (\ref{eq:res1}) and (\ref{eq:res2})  which admit   for the present ensemble a solution   of the form 
\begin{eqnarray}
\mathsf{G}_j &=& \mathsf{g} \\  
\mathsf{G}_k^{(j)} &=& \mathsf{g}_{+} ,\quad {\rm if}\quad C_{jk} = 1, \\ 
\mathsf{G}_k^{(j)} &=& \mathsf{g}_{-} ,\quad {\rm if}\quad C_{kj} = 1. 
\end{eqnarray} 
The quantities  $\mathsf{G}_j$ and $\mathsf{G}_k^{(j)}$  are thus  independent of the vertex label $j$.  
Alternatively, if we interpret the relations  (\ref{eq:res1}) and (\ref{eq:res2}) as a set of recursive distributional equations, then the random variables $\mathsf{G}_j$ and $\mathsf{G}_k^{(j)}$ are deterministic.  
To find the values of $\mathsf{g}$,  $\mathsf{g}_{+}$ and $\mathsf{g}_{-}$, we need to solve 
 \begin{eqnarray}
   \mathsf{g}^{-1} &=
   -i\eta \mathbf{1}_2 + \mathsf{z}  - c \: \sigma_+\mathsf{g}_{+} \sigma_- - c \:\sigma_-\mathsf{g}_{-} \sigma_+, \label{eq:res1reg}\\
      \mathsf{g}_{+}^{-1} &= 
   -i\eta \mathbf{1}_2 + \mathsf{z} - c  \: \sigma_+\mathsf{g}_{+}\sigma_- - (c-1)\:\sigma_- \mathsf{g}_{-} \sigma_+, \label{eq:res2reg}\\
      \mathsf{g}_{-}^{-1} &= 
   -i\eta \mathbf{1}_2 + \mathsf{z}  - (c-1) \: \sigma_+\mathsf{g}_{+}\sigma_- - c \:\sigma_-\mathsf{g}_{-} \sigma_+, \label{eq:res3reg}
 \end{eqnarray}
 where we have used the notation
 \begin{equation}
\sigma_+ = \left(\begin{array}{cc} 0 & 1 \\ 0 & 0  
 \end{array}\right),\quad \sigma_- = \left(\begin{array}{cc} 0 & 0 \\ 1 & 0  
 \end{array}\right).
 \end{equation}  
Equations~(\ref{llp}) and  (\ref{off})  imply the following expressions for the spectral distribution    $\rho(z)$ and the left-right eigenvector correlator $\mathcal{C}$
 \begin{equation}
   \rho(z) =  \frac{1}{\pi}\lim_{\eta \rightarrow 0^+}    \partial_{z^{*}} \left[ \mathsf{g} \right]_{21} , \quad \mathcal{C}(z) =- \frac{1}{\pi}\left[ \mathsf{g} \right]_{11}\left[ \mathsf{g} \right]_{22}.
   \label{llpxxx}
 \end{equation}

Equations (\ref{eq:res1reg})-(\ref{eq:res3reg}) admit two solution branches: a first solution branch holds for values $|z|>\sqrt{c}$ and  provides us with  the  trivial solution  (\ref{eq:trifvialoriented}), which here becomes
  \begin{eqnarray}
   \mathsf{g} &= \mathsf{g}_+ = \mathsf{g}_- = \left(\begin{array}{cc} 0& 1/z^\ast \\ 1/z & 0 \end{array}\right)+ O(\eta) \label{eq:solTriv}.
 \end{eqnarray} 
 Thus, the trivial solution has $\rho(z) =\mathcal{C}(z) =  0$.

A second solution branch to the relations  (\ref{eq:res2reg}) and (\ref{eq:res3reg}) holds for values $|z|\leq \sqrt{c}$, and it is given by  
 \begin{eqnarray}
    \mathsf{g}_+ =  \frac{1}{c} \left(\begin{array}{cc}  \sqrt{ \left(\frac{c-1}{c}\right) \left( c - |z|^2 \right)} & z \\ z^*
     &  - \sqrt{ \left(\frac{c}{c-1}\right)  \left( c - |z|^2 \right)}
\end{array}\right), \\
   \mathsf{g}_- = \frac{1}{c} \left(\begin{array}{cc}  \sqrt{ \left(\frac{c}{c-1}\right) \left( c - |z|^2 \right)} & z \\ z^*
     &  - \sqrt{ \left(\frac{c-1}{c}\right)  \left( c - |z|^2 \right)}
\end{array}\right),
 \end{eqnarray} 
 which yields the following analytical expression for the resolvent
 \begin{equation}
 \mathsf{g} =  \frac{(c-1)}{\left( c^2 - |z|^2 \right)} \left(\begin{array}{cc}  \sqrt{ \left(\frac{c}{c-1}\right) \left( c - |z|^2 \right)} & z \\ z^*
     &  - \sqrt{ \left(\frac{c}{c-1}\right)  \left( c - |z|^2 \right)}
\end{array}\right).
 \end{equation}  
 Substituting the above expression for $\mathsf{g}$ in Eq.~(\ref{llp}) we
 find for $c>1$ the spectral
 distribution \cite{Rogers2010thesis,Neri2012}  
\begin{equation}
  \rho(z)=\frac{(c-1)}{\pi}\left(\frac{c }{c^2 -|z|^2}\right)^2 , \quad |z| \leq \sqrt{c}\,,
    \label{sqp11}
\end{equation}  
and the correlator between  right and left eigenvectors
\begin{equation}
  \mathcal{C}(z)=\frac{c(c-1)}{\pi}\frac{\left( c-|z|^2 \right)}{\left( c^2 - |z|^2 \right)^2} , \quad |z| \leq \sqrt{c}\,.
    \label{sqp11xx}
\end{equation} 
diagonalisation results in Figure~\ref{fig:Poisson}(b) show that the empirical spectral distribution converges to the result~(\ref{sqp11}) in the limit of large $n$.  For $|z| \rightarrow 0$, there is a discrepancy between
Eq.~(\ref{sqp11}), valid strictly in the limit $n \rightarrow \infty$, and numerical diagonalisation for the spectral
distribution.   Such difference arises due to the finite values of $n$ employed in the numerical calculations.

We discuss  some  limiting cases of the formulas (\ref{sqp11}) and     (\ref{sqp11xx}).
We first consider the limit $c\rightarrow \infty$ of sparse matrices of high degree $c$.  This limit can be taken with  the transformation $z  = \sqrt{c} \kappa$, such that the support of the spectral distribution is confined to a unit circle $|\kappa|\leq 1$. Without this rescaling, the support of the spectral distribution would cover the whole complex plane for $c\rightarrow \infty$.  The resultant spectral distribution on the $\kappa$-space is     $\rho(\kappa)=1/\pi$ if $|\kappa|\leq 1$,
and $\rho(\kappa) = 0$  otherwise.   In other words, the spectral distribution of the adjacency matrices of sparse oriented regular random graphs converges to the circular law when the connectivity $c$ is large but nevertheless small with respect to the matrix size $n$.   Analogously, we have that $\mathcal{C}(\kappa) = (1-|\kappa|^2)/\pi$ for $c\rightarrow \infty$, which is  the same formula as for the Ginibre ensemble   \cite{chalker1998eigenvector, mehlig2000statistical, PhysRevE.60.2699}.

Equation (\ref{sqp11})  reveals that $c=1$ is also a particular case:  the spectral distribution  $\rho (z)$ is then zero  for all values of $z$ except for $|z|=1$, for which Eq.~(\ref{sqp11}) diverges. Indeed, for $c=1$ all eigenvalues of the adjacency matrix lie on the unit circle.     
From a topological point of view, graphs from the $c=1$ regular ensemble consist of a large ring together with several smaller rings.   The large ring is critical, in the sense that its length $\ell$ is of the order  $O(n^{\alpha})$ with an exponent $0<\alpha<1$.    The adjacency matrix  $\mathbf{R}$ of an oriented  ring  of length $\ell$   has eigenvalues $\lambda_j = e^{\frac{2\pi j}{\ell}i}, j\in \left\{0,1,\ldots, \ell-1\right\}$. This follows  from the observation that   ${\rm Tr}[\mathbf{R}^m] = 0$ for $1 \leq m<\ell$  and ${\rm Tr}[\mathbf{R}^\ell] = 1$.   

An interesting feature of (\ref{sqp11}) is its circular symmetry in the complex plane: the spectral distribution is symmetric under rotations $\rho(e^{i\theta}z) = \rho(z)$ for all values of $\theta\in\mathbb{R}$.  The circular symmetry of the spectrum is a generic feature of adjacency matrices of oriented locally tree-like graphs.  Indeed, the adjacency matrix $\mathbf{A}$ of an oriented tree graph has the property ${\rm Tr}[\mathbf{A}^m] = 0$ for all $m\in \mathbb{N}$.   As a consequence, all eigenvalues of  $\mathbf{A}$ are equal to zero.  At finite values of $n$, random regular graphs have  cycles of length $O(\log n)$, hence, we observe non-zero eigenvalues in Figure~\ref{fig:Poisson}(a).  In the limit of $n\rightarrow \infty$ the spectrum of adjacency matrices of  oriented regular graphs converges  to the spectrum of an oriented tree operator.   The limiting operator has a continuous component in its spectrum which is circularly symmetric, i.e., $\int dz \rho(z) z^m = 0$ for all $m\in \mathbb{N}$ and  ${\rm Tr}[\mathbf{A}^m] \rightarrow \int dz \rho(z) z^m$ for large values of $m$.    For non-oriented ensembles that are locally tree-like or for oriented ensembles that are  not locally tree-like  the circular symmetry of the spectrum will be broken, see references~\cite{Neri2012, Metz2011, bolle2013spectra} for examples.   

We continue our analysis of the spectrum of adjacency matrices of random oriented regular graphs 
with a study of  the outlier eigenvalues and statistical properties of eigenvectors.    As mentioned before, for $c$-regular graphs  the uniform vector is an eigenvector associated with an outlier $\lambda_{\rm isol} = c$.    We show how this result can be recovered from the generic   relations  (\ref{jk11}), (\ref{jk21}), (\ref{Rk11}) and (\ref{Lk11}).   
For adjacency matrices of regular oriented graphs,  these relations simplify into
\begin{eqnarray}
  R_j = \frac{1}{z} \sum_{k \in \partial_j^{\rm out}} R_k,  \label{eq:r1}\\
  L_j = \frac{1}{z^*} \sum_{k \in \partial_j^{\rm in}} L_k. \label{eq:r2}
\end{eqnarray}
   Outlier eigenvalues  are values of $z\notin \sigma_{\rm ac}$ for which the recursion relations (\ref{eq:r1})-(\ref{eq:r2})  admit normalisable solutions.  We readily verify that if $z=c$, then $R_j = 1$ and $L_j = 1/n$ solve eqs. (\ref{eq:r1})-(\ref{eq:r2}) and preserve the biorthogonality property.

Can we find other values of $z$ that admit normalisable solutions to the relations (\ref{eq:r1}-\ref{eq:r2})?    
In order to find  normalisable solutions to the recursive relations  (\ref{eq:r1})-(\ref{eq:r2}) we derive  equations for
the average value of $R_j$ and the  second moment of $R_j$ 
\begin{eqnarray}
  \langle R \rangle = \frac{ c}{z} \langle R  \rangle, \label{ff1} \\
  \langle R^2  \rangle = \frac{c}{z^2} \langle R^2  \rangle + \frac{c (c-1)}{z^2}\langle R
     \rangle^2,  \label{ff2} \\
   \langle |R|^2  \rangle = \frac{c}{|z|^2} \langle |R|^2  \rangle + \frac{c (c-1)}{|z|^2}|\langle R   \rangle|^2.  \label{ff3}
\end{eqnarray} 
We identify a normalisable solution as a solution for which the mean norm fulfils $0 < \langle |R|^2\rangle < \infty$. Notice that for $z=c$, the relations (\ref{ff1}), (\ref{ff2}) and (\ref{ff3}) admit a normalisable solution with $\langle R^2  \rangle = \langle |R|^2  \rangle = \langle R   \rangle^2$, consistent with $R=1$ or with the peaked eigenvector distribution $p_R(r) = \delta(r-1)$.   Although this is a trivial exercise, it tells us
what is the general strategy to find the  location of the outlier from the eigenvector moments: one has to find the value of $z$ giving
a nontrivial solution to $\langle |R|^2  \rangle$.    

Interestingly, from the ensemble averaged equations (\ref{ff1} - \ref{ff3}) we can also identify other normalisable solutions and thus find the statistics of eigenvectors  of the ensemble that are not outliers.  Namely, if we set   $\langle R  \rangle = 0$ and $|z| = \sqrt{c}$, then we find a solution with $\langle |R|^2 \rangle \neq 0$.   Note that 
$|z| = \sqrt{c}$ is the boundary of $\sigma_{\rm ac}$, and therefore our method  provides also information about the statistics of eigenvectors at the boundary of the support.  
In this case, Eq. (\ref{ff2}) is fulfilled by setting either $\langle R^2  \rangle = 0$ or by setting $z = \sqrt{c}$. In the latter situation, we can have a solution with  $\langle R^2  \rangle \neq 0$, corresponding with a real eigenvector.   
  This simple analysis suggests that the distribution of
right eigenvector components,  corresponding with
eigenvalues at the boundary of $\sigma_{\rm ac}$, is circular symmetric in the plane $({\rm Re}[r],{\rm Im}[r])$, whereas rotational
invariance can be broken when $z$ and $J$ have the same complex argument. A similar analysis can be performed
for the distribution of the left eigenvector components at the boundary of $\sigma_{\rm ac}$, with similar results. If we define the spectral
gap $\zeta$ as the shortest distance between the outlier and the boundary of $\sigma_{\rm ac}$, we
find the finite value $\zeta =  c - \sqrt{c} $ for  oriented regular random graphs.
Below we consider a random graph model with fluctuating edges, which clarifies further the mechanism underlying the
formation of a spectral gap.

\subsection{Adjacency Matrices of Oriented  Erd\H{o}s-R\'{e}nyi  Graphs}
\label{oripois}
The theory of random graphs traces its roots back to the seminal work of Erd\H{o}s and R\'{e}nyi, who introduced two simple random graph models in which edges are placed uniformly at random (fixing either the total number of edges \cite{erdos1959} or the probability of an edge \cite{erdos1960}). In either of these models, if one takes the limit of large graph sizes, keeping the average number of edges per node fixed and finite, one obtains a \emph{Poisson} ensemble, in which the degrees follow a Poisson distribution $p_K(k)=c^ke^{-c}/k!$. Erd\H{o}s-R\'{e}nyi models are undirected and thus have symmetric adjacency matrices. 

In this section we consider the generalisation of these models to the oriented case, so that nodes have in-degrees and out-degrees both distributed according to a Poisson rule. Adjacency matrices from this ensemble can be obtained for example by randomly placing  $2cn$ entries with value one in an $n\times n$ matrix filled otherwise with zeros, where $c\in\mathbb{R}^+$. It is easy to check that, for $c$ fixed and finite, the probability of having symmetric edges  $A_{ij}=A_{ji}=1$ is vanishingly small in the large $n$ limit, and hence we refer to this model as the oriented  {\it Poisson} random graph or {\it oriented Erd\H{o}s-R\'{e}nyi} graph.  Since degrees fluctuate in oriented Erd\H{o}s-R\'{e}nyi  graphs, we refer to these graphs as {\it irregular}.  We use  the convention that $c$ stands for the mean indegree and outdegree, just as in the case of regular graphs considered before. Hence,  the mean degree per node is~$2c$.

\begin{figure}[H]
\begin{center}
\includegraphics[scale=3.5]{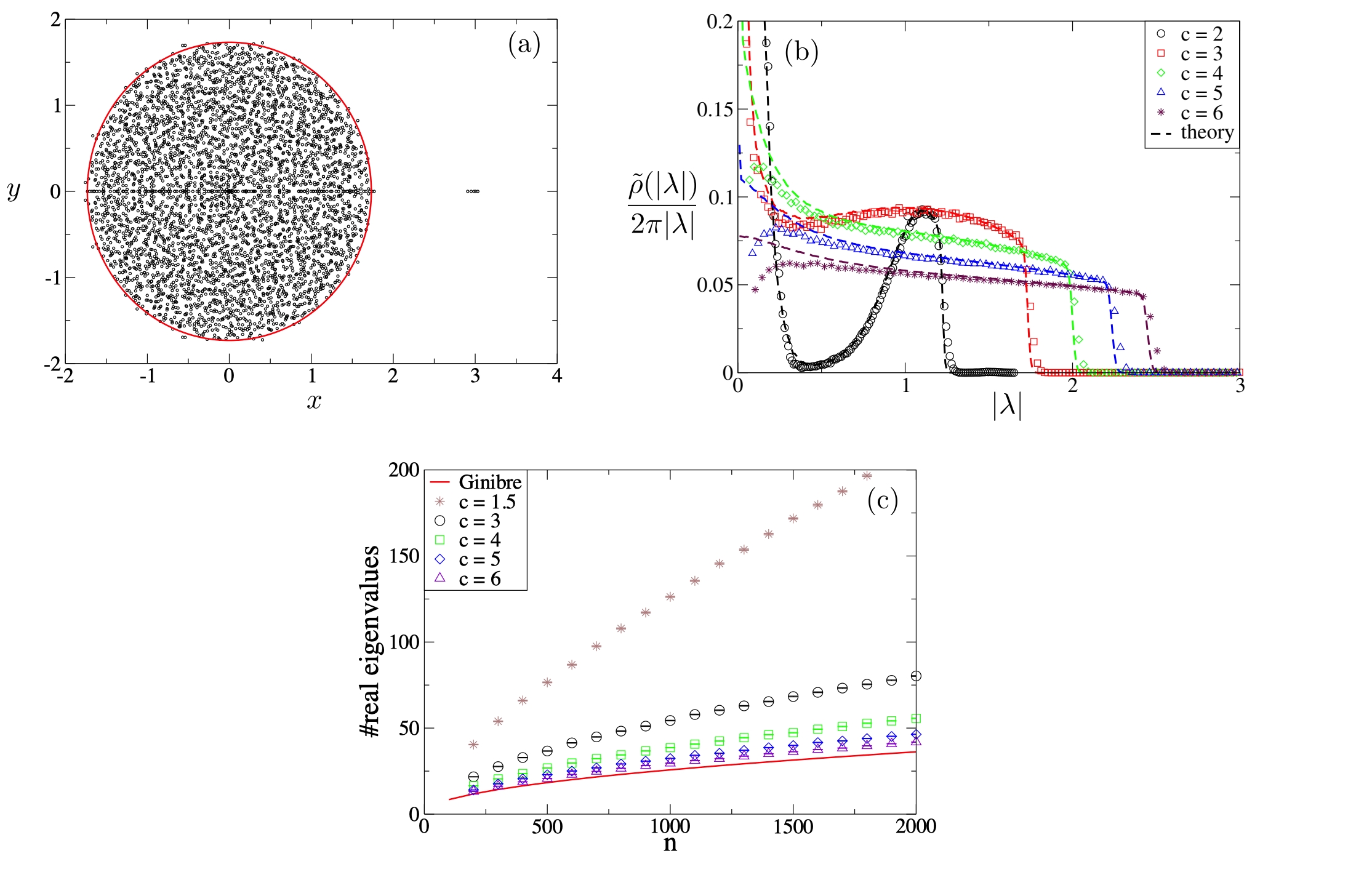}
\caption{(a) Spectrum of the adjacency matrices of oriented  Poisson random graphs with $c=3$ and $J_{ij}=1$, for any pair of adjacent nodes $i$ and $j$. We show the eigenvalues for five matrix realisations of size $n=1000$ drawn from the ensemble, together with the theoretical result $|z|=\sqrt{c}$ (red solid line) for the boundary of the support.    (b) The spectral distribution as a function of $|z|$ is plotted for different values of $c$. Because of the circular symmetry of $\rho(\lambda)$ in the complex plane, we have that $\tilde{\rho}(|\lambda|)/(2\pi|\lambda|) = \rho(\lambda)$.  The markers are  histograms generated from directed diagonalisation results of $1000$  matrix realisations of size $n=2000$.   The dashed lines are theoretical results  for $n\rightarrow \infty$ obtained by solving the equations   (\ref{eq:gPop1}) using population dynamics.  (c) Average number of real eigenvalues in a given matrix  from the ensemble as a function of $n$.   The red line is the analytical result  $\sqrt{2n/\pi}(1-3/(8n))+1/2$ for the Ginibre ensemble \cite{edelman1994many}.     Each data point is an   average over $1000$  realisations.   We consider an eigenvalue real if the absolute value of its imaginary part is smaller than 1e-20.   Matrices are diagonalised using the  subroutine {\it gsl\_eigen\_nonsymmv} of the  GNU Scientific Library (GSL)  (https://www.gnu.org/software/gsl/).    }  \label{fig:Poisson3}  
\end{center}
\end{figure}

Before addressing the theoretical results for the adjacency matrix of oriented Poisson random graphs in the limit $n\rightarrow \infty$, let us observe experimental results from direct diagonalisations of matrices of finite size.  Figure~\ref{fig:Poisson3}(a)   shows the eigenvalues of five matrices randomly drawn from the ensemble of Poisson matrices with  mean indegree and outdegree $c = 3$ and  size $n=1000$. Figure~\ref{fig:Poisson3}(a) for irregular graphs can  be compared with Figure~\ref{fig:Poisson}(a) for  regular graphs.    A striking difference between Figures~\ref{fig:Poisson3}(a) and~\ref{fig:Poisson}(a) is that for irregular oriented random graphs the outlier eigenvalues are random variables. Nevertheless, for large values of $n$ the fluctuations of the outlier become irrelevant and the outlier converges to a deterministic limit.  
  Analogously as in Figure~\ref{fig:Poisson}(a), we   observe in Figure~\ref{fig:Poisson3}(a) a bulk component in the spectrum that is localised in a circle of radius $\sqrt{c}$. Eigenvalues occupy the surface of the circle in a random manner, except for an accumulation of eigenvalues on the real axis.    Figure~\ref{fig:Poisson3}(c) shows that for $c\rightarrow \infty$  the number of real eigenvalues converges  to the analytical expression $\sqrt{2n/\pi}(1-3/(8n))+1/2$ for the Ginibre ensemble \cite{edelman1994many}. 

The topology of irregular graphs at small values of $c$ is significantly different than the topology of regular graphs.  Since these topological differences have an influence on the spectrum of the associated adjacency matrix we  briefly discuss them here.  
One of the important differences between regular  random graphs and irregular  random graphs is that irregular graphs consist of a giant connected component that coexists with  disconnected clusters of finite size, whereas for regular graphs all vertices are connected.    Another difference is that for irregular graphs the giant connected component is not strongly connected but it is instead weakly connected, see  
references  \cite{newman2001random, Ivan2016, timar2017mapping} for a clear discussion of the different giant connected components of directed graphs.    Oriented graphs can also have a giant strongly connected component.  Indeed,   oriented Erd\H{o}s-R\'{e}nyi  graphs typically contain a  giant weakly (strongly) connected component of size $O(n)$ when $c\geq 0.5$  ($c\geq 1$),  whereas for $c\leq 0.5$ ($c\leq 1$) the graph consists of a collection of weakly (strongly) disconnected clusters each of   size~$O(\log n)$.

The topological features of irregular graphs   have an influence on the spectrum of adjacency matrices of oriented Erd\H{o}s-R\'{e}nyi  graphs and lead to a spectrum that looks different than for regular graphs.     Disconnected clusters in the ensemble contribute to a pure-point component in the spectrum and also parts of the weakly connected and strongly connected components of the graph  contribute to the pure-point part of the spectrum.   On the other hand, the continuous part of the spectrum and the outlier  are generated by the giant strongly connected component of the graph.   For directed graphs there is no study  yet that carefully analyses the contribution of the different graph components to the spectrum,   in contrast to the case of  adjacency matrices of symmetric graphs \cite{semerjian2002sparse, bauer2001random, khorunzhy2004eigenvalue}.

\begin{figure}[!ht]
\begin{center}
\includegraphics[scale=0.29]{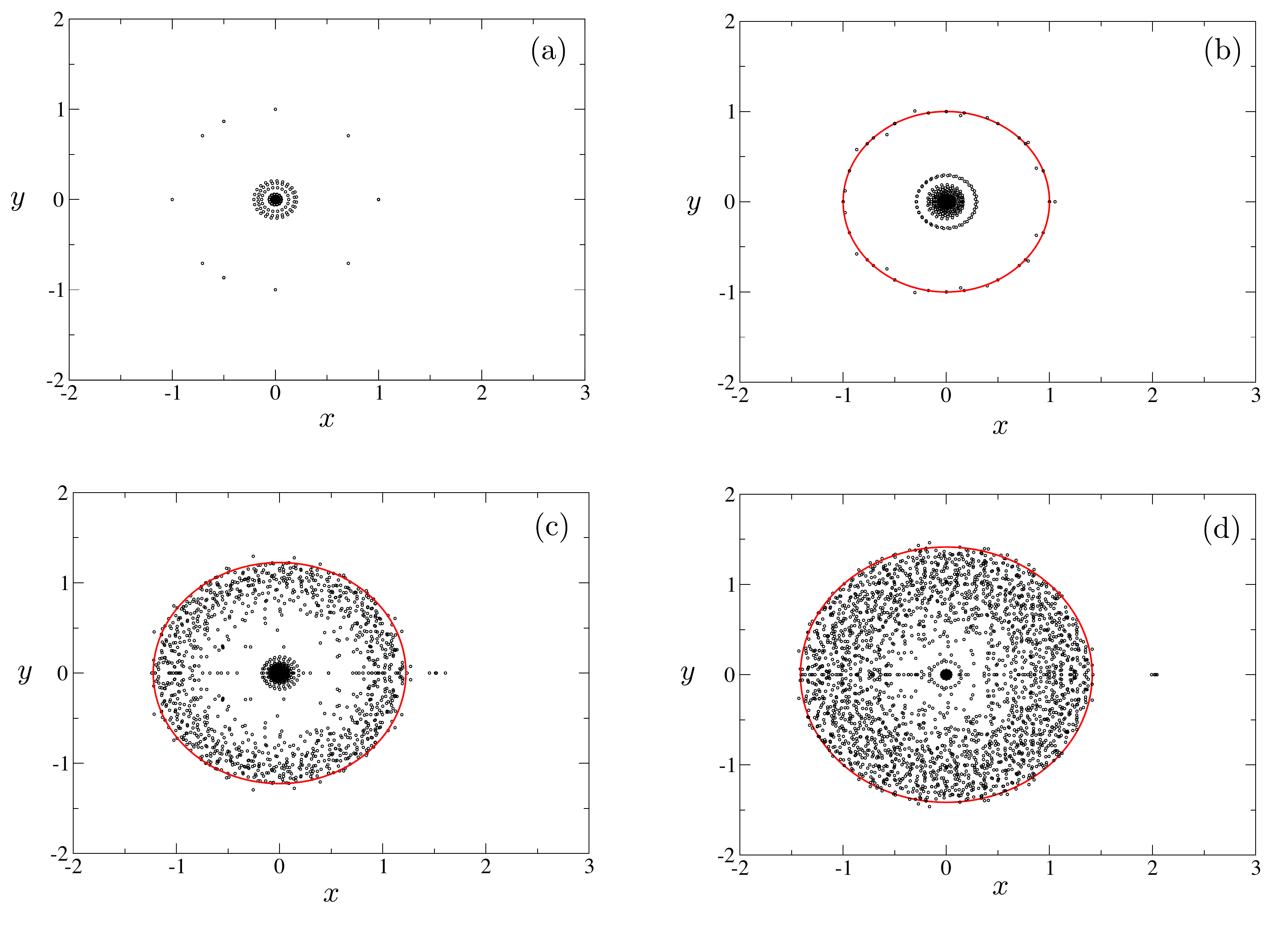}
\caption{ Spectrum of the adjacency matrices of oriented Poisson graphs at low connectivities $c$, viz. $c=0.9$ (a),  $c=1$ (b), $c=1.5$ (c) and $c=2$ (d).    Each subfigure presents the eigenvalues of five matrix realisations of size $n=1000$ randomly drawn from the ensemble.   The red solid line is the theoretical boundary $|\lambda|<\sqrt{c}$ that confines the absolute continuous part of the spectrum in the limit $n\rightarrow \infty$, and which only exists for $c>1$.    }  \label{fig:low}  
\end{center}
\end{figure}

  In order to  observe the different parts of the spectrum, we plot in Figure~\ref{fig:low} the spectra of adjacency matrices of oriented Poisson graphs at low values of the mean indegree $c$.   We observe that for $c\geq 1$  eigenvalues occupy randomly the space within  the red circle of radius $\sqrt{c}$.  For $c\rightarrow 1^+$ eigenvalues are gradually pushed towards the centre of the circle or towards its boundary.  Finally, at $c=1$ eigenvalues either localise at the boundary of the circle or at the centre.  We observe that the eigenvalues at the centre are organised in a regular manner, and do not exhibit the random spacing we observed in Figure~\ref{fig:Poisson3}(a).  The regular spacing of eigenvalues indicates  that these eigenvalues   form    the pure point part of the spectrum in the limit of large $n$, and are thus associated to isolated clusters.   For $c<1$ we observe that all eigenvalues are spaced in a regular manner, and therefore for large $n$  there will be  only a pure point component to the spectrum.    Notice that $c=1$ is the threshold for which the giant strongly connected component in the spectrum disappears \cite{dorogovtsev2001giant, timar2017mapping}, and therefore we can conclude that the  giant strongly connected component is essential for the presence of an absolute continuous component in the spectrum.

  We can partly understand the patterns of eigenvalues for $c<1$ using generic arguments.     Isolated clusters   typically consist of  finite oriented tree graphs, and therefore their adjacency matrices $\mathbf{T}$ satisfy the property ${\rm Tr}[\mathbf{T}^m] = 0$ for all values of $m\in \mathbb{N}$.  As a consequence, all the eigenvalues of $\mathbf{T}$ are equal to zero.   

Let us now focus on developing a theoretical understanding of the spectrum of oriented irregular graphs for $n\rightarrow \infty$.  
We first analyse the spectral distribution $\rho$ with the generalised resolvent relations  (\ref{eq:res1}) and (\ref{eq:res2}).   In the present ensemble, these relations simplify into 
 \begin{eqnarray}
   \mathsf{G}_{j} &= \left(
    \mathsf{z} -i\eta \mathbf{1}_2 - \sigma_- \sum_{k\in\partial^{\rm in}_j}\mathsf{G}^{(j)}_{k}\sigma_+  - \sigma_+\sum_{k\in\partial^{\rm out}_j}\mathsf{G}^{(j)}_{k}\sigma_- \right)^{-1}, \label{eq:res1xxx}
 \end{eqnarray}
 and 
\begin{eqnarray}
       \mathsf{G}^{(\ell)}_{j}  &= \left( \mathsf{z}  -i\eta \mathbf{1}_2  - \sigma_- \sum_{k\in\partial^{\rm in}_j\setminus {\ell}}\mathsf{G}^{(j)}_{k}\sigma_+  - \sigma_+\sum_{k\in\partial^{\rm out}_j\setminus {\ell}}\mathsf{G}^{(j)}_{k}\sigma_- \right)^{-1} \label{eq:res2xxx}.
 \end{eqnarray}
For values  $|z|>\sqrt{c}$, the relations (\ref{eq:res1xxx})  and (\ref{eq:res2xxx}) admit the trivial solution  (\ref{eq:trifvialoriented}) for all values of $j$,  
\begin{eqnarray}
\mathsf{G}_j = \mathsf{G}^{(\ell)}_{j} =  \left(\begin{array}{cc} 0& 1/z^\ast \\ 1/z & 0 \end{array}\right)+ O(\eta) .
\end{eqnarray}
  Hence, for $|z|>\sqrt{c}$ the spectral distribution and the correlator are given by $\rho(z) =  \mathcal{C}(z) = 0$.

\begin{figure}[ht!]
\begin{center}
\includegraphics[scale=0.7]{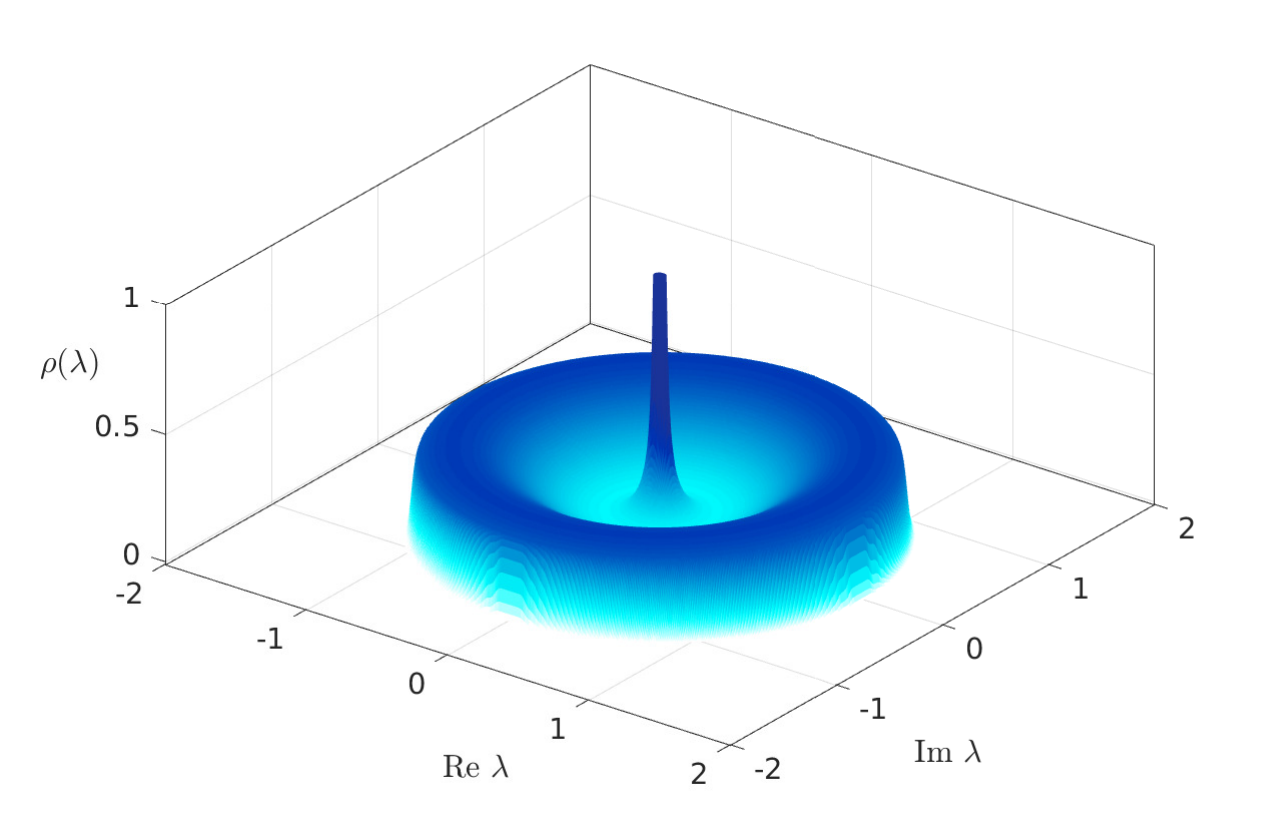}
\caption{Spectral distribution $\rho$ for the adjacency matrices of oriented Poisson graphs with mean indegree $c=2$.  Numerical solution of the equations (\ref{eq:gPop}), (\ref{eq:gPop1}) and (\ref{eq:rhoZast}) using a population dynamics algorithm.}  \label{fig:poppois}  
\end{center}
\end{figure}

For values $|z|\leq \sqrt{c}$ we  obtain a non-trivial solution branch to the relations (\ref{eq:res1xxx}) and  (\ref{eq:res2xxx}).  Indeed, the relations (\ref{eq:boundary}) for the stability of the trivial solution  reveal that for $|z|\leq \sqrt{c}$ the trivial solution is unstable under small perturbations along the diagonal elements of $\mathsf{G}_j$.     Unfortunately, for oriented Poisson graphs, we do not know an explicit  analytical expression  for the non-trivial solution.  But we can  write the  relations (\ref{eq:res1xxx}) and (\ref{eq:res2xxx})  in terms of recursive distributional equations, and solve the latter  numerically. Thus, we define the following two distributions 
\begin{eqnarray}
  \fl
p_{\mathsf{G}}(\mathsf{g}) = \lim_{n \rightarrow \infty} \frac{1}{n} \sum^n_{j=1}\delta(\mathsf{g}-\mathsf{G}_j) ,\quad p_{\mathsf{G}^c}(\mathsf{g})= \lim_{n \rightarrow \infty} \frac{1}{2nc} \sum^n_{j=1}\sum_{\ell\in \partial_j}\delta(\mathsf{g}-\mathsf{G}^{(\ell)}_j).
\end{eqnarray}
 For the present ensemble the two distributions are equal $p_{\mathsf{G}} = p^{c}_{\mathsf{G}}$, following from the fact that for Poisson graphs $p_K^{\rm in}(k)k/c =  p_K^{\rm in}(k-1)$ and  $p_K^{\rm out}(k)k/c = p_K^{\rm out}(k-1)$.  Thus, in the limit $n\rightarrow \infty$, the generalised resolvent relations  (\ref{eq:res1xxx}) and  (\ref{eq:res2xxx}) imply the  following recursive distributional equation
\begin{eqnarray}
\fl p_{\mathsf{G}}(\mathsf{g}) = \sum^{\infty}_{k^{\rm in}=0}\frac{e^{-c}c^{k^{\rm in}}}{k^{\rm in}!}\sum^{\infty}_{k^{\rm out}=0}\frac{e^{-c}c^{k^{\rm out}}}{k^{\rm out}!} \int \prod^{k^{\rm in}}_{j=1}d^2\mathsf{g}^{\rm in}_j p_{\mathsf{G}}(\mathsf{g}^{\rm in}_j) \prod^{k^{\rm out}}_{\ell=1}d^2\mathsf{g}^{\rm out}_\ell p_{\mathsf{G}}(\mathsf{g}^{\rm out}_\ell) \\ 
\times \delta\left[\mathsf{g} - \left(
    \mathsf{z} -i\eta \mathbf{1}_2 - \sigma_- \sum^{k^{\rm in}}_{j=1}\mathsf{g}^{\rm in}_{j}\sigma_+  - \sigma_+ \sum^{k^{\rm out}}_{\ell=1}\mathsf{g}^{\rm out}_{\ell}\sigma_- \right)^{-1} \right],   \label{eq:gPop}
\end{eqnarray} 
whose solution provides us with the spectral distribution and the correlator,  
\begin{eqnarray}
\fl \rho(z) =\frac{1}{\pi}  \lim_{\eta \rightarrow 0^+}  \partial_{z^{*}}  \int d^2\mathsf{g}\:p_{\mathsf{G}}(\mathsf{g})    \left[ \mathsf{g} \right]_{21}, \quad  \mathcal{C}(z) =- \frac{1}{\pi} \int d^2\mathsf{g} \: p_{\mathsf{G}}(\mathsf{g})   \left[ \mathsf{g} \right]_{11} \int d^2\mathsf{g} \: p_{\mathsf{G}}(\mathsf{g}) \left[ \mathsf{g} \right]_{22}. \nonumber\\\label{eq:adfd}
\end{eqnarray}    
We solve the relation  (\ref{eq:gPop})  with the population dynamics algorithm of \cite{abou1973selfconsistent, mezard2001bethe}, as discussed in subsection 3.4. However, the derivative $\partial_{z^{*}}$ in (\ref{eq:adfd}) leads to a large statistical error in the resultant spectral distribution.  Therefore, we are led to solve a recursive distributional equation for the joint distribution op $\mathsf{G}_j$  and $\partial_{z^\ast}\mathsf{G}_j$ \cite{Tim2009}
\begin{eqnarray}
\fl p(\mathsf{g},\mathsf{h}) = \sum^{\infty}_{k^{\rm in}=0}\frac{e^{-c}c^{k^{\rm in}}}{k^{\rm in}!}\sum^{\infty}_{k^{\rm in}=0}\frac{e^{-c}c^{k^{\rm out}}}{k^{\rm out}!}
\nonumber\\ 
 \times \int \prod^{k^{\rm in}}_{j=1}d^2\mathsf{g}^{\rm in}_j d^2\mathsf{h}^{\rm in}_j\: p(\mathsf{g}^{\rm in}_j, \mathsf{h}^{\rm in}_j) \int \prod^{k^{\rm out}}_{\ell=1}d^2\mathsf{g}^{\rm out}_\ell  d^2\mathsf{h}^{\rm out}_\ell p(\mathsf{g}^{\rm out}_\ell, \mathsf{h}^{\rm out}_\ell)   \nonumber\\
 \times\delta\left[\mathsf{g} - \left(
    \mathsf{z} -i\eta \mathbf{1}_2 - \sigma_- \sum^{k^{\rm in}}_{j=1}\mathsf{g}^{\rm in}_{j}\sigma_+  - \sigma_+ \sum^{k^{\rm out}}_{\ell=1}\mathsf{g}^{\rm out}_{\ell}\sigma_- \right)^{-1} \right]
    \nonumber\\
 \times  \delta\left[\mathsf{h}+ \mathsf{g} \left( \partial_{z^*}\mathsf{z} -  \sigma_- \sum^{k^{\rm in}}_{j=1}\mathsf{h}^{\rm in}_{j}\sigma_+  - \sigma_+ \sum^{k^{\rm out}}_{\ell=1}\mathsf{h}^{\rm out}_{\ell}\sigma_- \right)\mathsf{g} \right].\label{eq:gPop1}
 \end{eqnarray}  
The spectral distribution is then simply 
\begin{eqnarray}
\rho(z) = \frac{1}{\pi} \lim_{\eta \rightarrow 0^+}   \int d^2\mathsf{g}\int d^2\mathsf{h}\:p (\mathsf{g}, \mathsf{h})    \left[ \mathsf{h} \right]_{21}. \label{eq:rhoZast}
 \end{eqnarray}  
In Figure~\ref{fig:poppois} we present the spectral distribution $\rho$ for the adjacency matrices of oriented Poisson graphs
with mean indegree and outdegree $c=2$ in the limit $n \rightarrow \infty$ which results from solving (\ref{eq:gPop}),  (\ref{eq:gPop1}) and (\ref{eq:rhoZast}).  In Figure~\ref{fig:Poisson3} we compare direct diagonalisation results at finite size $n$ with population dynamics results for $n\rightarrow \infty$ and find a good agreement between both approaches.      Nevertheless, at small values of $|\lambda|$ we find some disagreement between theory and direct diagonalisation, which we believe are due to finite size effects.  

 If we compare the spectral distribution in Figure~\ref{fig:poppois} with Eq.~(\ref{sqp11}) for regular graphs, we observe a striking difference: for adjacency matrices of oriented  Erd\H{o}s-R\'{e}nyi graphs the spectrum is a non-monotonous function of $|z|$ and it diverges for $|z|\rightarrow 0$.  
The divergence of the spectral distribution for  $|z|\rightarrow 0$ can be understood by analysing the relation (\ref{eq:gPop})  for small values of $|z|$.     We recall that  irregular random graphs consist of a giant component together with a large number of  isolated clusters of finite size that are not connected to the giant component. Therefore, we can partition the set $\left\{1,2,\ldots, n\right\} = \mathcal{V}_{\rm wc} \cup \mathcal{V}_{\rm dc}$, with $\mathcal{V}_{\rm wc}$ the set of nodes that belong to the giant component and $\mathcal{V}_{\rm dc}$ the set of nodes that belong to an isolated finite cluster.     Since the isolated clusters have with probability one a tree topology, we have the solution 
\begin{eqnarray}
\mathsf{G}_j  = \mathsf{G}^{(\ell)}_j = \left(\begin{array}{cc} 0& 1/z^\ast \\ 1/z & 0 \end{array}\right)+ O(\eta) ,\quad j \in \mathcal{V}_{\rm dc}.  \label{eq:isolDC}
\end{eqnarray}
The variables $\mathsf{G}_j$ for $j\in \mathcal{V}_{\rm wc}$ are independent of (\ref{eq:isolDC}),  because there is no edge in the graph that connects an isolated cluster to the giant component.   We obtain the distribution
\begin{eqnarray}
p_{\mathsf{G}}(\mathsf{g}) = \beta p^{\rm wc}_{\mathsf{G}} + (1-\beta)  p^{\rm dc}_{\mathsf{G}}(\mathsf{g}), \label{eqxadsfasd}
\end{eqnarray}
with 
\begin{eqnarray}
p^{\rm dc}_{\mathsf{G}}(\mathsf{g}) = \delta(\mathsf{g} - \mathsf{z}), \label{eq:ppadf}
\end{eqnarray}
and 
\begin{eqnarray}
1-\beta=e^{-2c\beta}\label{eq:weakly}
\end{eqnarray}
 the fraction of nodes  $\lim_{n\rightarrow \infty}|\mathcal{V}^{\rm wc}|/n$ that  contribute to  the giant weakly connected component \cite{newman2001random, dorogovtsev2001giant, timar2017mapping}.      
Equations (\ref{eq:adfd}), (\ref{eqxadsfasd}) and (\ref{eq:ppadf}) imply that 
\begin{eqnarray}
\rho(z) = \beta \rho_{\rm wc}(z) + (1-\beta)\delta(z)
\end{eqnarray}
since $[\mathsf{z}]_{21} = 1/z$ and  $\partial_{z^\ast}(1/z) = \pi\delta(z)$.    The spectral distribution $\rho_{\rm wc}(z)$ is not explicitly known, but can be computed using 
\begin{eqnarray}
\rho_{\rm wc}(z) =\frac{1}{\pi}  \lim_{\eta \rightarrow 0^+}  \partial_{z^{*}}  \int d^2\mathsf{g}\:p^{\rm wc}_{\mathsf{G}}(\mathsf{g})    \left[ \mathsf{g} \right]_{21}.
\end{eqnarray}
  We  thus conclude that the spectral distribution $\rho$ contains a delta peak at the origin of the complex plane. The size of the delta peak is equal or larger than  $1-\beta$, where $\beta$ is the relative size of the giant weakly connected component of the graph.

We continue our analysis of the adjacency matrices of oriented Erd\H{o}s-R\'{e}nyi   graphs with a study of the  outlier.   
For oriented Erd\H{o}s-R\'{e}nyi   random graphs, the first and second moments of the right
eigenvectors fulfil formally the same relations as for regular oriented random graphs.  This means that the relations (\ref{eq:r1}),  (\ref{eq:r2}) and (\ref{ff1}-\ref{ff3}) also hold for adjacency matrices of oriented Erd\H{o}s-R\'{e}nyi   graphs.  However, note that now  the neighbourhoods $\partial^{\rm in}_j$ and $\partial^{\rm out}_j$ are those of oriented Erd\H{o}s-R\'{e}nyi   graphs.     Hence, the ensemble of adjacency matrices of  oriented Erd\H{o}s-R\'{e}nyi   graphs  has one outlier located at $\lambda_{\rm isol} = c$ and the distribution $p_R$ of the right-eigenvector elements of the outlier is such that $\langle R^2  \rangle = \langle |R|^2  \rangle = \langle R   \rangle^2$.   This is the same result as for regular oriented graphs and  this is in fact generally true:  the outlier $\lambda_{\rm isol}$ and the first two moments $\langle R^2  \rangle$  and $\langle R \rangle$  of the  distribution of the eigenvector elements associated with the outlier are   universal for   adjacency matrices of locally tree-like oriented  random graphs, in the sense that they only depend on the mean degree.      This universality follows from the fact that the relations  (\ref{ff1}-\ref{ff3}) only depend on $c$  and not on the degree fluctuations in the ensemble.   The higher order moments of $p_R$ will depend on the fluctuations of the degrees and are thus non-universal quantities. As a consequence, the distribution $p_R$ in the case of oriented Poisson graphs is nontrivial, in contrast to the Dirac distribution for oriented regular graphs.

\subsection{Adjacency Matrices of Weighted Oriented  Erd\H{o}s-R\'{e}nyi  Graphs}
We now consider an additional degree of complexity and add weights to the edges of  oriented  Erd\H{o}s-R\'{e}nyi  graphs.     We consider the case where  the  diagonal weights $D_j=0$ and the off-diagonal weights $J_{jk}$ are complex-valued  i.i.d.~variables drawn from a distribution $p_J$.   

   In this case the theoretical analysis of the previous section remains largely unchanged.     For large values of $|z|$ 
    the trivial solution  (\ref{eq:trifvialoriented}) solves the generalised resolvent equations and hence the spectral distribution and the correlator reads $\rho(z) =\mathcal{C}(z)= 0$. Equation (\ref{eq:boundary}) reveals that the trivial solution becomes unstable for  $|z| \leq \sqrt{c \langle |J|^2\rangle} $.   The solution to the resolvent equations for    $|z| \leq \sqrt{c \langle |J|^2\rangle} $ has to be determined numerically using the population dynamics algorithm.
    
    In what follows we focus on the properties of outliers and eigenvectors of adjacency matrices of weighted oriented  Erd\H{o}s-R\'{e}nyi  graphs.     Our starting point are the generic equations (\ref{jk11}), (\ref{jk21}), (\ref{Rk11}) and  (\ref{Lk11}), which hold for eigenvectors of outliers of oriented random matrices that are locally tree-like.   We apply these formulas to the current ensemble, and derive the following relations for the 
the first and second moments of the right
eigenvectors 
\begin{eqnarray}
  \langle R  \rangle = \frac{c \langle J \rangle}{z} \langle R  \rangle, \label{fq1} \\
  \langle R^2  \rangle = \frac{c  \,\langle J^2 \rangle  }{z^2} \langle R^2  \rangle
  + \frac{ c^2 \langle J \rangle^2  }{z^2}\langle R   \rangle^2,  \label{fq2} \\
  \langle |R|^2  \rangle = \frac{c \, \langle |J|^2 \rangle }{|z|^2} \langle |R|^2  \rangle
  + \frac{c^2 |\langle J \rangle |^2  }{|z|^2}|\langle R   \rangle|^2. \label{fq3}
\end{eqnarray}  
Notice that for $p_J(a) = \delta(a-1)$ the relations  (\ref{fq1})-(\ref{fq3}) are equivalent to the relations (\ref{ff1}-\ref{ff3}) for adjacency matrices of oriented random graphs. 
 The  relations  (\ref{fq1})-(\ref{fq3}) admit a normalisable solution  if $z=\lambda_{\rm isol} = c \langle J \rangle_J$, which
is the outlier in the complex plane for this ensemble.  The equations (\ref{fq2}) and (\ref{fq3}) at $z=c \langle J \rangle_J$ assume the form
\begin{eqnarray}
  \frac{|\langle R   \rangle|^2  }{\langle |R|^2  \rangle} = \frac{ c |\langle J \rangle_J |^2 -  \langle |J|^2 \rangle_J  }
       {c |\langle J \rangle_J |^2   } , \label{lln1} \\
        \frac{\langle R   \rangle^2  }{\langle r^2  \rangle} = \frac{ c \langle J \rangle_J^2 -  \langle J^2 \rangle_J  }
       {c \langle J \rangle_J^2  }. \label{lln2} 
\end{eqnarray}  
The above relations  are valid if  the right hand side of Eq. (\ref{lln1}) is positive, which leads
to the following condition
\begin{equation}
  \frac{|\langle J \rangle_J |^2  }{\langle |J|^2 \rangle_J  } > \frac{1}{c}
  \label{lpppp}
\end{equation}  
for the existence of the outlier $\lambda_{\rm isol} = c \langle J \rangle_J$.  
Hence, the spectral gap of oriented Poisson random graphs equals to 
$\zeta = c |\langle J  \rangle| - \sqrt{c \langle |J|^2  \rangle }$. This result holds also for 
weighted oriented random graphs that are not Poissonian,  and it shows that the rescaled gap $\zeta/c |\langle J  \rangle|$ is an universal function
of the parameter $y = \langle |J|^2 \rangle/c |\langle J \rangle|^2$ \cite{Neri2016}. 
Equation (\ref{lpppp}) allows us to draw some general conclusions. It 
shows that weighted oriented random graphs with non-fluctuating edges always have an outlier if $c>1$, whereas 
models with $\langle J \rangle_J = 0$ do not have outliers.
Equations (\ref{lln1}) and (\ref{lln2}) are the analytical expressions for the first two moments of the right eigenvector
corresponding with the outlier $c \langle J \rangle_J$. Since $\langle R^2  \rangle \neq 0$, the distribution of the right eigenvector is not
rotationally invariant in $({\rm Re}r,{\rm Im}r)$.

Remarkably, the relations (\ref{fq1})-(\ref{fq3})  also admit  normalisable solutions that correspond with right eigenvectors located at the boundary of $\sigma_{\rm ac}$, i.e., $|z| = \sqrt{c\langle |J|^2\rangle}$.   Indeed,  setting  
$\langle R \rangle = 0$ in  Eq.~(\ref{fq3})
and $\langle |R|^2  \rangle \neq 0$ we find that $|z| = \sqrt{c \langle |J|^2  \rangle}$, which
marks the boundary of the continuous component of the spectral distribution.
  There are in fact two  types  of solutions for equations (\ref{fq1})-(\ref{fq3}) at $|z| = \sqrt{c\langle |J|^2\rangle}$.  A first class of solutions has 
$\langle R \rangle = \langle R^2 \rangle = 0$ and  corresponds with  a distribution $p_R$ that is circularly symmetric in the complex plane.   Another class of solutions has $\langle R \rangle =  0 $, but $\langle  R^2 \rangle \neq 0 $. 
To identify this solution in (\ref{fq1})-(\ref{fq3})  we first
write 
$J = |J|e^{i \theta_J}$ ($ -\pi < \theta_J \leq \pi$) and $z = |z|e^{i \theta_z}$ ($ -\pi < \theta_z \leq \pi$).  Second,  we insert
$|z| = \sqrt{c \langle |J|^2  \rangle}$ in Eq. (\ref{fq2}) and consider models for which $|J|$ and $\theta$ are
statistically independent.  Finally, we observe that there is a  normalisable solution $\langle R^2  \rangle \neq 0$ provided
the argument $\theta_z$ of $z$ is equal to 
\begin{equation}
e^{2 i \theta_z} =  \langle e^{2 i \theta_J} \rangle_{\theta}.
\end{equation}  
Consider the following example of a deterministic $\theta_J$, i.e.,  $p_{\theta_J} = \delta(\theta_J-\theta_0)$.  Then there exists a solution with   
$\langle R^2 \rangle \neq 0$ provided  $z=\sqrt{c \langle |J|^2  \rangle} e^{i \theta_0}$
or $z= \sqrt{c \langle |J|^2  \rangle} e^{i (\theta_0 - \pi)}$.   For all other values of $\theta_z$ on the boundary, the distribution $p_R$ will be  circularly symmetric in the complex plane.

We end this section by illustrating how one can derive numerical results for the
 distributions of eigenvector elements $p_R$ in the limit $n \rightarrow \infty$.    This can be done by interpreting the relations (\ref{jk11}), (\ref{jk21}), (\ref{Rk11}) and  (\ref{Lk11})  in a distributional sense. 
The distributions
of the right and left eigenvector components, defined respectively as $p_{\rm R}(r)$ and  $p_{\rm L}(l)$, are formally
given by
\begin{eqnarray}
p_{\rm R}(r) = \lim_{n \rightarrow \infty} \frac{1}{n} \sum_{j=1}^n \delta\left( r - R_j  \right) , \quad
p_{\rm L}(l) = \lim_{n \rightarrow \infty} \frac{1}{n} \sum_{j=1}^n \delta\left( l - L_j  \right) .\nonumber
\end{eqnarray}  
For oriented and weighted Poisson random graphs, the right and left hand sides of Eqs.~(\ref{jk11}) and (\ref{jk21})
are equal in distribution. Since all random variables appearing in these equations are
statistically independent, one can immediately write down the following
self-consistent equations for  $p_{\rm R}(r)$ and $p_{\rm L}(l)$
\begin{eqnarray}
  \fl p_{\rm R}(r) = 
  \left\langle \delta\left(r- \frac{1}{ z } \sum^{K}_{\ell=1} J_{\ell} R_\ell \right) \right\rangle ,  \quad
  p_{\rm L}(l) =
  \left\langle \delta\left(l - \frac{1}{z^{*} } \sum^{K}_{\ell=1} J_{\ell}^{*} L_\ell \right) \right\rangle ,
  \label{PLx}
\end{eqnarray}
where $J_j$, $R_j$, $L_j$ and $K$ are i.i.d.~random variables drawn from the distributions $p_J$, $p_R$, $p_L$ and $p_K(k) = e^{-c} c^k/k!$, respectively.     We can write the self-consistent equations (\ref{PLx}) also as
\begin{eqnarray}
  \fl
  p_{\rm R}(r) = \sum_{k=0}^{\infty} \frac{e^{-c} c^k}{k!} \int
  \left( \prod^{k}_{\ell=1} d a_\ell \, d r_\ell \, p_J(a_\ell)  p_{\rm R}(r_\ell)  \right)\delta\left(r- \frac{1}{ z } \sum^{K}_{\ell=1} a_{\ell} r_\ell \right) ,\nonumber\\  \label{PR} \\
  \fl
  p_{\rm L}(l) = \sum_{k=0}^{\infty}  \frac{e^{-c} c^k}{k!} \int 
  \left( \prod^{k}_{\ell=1} d a_\ell \, dl_\ell \, p_J(a_\ell) p_{\rm L}(l_\ell) \right)
\delta\left(l - \frac{1}{ z^{*} } \sum^{K}_{\ell=1} a_{\ell}^{*} l_\ell\right) .\nonumber  \\ .
  \label{PL}
\end{eqnarray}
The above equations enable us to determine $p_{\rm R}(r)$ and $p_{\rm L}(l)$ when $z$  is equal
to an outlier $\lambda_{\rm isol}$ or when $z$ lies at the boundary of $\sigma_{\rm ac}$.

\begin{figure}[H]
\begin{center}
\includegraphics[scale=0.48]{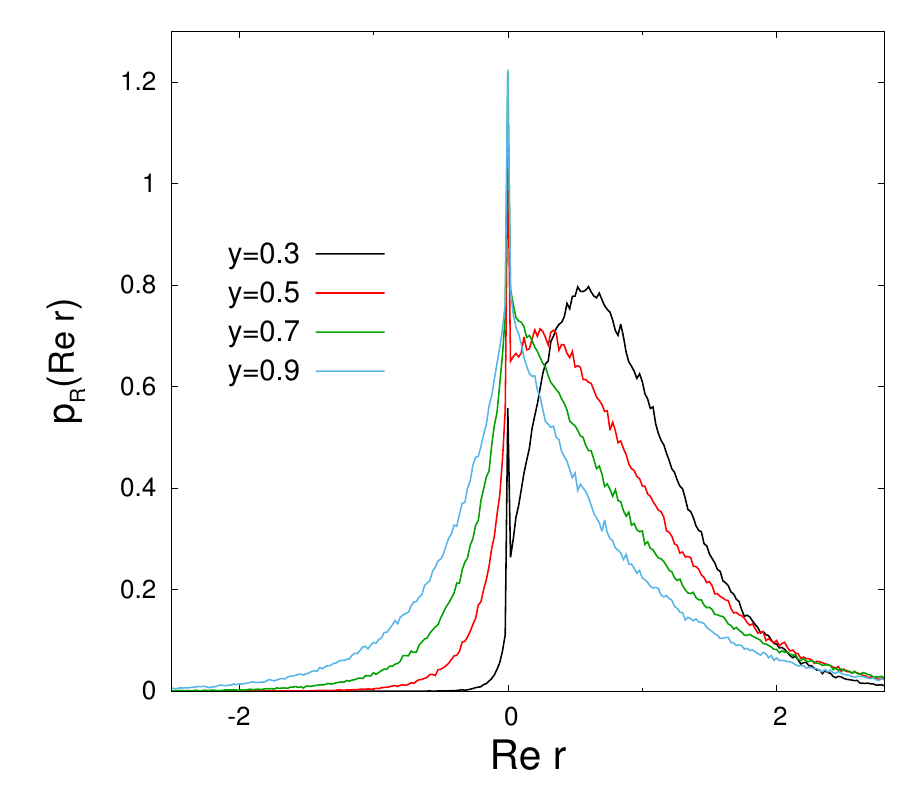}
  \includegraphics[scale=0.48]{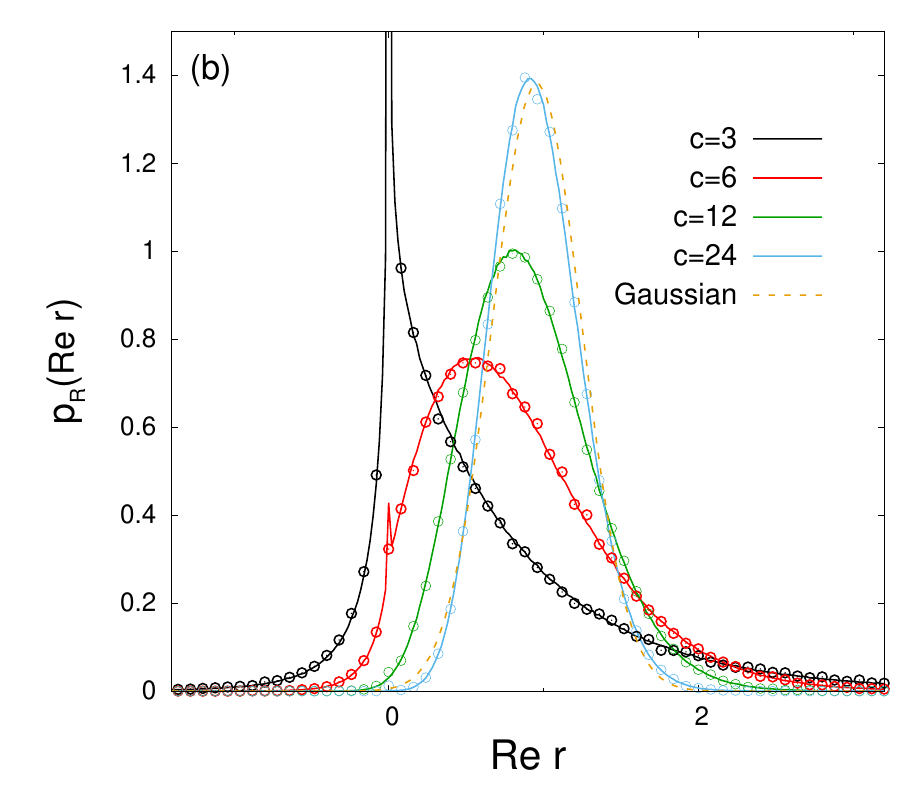}
  \caption{Distribution of the real part of the right eigenvector components at
    the outlier $z = c \langle  J \rangle$ of oriented Poisson random graphs with nonzero edges drawn from a Gaussian distribution
    with mean $\langle  J \rangle = 1$ and mean indegree (outdegree) $c$. Results are shown
    for (a) $c=5$ and different $y =\langle J^2 \rangle/c$, and for (b) fixed $\langle J^2 \rangle =2$ and increasing $c$. The
    distribution of the imaginary parts of the right eigenvector components is given by $p_{\rm R}({\rm Im}r)  =  \delta({\rm Im}r)$. The
    theoretical results for $n\rightarrow \infty$ (solid lines) are obtained from the numerical solution of Eq. (\ref{PR}) using the population dynamics method.
   The symbols are direct diagonalisation results derived from an ensemble of $100$ random matrices of size $n=1000$. 
  }
\label{fig:PoissonDistrEig}  
\end{center}
\end{figure}

Equations (\ref{PR}) and (\ref{PL})
do not admit analytical closed solutions in the general case and one has to resort to a numerical method.
Here we derive numerical solutions for $p_{\rm R}(r)$ using the population dynamics algorithm, as described in subsection 3.4. Figure \ref{fig:PoissonDistrEig} shows
 results for $p_{{\rm Re}[R]}(r)$ at the outlier $z = c \langle  J \rangle$,  derived from the numerical solution of Eq.~(\ref{PR}) in the case of edges $J \in \mathbb{R}$ drawn from a Gaussian distribution. 
For fixed average degree $c$, the right eigenvector distribution becomes gradually more symmetric
as $y = \langle J^2 \rangle/c \langle J \rangle^2$ approaches $y \rightarrow 1$, which corresponds
to a vanishing spectral gap and the merging of the outlier in the boundary of the spectral distribution.
The eigenvector distribution at the outlier is nonuniversal, since it depends on the
details of the random graph model under study, such as the degree distribution and the distribution of the
nonzero edges \cite{Neri2016}. Figure
\ref{fig:PoissonDistrEig}(b) illustrates the behaviour of  $p_{{\rm Re}[R]}(r)$ for fixed ratio
$\langle J^2 \rangle/ \langle J \rangle^2=2$ and increasing values of $c$. For $c \rightarrow \infty$,  $p_{{\rm Re}[R]}(r)$ converges
to   $p_{{\rm Re}[R]}(r) = \delta({\rm Re}\,r -1 )$, whereas the eigenvector components follow
a Gaussian distribution for large but finite $c$. The speed of convergence to the $c \rightarrow \infty$
solution is controlled by the variance of ${\rm Re}\,[R]$, which behaves as
$\langle ({\rm Re}[R])^2 \rangle - \langle {\rm Re}[R] \rangle^2 = \langle J^2 \rangle/c \langle J \rangle^2$.
Figure \ref{fig:PoissonDistrEig} also compares our theoretical results with data obtained from diagonalizing large
random matrices: the agreement
between these two independent approaches is very good in all different situations.

\section{Summary and outlook}

In this article we have reviewed some recent developments to study the spectra of sparse
non-Hermitian random matrices, whose main feature is the
presence of a large number of zero entries. Sparse non-Hermitian random matrices can be seen as the adjacency matrices of
directed random graphs, where edges have an orientation and each node of the graph is connected to a finite
number of others. Alternatively, sparse random matrices can be seen as generators of  the dynamics of complex systems.     We have discussed the general properties characterising the spectra of infinitely large non-Hermitian
random matrices and the main difficulties with the application of the resolvent method, due to the presence of complex eigenvalues.
In order to surmount some of these problems, we have shown how to implement the so-called Hermitisation procedure, which
allows us to regularise the resolvent of a non-Hermitian random matrix by means of an enlarged
normal matrix. The problem of determining certain spectral properties of non-Hermitian sparse random matrices
is then reduced to the computation of the inverse of such enlarged matrix.

Making use of the local tree-like structure of infinitely large random graphs \cite{bordenave2010}, we
have carefully explained an unified approach to invert the regularised resolvent and derive analytic results for the spectrum, 
the spectral distribution and its support, the location of eventual outliers, the
 spectral gap,  the full distribution of the eigenvectors corresponding to an outlier, and the correlations between left and right eigenvector elements.    In order to
illustrate the theory at work, we have applied it to three examples of adjacency matrices of oriented random graphs: regular random graphs,  Erd\H{o}s-R\'{e}nyi  random graphs and weighted Erd\H{o}s-R\'{e}nyi  random graphs. In these cases, we have derived the behaviour of the aforementioned
spectral observables, including a couple of analytical results.  Our theoretical results are corroborated by numerical results from direct diagonalisations of large random matrices using standard numerical routines.

We close this review by pointing out some research lines where a more complete understanding
is needed.   A first interesting research direction is the  development  of  new methods to  sparse non-Hermitian random matrices in order to provide tools to study spectral properties that are not discussed in this paper.     The theory of sparse non-Hermitian random matrices is still in its infancy when compared
to classic random matrix theory. In the latter case, one essentially deals with invariant and
fully-connected random matrices, for which the eigenvalues are decoupled from the eigenvectors and generic analytical approaches
are available. As examples, the behaviour of various spectral observables, including the spectral distribution, the
two-point eigenvalue correlation function, and the eigenvalue fluctuations, are derived from a single
mathematical object, namely the joint probability density of all eigenvalues \cite{livan2018introduction}. The situation is much more
difficult in the case of sparse random matrices, since the statistical properties of the eigenvalues 
are usually coupled to those of the eigenvectors, which prevents the analytical knowledge of the joint
probability density of eigenvalues. Thus, there is typically no unifying approach to study non-Hermitian
sparse random matrices, and one has to tailor
specific techniques to address the behaviour of specific observables, although
there have been important developments towards the understanding of the spectral distribution and the outlier eigenvalues, as demonstrated in this paper.  The behaviour of various other
spectral observables of non-Hermitian sparse random matrices have to our knowledge  barely been studied,  including 
the two-point eigenvalue correlation function, the distribution of real eigenvalues, the localisation of eigenvectors \cite{amir2016, PhysRevE.94.063305}, the study of the statistics of eigenvectors,  the identification of  contributions from the giant component and   from finite clusters,  the study of   finite size corrections to the spectrum, and  the computation of large deviation functions of spectral properties.  In this regard, it would be very interesting to develop other methods for sparse non-Hermitian matrices such as the supersymmetric \cite{fyodorov1997almost, efetov1999supersymmetry, fyodorov2003random, verbaarschot2004supersymmetric} or the   replica method \cite{rodgers1988density, kuhn2008spectra}, which have been very successful to study  properties of symmetric random matrices  \cite{metz2014finite, metz2015index, Metz2016, Reimer2016, castillo2018large, Castillo2018}.

A second interesting research direction is to provide a more detailed study of the implications of random matrix theory  for the dynamics of complex systems.  The  research in sparse random matrices has been boosted in the last two decades by the increasing
theoretical and empirical progress in the field of complex networks, which aims to understand
the structure and dynamics
of complex systems arranged in networks or graphs \cite{Doro2003, newman2018networks, dorogovtsev2013evolution, barabasi2016network, BarratBook}, being them networks of chemical reactions, social
networks of acquaintances, or the air transportation network.  The dynamics of   systems on random networks can be modelled in terms of differential equations of degrees of freedom coupled by a random matrix.  As discussed in Section~\ref{SparseDef}, the stability of fixed point solutions, the relaxation to the steady state, and the nature of the steady state are properties of dynamical systems that follow from the spectral properties of random matrix ensembles.    The fact that these properties are often universal and independent of the details of the ensemble makes that generic arguments on, for instance, the stability of ecosystems can be explored using random matrix theory \cite{may1972will}.   Random matrices provide therefore a  natural  framework to model such
large interacting complex systems and, consequently, it would be interesting to apply sparse random matrix theory to gain generic insights about  principles in the dynamics of complex systems.   

We also point out that the eigenvalues of random matrices  provide an incomplete picture of
the resilience of complex systems to perturbations.
It has been recently found that the non-normality of the adjacency matrices characterising some
complex systems amplifies the network responses to small perturbations \cite{Verdy2008,Henn2012,Malbor2018}, which may even drive the
system away from an equilibrium point. Thus, characterising the non-normality of sparse non-Hermitian
random matrices and exploring its effects on the stability of complex networks are important topics of
future research.

A third interesting research direction is to study the spectral properties of random graph ensembles that have more structure
than regular or  Erd\H{o}s-R\'{e}nyi  graphs. A crucial problem in this direction concerns  to
identify how  the network heterogeneous structure influences 
 the spectral properties of  matrices defined on the network.
Although there are many
partial results linking the network structure to the eigenvalues in an explicit
way, see for example references \cite{Doro2003,rogers2010a, Reimer2016, Peron2018,Marrec2017},  it would
be desirable to have an analytical approach to take into account different structural features  in a controllable manner.   One of the big challenges here is to understand  what is the influence of cycles on the spectra of sparse matrices 
\cite{Metz2010, bolle2013spectra,Coolen2016}, since this breaks the underlying assumption  of locally tree-likeness, which lies at the heart of the cavity method.

A final interesting research direction concerns  the development of novel  fundamental insights in the features of sparse random matrices.   In this regard, one  feature we believe is worthwhile to study in more depth are  universality properties of sparse non-Hermitian random matrices. The universal behaviour
of spectral observables has been one of the driving forces behind the enormous interest in random matrix theory
\cite{mehta2004random}.
 As argued by May in his seminal work, the stability of complex dynamical systems can be studied using random matrix theory, since spectral properties of large random matrices are to a large extent independent of the particular realisation of the
randomness.  
In the case of sparse random matrices, it is well-known that the eigenvalue distribution in the bulk
of the spectrum depends on the structural details
of the corresponding random graph model \cite{Tim2008,Tim2009,Metz2011}, such as the degree distribution
and the distribution of the weights, which seems to imply that universality  is lost for sparse systems and May's arguments do not apply.       However, as shown in \cite{Neri2016} and addressed in this review,  matrices or graphs, in which bidirected edges are absent, do have universal spectral properties outside the bulk, such as
the  spectral gap, the eigenvalue with the largest real part and the eigenvector moments corresponding to this eigenvalue \cite{Neri2016}.   These are precisely the spectral properties that  determine the stability and steady state dynamics of complex systems. 
These results show that it is possible to obtain universal behaviour in the case of sparse matrices, but
one has to look at the right spectral observables.   We have barely touched on the issue of universality, but it would be very interesting to explore this avenue to study complex systems.

\ack

FLM thanks CNPq (Edital Universal 406116/2016-4) and London Mathematical Laboratory for financial support.   
IN would like to thank  Chiara Cammarota, Silvio Franz, Yan Fyodorov,  Reimer K\"uhn,  Andrea Mambuca, and Pierpaolo Vivo  for  fruitful discussions.  
TR thanks the Royal Society for financial support. 

\section{Appendix}

In this appendix we explain an  approach to
derive analytical results for certain  spectral properties of sparse non-Hermitian
random matrices, namely, the spectral distribution $\rho$ and the outlier eigenvalues $\lambda_{\rm isol}$ and its corresponding left and right eigenvectors $\langle u_{\rm isol}|$ and $|v_{\rm isol}\rangle$.    Recall that in Section (\ref{HermMet})  we have shown that these properties are given by the  elements of the inverse of the matrix $\mathbf{B}$ in (\ref{eq:matrixB}).   In this appendix we show that the elements of the inverse of  $\mathbf{B}$  can be expressed in terms of expectation values over a complex-valued measure.   This measure is similar to the measures studies in spin-glass theory, and we can therefore use mathematical methods used in context of spin-glass theory.   
We use Gaussian belief
propagation or \cite{weiss2000correctness, bickson2008gaussian}, in the physics jargon, the cavity method \cite{abou1973selfconsistent,mezard1988spin, mezard2001bethe, MezardBook}, in order to determine the expectation values over the complex-valued measure.

Let us introduce the complex-valued Gaussian function
\begin{equation}
  \mathcal{P}_{\alpha} (\boldphi,\boldphi^{\dagger};\eta) = \frac{1}{Z_{\alpha}(\eta)}
  \exp{\left[- \sum_{k,j=1}^{2n} \phi_j^{*} B_{kj} \phi_k + i \eta \alpha \sum_{j=1}^{2 n} \left( \phi_j + \phi_j^{*} \right)    \right] } ,
  \label{lp1a}
\end{equation}
with $\boldphi \in \mathbb{C}^{2 n}$, the regularizer $\eta > 0$, and the arbitrary parameter $\alpha \in \mathbb{R}$.
The factor $Z_{\alpha}(\eta)$ is such that the function $\mathcal{P}_{\alpha} (\boldphi,\boldphi^{\dagger};\eta)$ is normalised.
The random variables $B_{kj}$ are the elements of the matrix 
$\mathbf{B}$ in (\ref{eq:matrixB}).  If the regularizer  $\eta > 0$, then the matrix  $\mathbf{B}$   is  positive-definite in the sense that  the real part of $\langle \phi|\mathbf{B}|\phi\rangle$ is positive.  This ensures that the complex valued function in (\ref{lp1a}) is normalisable and $Z_{\alpha}\in \mathbb{C}$.   Note that the condition $Z_{\alpha}\in \mathbb{C}$ provides a mean to find meaningful solutions in the limit of $n\rightarrow \infty$.  

The elements of the inverse of the matrix $\mathbf{B}$ can be expressed in terms of integrals over the function (\ref{lp1a}), 
\begin{eqnarray}
  \int d \boldphi d \boldphi^{\dagger} \: \phi_j \mathcal{P}_{1} (\boldphi,\boldphi^{\dagger};\eta) = i \eta
  \left( \mathbf{B}^{-1} |1\rangle   \right)_j, \label{eq:rel80} \\
  \int d \boldphi d \boldphi^{\dagger} \:\phi_j \phi_k^{*} \mathcal{P}_{0} (\boldphi,\boldphi^{\dagger};\eta) =
   \left( \mathbf{B}^{-1} \right)_{jk},\label{eq:rel81}
\end{eqnarray}  
with $ d \boldphi d \boldphi^{\dagger} = \prod_{j=1}^{2 n} d \phi_j d \phi_j^{*}$ and
the $2 n$-dimensional uniform vector $ |1\rangle   = (1 \,\, 1 \,\, \dots \,\, 1)^{T}$.  The relations 
(\ref{eq:rel80}) and (\ref{eq:rel81}) allow us to express the spectral properties of interest in terms of expectation values over $\mathcal{P}_{1}$ and $\mathcal{P}_{0}$.  
Using the formal marginalisation of $\mathcal{P}_{\alpha} (\boldphi,\boldphi^{\dagger};\eta)$
\begin{equation}
  \mathcal{P}_{\alpha} (\phi_j ,\phi_j^{*} ;\eta) = \int \left( \prod_{k=1 (k \neq j)}^{2n}  d \phi_k d \phi_k^{*}   \right)
  \mathcal{P}_{\alpha} (\boldphi,\boldphi^{\dagger};\eta),
\end{equation}  
and the relations (\ref{llpxx}) and (\ref{llp}) we find for the spectral distribution
\begin{equation}
  \fl
  \rho(z) = \lim_{n \rightarrow \infty} \frac{i}{n \pi} \lim_{\eta \rightarrow 0^{+}} \frac{\partial}{\partial z^*}
  \sum_{j = 1}^{n} \int d \phi_j d \phi_j^{*} d \phi_{j+n} d \phi_{j+n}^{*} \: \phi_j^{*} \phi_{j+n}
  \mathcal{P}_{0} (\phi_j ,\phi_j^{*}, \phi_{j+n} ,\phi_{j+n}^{*} ;\eta).
  \label{ghj1}
\end{equation}  
Using (\ref{eigB}), we find that the elements of the  left and right eigenvectors  of  the outlier   $\lambda_{\rm isol}$ are 
\begin{eqnarray}
  \fl
  \langle j |u_{\rm isol} \rangle = \frac{- i }{2\langle u_{\rm isol}|1\rangle} \lim_{\eta \rightarrow 0^{+}} \int d \phi_j d \phi_j^{*}\: \phi_j
  \mathcal{P}_{1} (\phi_j ,\phi_j^{*} ;\eta) \Big{|}_{z=\lambda_{\rm isol}}, \label{ghj2} \\
  \fl
   \langle j | v_{\rm isol} \rangle = \frac{- i }{2\langle v_{\rm isol}|1\rangle}  \lim_{\eta \rightarrow 0^{+}} \int d \phi_{j+n} d \phi_{j+n}^{*} \:\phi_{j+n}
  \mathcal{P}_{1} (\phi_{j+n} ,\phi_{j+n}^{*} ;\eta)\Big{|}_{z=\lambda_{\rm isol}}, \label{ghj3}
\end{eqnarray}
with $j=1,\dots,n$.  The relations (\ref{ghj1}-\ref{ghj3}) express the computation of $\rho(z) $, $\langle j |u_{\rm isol} \rangle$ and    $\langle j | v_{\rm isol} \rangle$ as a statistical physics problem, in the sense that we are left to compute marginals of a density associated with a Gaussian measure.  

Before proceeding, we make a suitable change of variables by defining the two-dimensional  column matrices
\begin{equation}
\mathsf{u}_j = \left(\begin{array}{cc} \phi_{j} \\ \phi_{j+n}   \end{array}\right), \quad j=1,\dots,n.
\end{equation}  
By substituting the explicit form of $\mathbf{B}$ in Eq. (\ref{lp1a}), we rewrite the multivariate
Gaussian function in terms of $\{ \mathsf{u}_j, \mathsf{u}_j^{\dagger} \}_{j=1,\dots,n}$
\begin{eqnarray}
  \fl
  \mathcal{P}_{\alpha} (\{ \mathsf{u}_j, \mathsf{u}_j^{\dagger} \}_{j=1,\dots,n};\eta)
  &=& \frac{1}{Z_{\alpha}(\eta)} \exp{\left[- i \sum_{j=1}^{n} \mathsf{u}_j^{\dagger}\left( \mathsf{z} - i \eta \mathbf{1}_2  \right)    \mathsf{u}_i
      + i \sum_{j,k=1}^{n} \mathsf{u}_j^{\dagger}\mathsf{A}_{jk} \mathsf{u}_k \right]} \\
  \fl
  &\times&
      \exp{\left[
      i \eta \alpha \sum_{j=1}^n \left( \mathsf{u} ^{\dagger} \mathsf{u}_j+   \mathsf{u}_j^{\dagger}\mathsf{u}  \right)
      \right]},
      \label{distr}
\end{eqnarray}
where $\mathsf{u} = (1 \,\,\, 1)^T$ and
\begin{eqnarray}
  \mathsf{A}_{kj} = \left(\begin{array}{cc} 0 & A_{kj} \\ A^\ast_{jk} & 0  
 \end{array}\right),\quad  \mathsf{z}= \left(\begin{array}{cc} 0 & z \\
   z^\ast & 0   
 \end{array}\right).
\end{eqnarray}  
In terms of $\{ \mathsf{u}_j, \mathsf{u}_j^{\dagger} \}_{j=1,\dots,n}$, Eqs. (\ref{ghj1}-\ref{ghj3}) assume the form
\begin{eqnarray}
 \fl
  \rho(z) = \lim_{n \rightarrow \infty} \frac{i}{n \pi} \lim_{\eta \rightarrow 0^{+}} \frac{\partial}{\partial z^*}
  \sum_{j = 1}^{n} \int d \mathsf{u}_j  d \mathsf{u}_j^{\dagger}
  \left[ \mathsf{u}_j^{\dagger}  \right]_1 \left[ \mathsf{u}_j \right]_2 
  \mathcal{P}_{0} ( \mathsf{u}_j,\mathsf{u}_j^{\dagger}  ;\eta), \label{j1} \\
  \fl
   \langle j | u_{\rm isol} \rangle = - \frac{i}{2    \langle u_{\rm isol} | 1 \rangle } \lim_{\eta \rightarrow 0^{+}} \int  d \mathsf{u}_j  d \mathsf{u}_j^{\dagger}
    \left[ \mathsf{u}_j \right]_1  \mathcal{P}_{1} (  \mathsf{u}_j , \mathsf{u}_j^{\dagger}  ;\eta)\Big{|}_{z=\lambda_{\rm isol}},  \label{j3} \\ 
    \fl
  \langle j | v_{\rm isol} \rangle = - \frac{i}{2    \langle v_{\rm isol} | 1 \rangle } \lim_{\eta \rightarrow 0^{+}} \int d \mathsf{u}_j  d \mathsf{u}_j^{\dagger}
  \left[ \mathsf{u}_j \right]_2  \mathcal{P}_{1} ( \mathsf{u}_j , \mathsf{u}_j^{\dagger} ;\eta) \Big{|}_{z=\lambda_{\rm isol}} \label{j2}.  
\end{eqnarray}
with $ d \mathsf{u}_j  d \mathsf{u}_j^{\dagger} = d \phi_j d \phi_j^{*} d \phi_{j+n} d \phi_{j+n}^{*}$. 

Equation (\ref{distr})
can be interpreted as a joint distribution of two-dimensional state vectors $\{ \mathsf{u}_j  \}_{j=1,\dots,n}$ interacting
through the elements of the random matrix $\bA$, the latter corresponding with a single instance of a sparse random graph.
According to Eqs. (\ref{j1}-\ref{j3}), the problem of determining the spectral properties of $\bA$ has been reduced to the calculation
of the local marginals $\mathcal{P}_{\alpha} (  \mathsf{u}_j , \mathsf{u}_j^{\dagger}  ;\eta)$ ($j = 1,\dots,n$) on the vertices of
a random graph. Below we show how to derive a set of recursive equations for  $\mathcal{P}_{\alpha} (  \mathsf{u}_j , \mathsf{u}_j^{\dagger}  ;\eta)$
using the cavity method.

By integrating Eq. (\ref{distr}) over all variables $\{ \mathsf{u}_k , \mathsf{u}_k^{\dagger}  \}_{k=1,\dots,n}$, except for $\mathsf{u}_j$ and
$\mathsf{u}_j^{\dagger}$, we obtain
\begin{eqnarray}
  \fl
  \mathcal{P}_{\alpha} (  \mathsf{u}_j , \mathsf{u}_j^{\dagger}  ;\eta)
  \sim \exp{\left[  - i \mathsf{u}_j^{\dagger}. \left( \mathsf{z} - i \eta \mathbf{1}_2  \right)    \mathsf{u}_j
      + i  \mathsf{u}_j^{\dagger}   \mathsf{A}_{jj}     \mathsf{u}_j
      + i \eta \alpha \left( \mathsf{u}^{\dagger}\mathsf{u}_j +   \mathsf{u}_j^{\dagger}\mathsf{u} \right)
      \right]} \nonumber \\
  \fl
  \times
  \int \left( \prod_{k \in \partial_j} d \mathsf{u}_k d \mathsf{u}_k^{\dagger}   \right)
  \exp{\left[  i \sum_{k \in \partial_j } \mathsf{u}_j^{\dagger} \mathsf{A}_{jk}  \mathsf{u}_k
    + i \sum_{k \in \partial_j } \mathsf{u}_k^{\dagger} \mathsf{A}_{kj}  \mathsf{u}_j    \right]}
  \mathcal{P}^{(j)}_{\alpha} (\{ \mathsf{u}_k, \mathsf{u}_k^{\dagger} \}_{k \in \partial_j };\eta),
  \label{hjpa}
\end{eqnarray}  
where $\partial_j$ is the neighbourhood of node $j$, i.e., the set of nodes adjacent to $j$ on the graph defined by $\bA$. The function
$\mathcal{P}^{(j)}_{\alpha} (\{ \mathsf{u}_k, \mathsf{u}_k^{\dagger} \}_{k \in \partial_j };\eta)$ denotes the distribution
of the state variables $\{ \mathsf{u}_k, \mathsf{u}_k^{\dagger} \}_{k \in \partial_j }$ on the cavity graph, consisting of
a graph with $n-1$ nodes obtained from $\bA$ by removing its $j$-th row and column. The approximation
symbol in the above equation means that we are neglecting the prefactor that normalises
$\mathcal{P}_{\alpha} (  \mathsf{u}_j , \mathsf{u}_j^{\dagger}  ;\eta)$; such prefactor is irrelevant for the
development of the cavity method. We assume that, for $n \rightarrow \infty$, our random graph models
have a local tree structure, i.e., the nodes in $\partial_j$ belong to disconnected branches. This
implies on the factorisation  in terms of local marginals
\begin{equation}
  \mathcal{P}^{(j)}_{\alpha} (\{ \mathsf{u}_k, \mathsf{u}_k^{\dagger} \}_{k \in \partial_j };\eta)
  = \prod_{k \in \partial_j}   \mathcal{P}^{(j)}_{\alpha} (\mathsf{u}_k, \mathsf{u}_k^{\dagger};\eta),
  \label{plo}
\end{equation}
which forms the essence of the cavity method or belief propagation. Equation (\ref{plo})
is a direct consequence of the absence of short loops on the graph when $n \rightarrow \infty$.
Inserting Eq.~(\ref{plo}) back in Eq.~(\ref{hjpa}) leads to
\begin{eqnarray}
  \fl
  \mathcal{P}_{\alpha} (  \mathsf{u}_j , \mathsf{u}_j^{\dagger}  ;\eta)
  \sim \exp{\left[  - i \mathsf{u}_j^{\dagger}\left( \mathsf{z} - i \eta \mathbf{1}_2  \right)    \mathsf{u}_j
      + i  \mathsf{u}_j^{\dagger}    \mathsf{A}_{jj}     \mathsf{u}_j
      + i \eta \alpha \left( \mathsf{u}^{\dagger}\mathsf{u}_j +   \mathsf{u}_j^{\dagger}.\mathsf{u}  \right)
      \right]} \nonumber\\
  \fl
  \times
   \prod_{k \in \partial_j} \int d \mathsf{u}_k d \mathsf{u}_k^{\dagger}   
  \exp{\left(  i  \mathsf{u}_j^{\dagger} \mathsf{A}_{jk}  \mathsf{u}_k
    + i  \mathsf{u}_k^{\dagger}. \mathsf{A}_{kj}  \mathsf{u}_j    \right)}
  \mathcal{P}^{(j)}_{\alpha} (\mathsf{u}_k, \mathsf{u}_k^{\dagger};\eta).
  \label{ggg}
\end{eqnarray}  
We still need to find self-consistent equations determining the distributions on the
cavity graph.
Let us consider $\mathcal{P}^{(\ell)}_{\alpha} (\mathsf{u}_j, \mathsf{u}_j^{\dagger};\eta)$, where $\ell \in \partial_j$. By
following the same steps as discussed above, we obtain 
\begin{eqnarray}
  \fl
  \mathcal{P}^{(\ell)}_{\alpha} (  \mathsf{u}_j , \mathsf{u}_j^{\dagger}  ;\eta)
  \sim \exp{\left[  - i \mathsf{u}_j^{\dagger}. \left( \mathsf{z} - i \eta \mathbf{1}_2  \right)    \mathsf{u}_j
      + i  \mathsf{u}_j^{\dagger}   \mathsf{A}_{jj}     \mathsf{u}_j
      + i \eta \alpha \left( \mathsf{u}^{\dagger} \mathsf{u}_j +   \mathsf{u}_j^{\dagger}\mathsf{u} \right)
      \right]} \nonumber \\
  \fl
  \times
   \prod_{k \in \partial_j\setminus \{ \ell \}} \int d \mathsf{u}_k d \mathsf{u}_k^{\dagger}   
  \exp{\left[  i  \mathsf{u}_j^{\dagger}\mathsf{A}_{jk}  \mathsf{u}_k
    + i  \mathsf{u}_k^{\dagger} \mathsf{A}_{kj}  \mathsf{u}_j    \right]}
  \mathcal{P}^{(j)}_{\alpha} (\mathsf{u}_k, \mathsf{u}_k^{\dagger};\eta),
  \label{fga}
\end{eqnarray}  

with $\partial_j \setminus  \{ \ell \}$ representing the neighbourhood of $j$ without node $\ell$. We
have been able to close Eq.~(\ref{fga}) for the single-site cavity marginals only by setting
$\mathcal{P}^{(i,\ell)}_{\alpha} (\mathsf{u}_j, \mathsf{u}_j^{\dagger};\eta) = \mathcal{P}^{(i)}_{\alpha} (\mathsf{u}_j, \mathsf{u}_j^{\dagger};\eta)$.
This is valid for sparse random graphs in the limit $n \rightarrow \infty$, since the local tree property
ensures that removing  $\ell \in \partial_i$ does not change the behaviour of the distributions on 
other nodes in $\partial_i$.

The sets of Eqs. (\ref{ggg}) and (\ref{fga}) form a closed system of recursive equations for the marginals on the nodes
of a graph. Since the functions $\mathcal{P}^{(i)}_{\alpha} (\mathsf{u}_j, \mathsf{u}_j^{\dagger};\eta)$
and $\mathcal{P}_{\alpha} (\mathsf{u}_j, \mathsf{u}_j^{\dagger};\eta)$ are marginals of a multivariate
Gaussian distribution, they can be parameterised as follows
\begin{eqnarray}
  \fl
  \mathcal{P}^{(j)}_{\alpha} (\mathsf{u}_k, \mathsf{u}_k^{\dagger};\eta) \sim
  \exp{\left[ - i\mathsf{u}_k^{\dagger} \left( \mathsf{G}_{k}^{(j)} \right)^{-1}   \mathsf{u}_k
      + i \alpha \left( \mathsf{u}_k^{\dagger} \mathsf{H}_{k}^{(j)}  + (\mathsf{H}_{k}^{(j)})^{\dagger}\mathsf{u}_k  \right) 
      \right]},  \label{jkhh1} \\
   \fl
  \mathcal{P}_{\alpha} (\mathsf{u}_k, \mathsf{u}_k^{\dagger};\eta) \sim
  \exp{\left[ - i \mathsf{u}_k^{\dagger}\mathsf{G}_{k}^{-1}    \mathsf{u}_k
      + i \alpha \left(  \mathsf{u}_k^{\dagger} \mathsf{H}_{k}  + \mathsf{H}_{k}^{\dagger}\mathsf{u}_k  \right) 
      \right]},   \label{jkhh2}
\end{eqnarray}  
where $\mathsf{G}_{k}$ and $\mathsf{G}^{(j)}_{k}$ are $2 \times 2$ matrices, while
$\mathsf{H}_{k}$ and  $\mathsf{H}_{k}^{(j)} $ are two-dimensional column vectors. Substituting
Eq.~(\ref{jkhh2}) in Eqs.~(\ref{j1}-\ref{j3}) we obtain
\begin{eqnarray}
 \fl
  \rho(z) = \lim_{n \rightarrow \infty} \frac{1}{n \pi} \lim_{\eta \rightarrow 0^{+}} \frac{\partial}{\partial z^*}
  \sum_{j = 1}^{n} \left[\mathsf{G}_{j}\right]_{21}, \label{kl1} \\
\fl
  R_j  = - i \lim_{\eta \rightarrow 0^{+}}  \left[\mathsf{G}_{j}\mathsf{H}_{j}\right]_{2}  ,  \label{kl2} \\
  \fl
   L_j = - i \lim_{\eta \rightarrow 0^{+}}  \left[\mathsf{G}_{j}\mathsf{H}_{j}\right]_{1},  \label{kl3}
\end{eqnarray}
where we have used the definitions  $R_j \equiv 2    \langle v_{\rm isol} | 1 \rangle \langle j | v_{isol} \rangle$ and $L_j \equiv 2    \langle u_{\rm isol} | 1 \rangle \langle j | u_{ isol} \rangle$.
The equations determining $\{ \mathsf{G}_{j} \}_{j=1,\dots,n}$ and $\{ \mathsf{H}_{j} \}_{j=1,\dots,n}$
are derived by plugging Eqs.~(\ref{jkhh1}) and (\ref{jkhh2}) in Eq. (\ref{ggg}) and  performing
the remaining Gaussian integrals on the right hand side

\begin{eqnarray}
  \mathsf{G}_{j}^{-1} = \mathsf{z} - i \eta \mathbf{1}_2
  -  \mathsf{A}_{jj}  - \sum_{k \in \partial_j} \mathsf{A}_{jk} \mathsf{G}_{k}^{(j)} \mathsf{A}_{kj} , \label{d1}  \\
  \mathsf{H}_{j} = \eta \mathsf{u} + \sum_{k \in \partial_j} \mathsf{A}_{jk} \mathsf{G}_{k}^{(j)} \mathsf{H}_{k}^{(j)}, \label{d2} 
\end{eqnarray}  
with $j=1,\dots,n$. The quantities $\{ \mathsf{G}_{j} \}_{j=1,\dots,n}$ and $\{ \mathsf{H}_{j} \}_{j=1,\dots,n}$, defined
on the graph corresponding to $\bA$, are given in terms of the parametrisation on the cavity graph.
Substituting Eq.~(\ref{jkhh1}) in (\ref{fga}) and  performing
the remaining Gaussian integrals on the right hand side, one derives
the following self-consistent equations
\begin{eqnarray}
  \left( \mathsf{G}_{j}^{(\ell)}\right)^{-1} = \mathsf{z} - i \eta \mathbf{1}_2
  -  \mathsf{A}_{jj}  - \sum_{k \in \partial_j \setminus \{ \ell \}} \mathsf{A}_{jk} \mathsf{G}_{k}^{(j)} \mathsf{A}_{kj} , \label{d3}  \\
  \mathsf{H}_{j}^{(\ell)} = \eta \mathsf{u} + \sum_{k \in \partial_j \setminus \{ \ell \}} \mathsf{A}_{jk} \mathsf{G}_{k}^{(j)} \mathsf{H}_{k}^{(j)}, \label{d4} 
\end{eqnarray}  
for $j=1,\dots,n$ and $\ell \in \partial_j$. Equations (\ref{d1}-\ref{d4}) form a closed set of recursive equations
that determine the spectral distribution and the outlier eigenpair of sparse non-Hermitian matrices through Eqs. (\ref{kl1}-\ref{kl3}).

The limit $\eta \rightarrow 0^{+}$ is implicit in Eqs.~(\ref{d1}-\ref{d4}). The spectral distribution
is determined from $\mathsf{G}_{i}$, obtained from the solution of Eqs.~(\ref{d1}) and (\ref{d3})
for  $z \in \sigma_{\rm ac}$.
The limit $\eta \rightarrow 0^{+}$ can be analytically performed only for some simple models. In general, for  $z \in \sigma_{\rm ac}$, one has
to solve numerically Eqs.~(\ref{d1}) and (\ref{d3}) for small but finite regularizer $\eta$.
The situation is different when $z$ lies outside the support of the continuous spectrum. In this case, the resolvent
matrices 
assume the form
\begin{eqnarray}
  \mathsf{G}_{j} = \left(\begin{array}{cc} 0 &  -G_j^{*} \\ -G_j & 0  
  \end{array}\right), \qquad
   \mathsf{G}_{j}^{(\ell)} = \left(\begin{array}{cc} 0 &  -(G_j^{(\ell)})^{*}  \\ -G_j^{(\ell)} & 0  
   \end{array}\right),
   \label{hjpa2}
\end{eqnarray}
for $\eta \rightarrow 0^+$. The above matrices solve Eqs.~(\ref{d1}) and (\ref{d3}) provided
 $G_j^{(\ell)}$  and $G_j$ fulfil
\begin{eqnarray}
  G_j = \frac{1}{A_{jj} - z  -  \sum_{k \in \partial_j} A_{jk}  A_{kj}   G_k^{(j)} },  \label{hj1} \\
  G_j^{(\ell)} = \frac{1}{A_{jj} - z  -  \sum_{k \in \partial_j \setminus \{ \ell \} } A_{jk} A_{kj}  G_k^{(j)}  }.
  \label{hj2}
\end{eqnarray}  
Substituting Eq.~(\ref{hjpa2}) in Eqs.~(\ref{d2}) and (\ref{d4}) and taking the limit $\eta \rightarrow 0^{+}$, we obtain
the following recursive equations for the eigenvector components $\{ R_j \}_{j=1,\dots.n}$ and $\{L_j \}_{j=1,\dots.n}$
\begin{eqnarray}
  R_j = - G_j  \sum_{k \in \partial_j} A_{jk} r_k^{(j)}, \label{n1} \\
  L_j = - G_j^{*}  \sum_{k \in \partial_j} A_{kj}^{*} L_k^{(j)},\label{n2}  \\
  R_j^{(\ell)} = - G_j^{(\ell)}   \sum_{k \in \partial_j \setminus \{ \ell \}} A_{kj} r_k^{(j)},\label{n3}  \\
  L_j^{(\ell)} = - \left( G_j^{(\ell)} \right)^* \sum_{k \in \partial_j\setminus \{ \ell \}} A_{kj}^{*} L_k^{(j)},\label{n4} 
\end{eqnarray}  
with $\ell \in \partial_j$, and the definitions $ R_j^{(\ell)} \equiv i G_j^{(\ell)} \left[ \mathsf{H}_{j}^{(\ell)} \right]_1$
and $ L_j^{(\ell)} \equiv i  \left( G_j^{(\ell)} \right)^*  \left[ \mathsf{H}_{j}^{(\ell)} \right]_2$. By comparing Eqs.~(\ref{kl1}) and (\ref{ghaap}), we
identify $G_j = \left[ \mathbf{G}_{\mathbf{A}}(z) \right]_{jj}$ ($j=1,\dots,n$). Thus, the solution of
Eqs. (\ref{hj1}) and  (\ref{hj2}) yields the diagonal part of the resolvent in the regime $z \notin \sigma_{\rm ac}$.
According to the graph definitions introduced in Section \ref{SparseDef}, we have that $A_{jj} = D_{jj}$ ($j=1,\dots,n$) and
$A_{jk} = J_{jk}$ ($j \neq k$), provided nodes $j$ and $k$ are adjacent. An outlier eigenvalue is located by finding $z$ such
that Eqs.~(\ref{n1}-\ref{n4}) have a nontrivial and normalisable solution for the eigenvector components.

\newpage
\section*{References}
\bibliographystyle{ieeetr} 
\bibliography{bibliography}

\end{document}